\documentclass[journal = jpcbfk, manuscript=article]{achemso}
\setkeys{acs}{articletitle = true}
\usepackage[version=4]{mhchem} 
\usepackage[T1]{fontenc}       

\usepackage{hyperref}
    \AtBeginDocument{}%
\usepackage{amssymb}
\usepackage{graphicx}
\usepackage{epstopdf}
\usepackage{cleveref}
\usepackage{booktabs, array}
\usepackage{multirow}
\usepackage{mathtools, bm, nicefrac, dsfont}
\usepackage{subcaption}
\usepackage{siunitx}
\usepackage{afterpage}
\usepackage{tikz}
	\usetikzlibrary{automata,positioning,shapes,decorations,tikzmark}
\usepackage{chemfig}
\usepackage{siunitx}
\usepackage{bigstrut}
\usepackage{calculator}
\usepackage{ifthen}
\usepackage{enumitem}

\definecolor{MyBlue}{RGB}{55,126,184}
\definecolor{MyRed}{RGB}{228,26,28}
\definecolor{MyGreen}{RGB}{77,175,74}

\newcommand{\np}[0]{NequIP}

\newcommand{\sinfo}[0]{Supporting Information}

\newcommand{\newil}[0]{\ce{[F-OMIM]^+[C_4F_9CO_2]^-}}

\newcommand{\oldil}[0]{\ce{[EMIM]^+[PF_6]^-}}

\newcommand{\lipf}[0]{\ce{Li^+[PF_6]^-}}

\newcommand{\angstrom}{\mbox{\normalfont\AA}}

\author{Zachary A. H. Goodwin}
\affiliation{John A. Paulson School of Engineering and Applied Sciences, Harvard University, Cambridge, MA 02138, USA}
\email{zgoodwin@seas.harvard.edu}

\author{Malia B. Wenny}
\affiliation{Department of Chemistry and Chemical Biology, Harvard University, Cambridge, MA 02138, USA}

\author{Julia H. Yang}
\affiliation{John A. Paulson School of Engineering and Applied Sciences, Harvard University, Cambridge, MA 02138, USA}
\alsoaffiliation{Harvard University Center for the Environment, 26 Oxford St., Cambridge, MA 02138}

\author{Andrea Cepellotti}
\affiliation{John A. Paulson School of Engineering and Applied Sciences, Harvard University, Cambridge, MA 02138, USA}

\author{Jingxuan Ding}
\affiliation{John A. Paulson School of Engineering and Applied Sciences, Harvard University, Cambridge, MA 02138, USA}

\author{Kyle Bystrom}
\affiliation{John A. Paulson School of Engineering and Applied Sciences, Harvard University, Cambridge, MA 02138, USA}

\author{Blake R. Duschatko}
\affiliation{John A. Paulson School of Engineering and Applied Sciences, Harvard University, Cambridge, MA 02138, USA}

\author{Anders Johansson}
\affiliation{John A. Paulson School of Engineering and Applied Sciences, Harvard University, Cambridge, MA 02138, USA}

\author{Lixin Sun}
\affiliation{John A. Paulson School of Engineering and Applied Sciences, Harvard University, Cambridge, MA 02138, USA}

\author{Simon Batzner}
\affiliation{John A. Paulson School of Engineering and Applied Sciences, Harvard University, Cambridge, MA 02138, USA}

\author{Albert Musaelian}
\affiliation{John A. Paulson School of Engineering and Applied Sciences, Harvard University, Cambridge, MA 02138, USA}

\author{Jarad A. Mason}
\affiliation{Department of Chemistry and Chemical Biology, Harvard University, Cambridge, MA 02138, USA}

\author{Boris Kozinsky}
\affiliation{John A. Paulson School of Engineering and Applied Sciences, Harvard University, Cambridge, MA 02138, USA}
\alsoaffiliation{Robert Bosch LLC, Research and Technology Center, Cambridge, MA 02142, USA}
\email{bkoz@seas.harvard.edu}

\author{Nicola Molinari}
\affiliation{John A. Paulson School of Engineering and Applied Sciences, Harvard University, Cambridge, MA 02138, USA}
\alsoaffiliation{Robert Bosch LLC, Research and Technology Center, Cambridge, MA 02142, USA}
\email{nmolinari@seas.harvard.edu}

\title{Transferability and Accuracy of Ionic Liquid Simulations with Equivariant Machine Learning Interatomic Potentials}
\abbreviations{}
\keywords{}

\begin{document}



\begin{abstract}

Ionic liquids (ILs) are an exciting class of electrolytes finding applications in many areas from energy storage to solvents, where they have been touted as ``designer solvents'' as they can be mixed to precisely tailor the physiochemical properties. As using machine learning interatomic potentials (MLIPs) to simulate ILs is still relatively unexplored, several questions need to be answered to see if MLIPs can be transformative for ILs. Since ILs are often not pure, but are either mixed together or contain additives, we first demonstrate that a MLIP can be trained to be compositionally transferable, i.e., the MLIP can be applied to mixtures of ions not directly trained on, whilst only being trained on a few mixtures of the same ions. We also investigate the accuracy of MLIPs for a novel IL, which we experimentally synthesize and characterize. Our MLIP trained on $\sim$200 DFT frames is in reasonable agreement with our experiments and DFT.
\end{abstract}

\newpage

Ionic liquids (ILs) are a unique and highly promising class of electrolytes that contain no solvent in their neat form~\cite{Welton1999,Hermann2008,Hallett2011}. This lack of solvent, such as organic carbonates or water, makes ILs advantageous for numerous applications from supercapacitors/batteries~\cite{Fedorov2014,Watanabe2017,Yao2022}, to gas storage~\cite{Hu2011,Wenny2022}, to themselves acting as ``green solvents'' for chemical reactions~\cite{Welton1999,Hallett2011}. These applications benefit from the use of ILs because ILs have extremely low vapor pressure, are non-flammable and can withstand large voltages without decomposing~\cite{Welton1999,Hermann2008,Hallett2011,Fedorov2014}, which are a set of properties that water and organic carbonates do not possess. The unique properties of ILs come, in part, from the large, highly asymmetric and ionic nature of the molecular species which comprise the IL~\cite{Welton1999,Hermann2008,Hallett2011,Fedorov2014}. Moreover, ILs can be mixed to tune desired physiochemical properties, creating a class of ``designer solvents'' with a huge chemical space~\cite{Niedermeyer2012}.


As ILs are often comprised of large, complicated, molecular ions, this has meant atomistic simulations are necessary for quantitative predictions of these concentrated electrolytes~\cite{Fedorov2014,Izgorodina2017,Dong2017,Jeanmairet2022,Yao2022}. Typically, classical molecular dynamics (MD) has been used to simulate ILs and calculate physiochemical properties~\cite{Fedorov2014,Jeanmairet2022,McEldrew2021corr,Yao2022}, largely because 1,000$+$ atoms can routinely be simulated for 1$+$~ns, which are the length and time scales typically required for ILs. The accuracy of the predictions from classical force fields is often under question, however~\cite{Dajnowicz2022}, which motivates some to use \textit{ab initio} MD (AIMD) to simulate ILs~\cite{Yao2022}. However, running density functional theory (DFT) is prohibitively expensive in comparison, which limits the simulations to short time scales ($\sim100$~ps) and system sizes ($\sim$100s of atoms)~\cite{Yao2022}. 


Machine learning interatomic potentials (MLIPs) have promised a way to bridge this efficiency-accuracy gap in atomistic simulations~\cite{Behler2019,Deringer2019,Unke2021,Deringer2021,Batzner2023,Yao2022}. In the context of ILs, there have been several works which develop MLIPs for specific electrolytes, such as the work of Montes-Campos \textit{et al.}~\cite{Campos2022}, Dajnowics \textit{et al.}~\cite{Dajnowicz2022} and Ling \textit{et al.}~\cite{Ling2023} for room temperature ILs, and the work of Tovey \textit{et al.}~\cite{Tovey2020}, Li \textit{et al.}~\cite{Li2021CP}, Mondal \textit{et al.}~\cite{Mondal2023} and Liang \textit{et al.}~\cite{Liang2020} for molten salts. These works often employ large datasets, e.g., with 1,000$+$ DFT calculations, mainly focus on the NVT ensemble, and typically only developed a model for one neat IL or molten salt. With the recent development of equivariant MLIPs, such as NequIP/Allegro~\cite{batzner2021se,Musaelian2023,Musaelian2023bio} and MACE~\cite{Batatia2023}, training a MLIP for a specific application has become significantly more accurate and data efficient. Hence, using equivariant MLIPs~\cite{batzner2021se,Musaelian2023,Musaelian2023bio,Batatia2023} to simulate ILs should provide an unprecedented accuracy/efficiency trade-off, but there are questions which remain to be answered if these MLIPs are to transform IL simulations and bridge this gap.


%
In this work, we aim to answer how to train a compositionally transferable MLIP, and establish the accuracy of a MLIP for a novel IL with relatively little training data. To test the compositional transferability, we choose to study a common salt-in-IL, \lipf{} in \oldil{}, which has recently attracted attention because the Li$^+$ has been observed to have a negative transference number at low salt molar fractions, as has been well studied from classical MD~\cite{molinari2019general,Molinari2019anomalies,McEldrew2021,Goodwin23PRXE} and experiments~\cite{gouverneur2018negative,Marc2021}. We demonstrate that \np{}/Allegro can learn to be transferable, i.e., having a model with similar accuracy for compositions not trained on as those that were trained on, whilst only being trained on several compositions, \textit{but only if these compositions are carefully chosen}. Of the routes tested, we find that at least 2 compositions are required, and the best approach is to train on high-entropy Li-salt molar fractions, e.g., 0.4-0.6, as these compositions ensure interactions are well sampled. Secondly, we synthesize and experimentally characterize a novel IL, \newil{}, which is similar to other ILs of interest for O$_2$ solubility~\cite{Wenny2022,Wenny2023}. We investigate how to train an accurate Allegro model with $\sim$\num{200} DFT frames, capable of capturing experimental densities, thermal expansion and isothermal compressibility of this novel \newil{}, with the main limitation being the approximations of our chosen DFT settings. For \newil{} we also investigate the role of how the structures are generated, and perform a hyperparameter search to optimize the model.



%
%
%
%

In the \sinfo{} (SI), we provide the experimental details for the synthesis and characterization of \newil{} (Section 1)~\cite{Wenny2022,Wenny2023}, and additional details for the computational methods employed that are briefly described below (Sections 2-5).



%
%
%

We use classical MD simulations to generate a diverse set of uncorrelated structures of the two ILs studied here, \newil{}, and \lipf{} in \oldil{}. Since the MD simulations for these two ILs are almost identical, we describe our method and point out any differences between them. The procedure to generate the initial, equilibrated structures is outlined in the SI Section 1~\cite{fadel2019role, molinari2019general, molinari2020chelation, Molinari2019anomalies,molinari2016molecular, molinari2019general}.


The classical MD simulations have periodic boundary conditions, use a timestep of $\delta t = \SI{1.0}{\femto\second}$, a velocity-Verlet algorithm to evolve the equations of motion, a Nos\'e-Hoover barostat (1000~$\delta t$ coupling) and thermostat (100~$\delta t$ coupling) to enforce pressure and temperature, respectively~\cite{hoover1985canonical, nose1984unified, hoover1986constant}, and are performed using the LAMMPS code~\cite{plimpton1995fast,LAMMPS2022}. The atomic interactions are modelled with OPLS-AA~\cite{jorgensen1988opls} with parameters obtained from Ref~\citenum{doherty2017revisiting}. For \lipf{} in \oldil{}, to reproduce realistic densities, we rescale the atomic charges to \SI{80}{\percent} of the original value~\cite{doherty2017revisiting, mogurampelly2017structure, pal2017effects, molinari2019general}. Whereas, for \newil{} we rescale the atomic charges by \SI{70}{\percent} to match our experimental data better (see SI Section 4). 


For \lipf{} in \oldil{} we investigate Li-salt molar fractions ranging from \num{0.0} to \num{1.0}, in increments of \num{0.1}, with 20 ion pairs in the simulation box for each composition. For each composition, two initial structures were generated, one for training and validation, and another for testing the trained models. To generate these datasets, each structure is evolved for \SI{80}{\nano\second} in the NVT ensemble at \SI{300}{\kelvin}. The atomic positions and force components are saved every \SI{20}{\pico\second}, to generate a total of \num{4000} snapshots for each composition. 


%
%

%
For \newil{}, we simulate three different temperatures (\num{300},\num{450},\SI{600}{\kelvin}), in the NPT ensemble with a pressure of 1~atm, with 10 ion pairs in the simulation box. For every new temperature, we first let the system equilibrate for \SI{1}{\nano\second}, and then run for \SI{6}{\nano\second} during which snapshots of the structure are saved every \SI{20}{\pico\second} to ensure sufficient decorrelation. Thus, we generate a total of \num{900} snapshots, \num{300} per temperature.






To obtain accurate energy, force and stress data, we performed density functional theory (DFT) calculations using Quantum Espresso~\cite{Giannozzi2009,Giannozzi2017}. We utilize the PBE exchange-correlation functional~\cite{Perdew1996}, D3~\cite{Grimme2010} dispersion correction for van der Waals interactions, and use the ultra-soft pseudo-potentials from Ref.~\citenum{Garrity2014}. We use $\Gamma$-point sampling, a cut-off of 50~Ry for the wavefunctions, a cut-off of 500~Ry for the charge density, and a convergence threshold of 10$^{-10}$~Ry. 


For \lipf{} in \oldil{}, we performed DFT calculations on 50 frames (equally spaced to maximize decorrelation) from each composition and run. For \newil{}, we initially selected 60 frames (again, equally spaced) at each temperature to generate a small DFT training dataset. Starting from these structures, we also performed fixed-cell AIMD on each structure for 50 steps (with a $\sim$1~fs timestep) at the corresponding temperatures (using the Verlet algorithm and a tolerance for rescaling velocities of 10~K) and also 10 steps of fixed-cell relaxation for each structure. We also calculated the single atom energies of all the elements in these frames, from only keeping each element in a box with length of $\sim$30~\angstrom.



We use the state-of-the-art graph neural network MLIPs, \np{}~\cite{batzner2021se} and Allegro~\cite{Musaelian2023,Musaelian2023bio}, to train, validate and run the simulations for these ILs. Initially, \np{} was used to perform the transferability test on the SiIL. \np{} is an equivariant message-passing graph neural network with unprecedented accuracy, but because of its message passing architecture it cannot be scaled to large systems efficiently. Allegro is a simplification of \np{}, without the message passing architecture because of the local cut-off, that can be scaled to large systems efficiently. Therefore, we also performed a transferability test with Allegro, and only used Allegro to run simulations in LAMMPS. We refer the readers to Refs.~\citenum{batzner2021se,Musaelian2023,Musaelian2023bio} for an extensive explanation of the network architecture and the importance of equivariant models~\cite{Batzner2023}.


%
%

For the classical dataset of \lipf{} in \oldil{}, to perform the initial transferability test with \np{}, we use $r_{max} = 5$, $l_{max} = 1$ with irreps 16x0o + 16x0e + 8x1o + 8x1e, with 5 layers, and use a learning rate of \num{0.05} with a batch size of \num{5}. For each composition we use 650 frames (equally spaced), 600 frames for training and 50 frames for validation, and when multiple compositions are trained on these datasets are combined, increasing the total number of frames trained on. The test set was an independent set of 650 frames at each composition. While clearly the models will not possess a greater accuracy than that of the classical force field; here we do not aim for accurate models, but rather aim to understand the intricacies of training a transferable model for ILs.


To perform the transferability tests with Allegro, we choose $r_{max} = 6$, $l_{max} = 2$, $n_{layers} = 2$, learning rate of 0.0005, batch size of 1, 80-20\% split of training validation data, and find that it is essential to employ single atom energies when training on multiple compositions with DFT data from Quantum Espresso. Only 50 frames (equally spaced) from each of the training/validation and test sets were selected to form the DFT calculations for the test. 

Note that this transferability investigation using test errors only allows us to identify which models are certainly failing from large errors. As good errors do not provide a conclusive measure of if a MLIP is stable and reliable when being run, further testing will be needed if these models are to be used to calculate physiochemical properties (see SI Section 5 for the protocol of \newil{}).


%
%
%


The results shown for \newil{} in the main text are based on a single Allegro model (see SI Section 5). We choose $r_{max} = 7$, $l_{max} = 2$, $n_{layers} = 2$, LR = 0.001, batch size of 4, stress weight of 100 (with force and energy of 1), and 80-20\% split of training validation data.


To run these models, we employ LAMMPS with a custom pair style. We use the Nos\'e-Hoover barostat (1000~$\delta t$ coupling) and thermostat (100~$\delta t$ coupling) to enforce pressure and temperature with a timestep of $\delta t = 0.5$~fs. To obtain the isothermal compressibility, we first equilibrated the system at 1~bar for 50~ps. Then the pressure was increased by 1~kbar and equilibrated again for 50~ps before running for another 50~ps at this pressure for production; this process was repeated 18 times to obtain how the equilibrium volume changed with pressure, which was then fit to the Tait equation of state to obtain the isothermal compressibility~\cite{Wenny2022}. Details of the other settings used for each calculation will accompany the results shown.

As outlined earlier, one of the most valuable properties of ILs is their ability to be mixed together to form ``designer liquids'' with physiochemical properties that can be precisely tuned~\cite{Niedermeyer2012}. Therefore, being able to simulate complex mixtures of ILs, and solutes, is a necessity. Typically, however, MLIPs are only trained and tested on a specific composition. This is because training older MLIP architectures built on invariant descriptors, such as GAP~\cite{Deringer2021} and DeePMD~\cite{Zhang2018}, require orders of magnitude more data, rendering them challenging enough to train for a single material. The recent advances in the accuracy and data-efficiency of MLIPs, largely from equivariant architectures, has allowed the scope of these methods to expand~\cite{Batzner2023}. One area that needs more quantitative attention is how to train a MLIP to be transferable, and still be accurate enough to reliably use for calculations on compositions that it was not trained on. With the recent interest in ``foundational'' MLIPs~\cite{Batatia2023FM}, this question is becoming even more important, and having quantitative, systematic tests of transferability on controlled datasets is valuable for these efforts.

\begin{figure*}
\centering
\includegraphics[width= 1\textwidth]{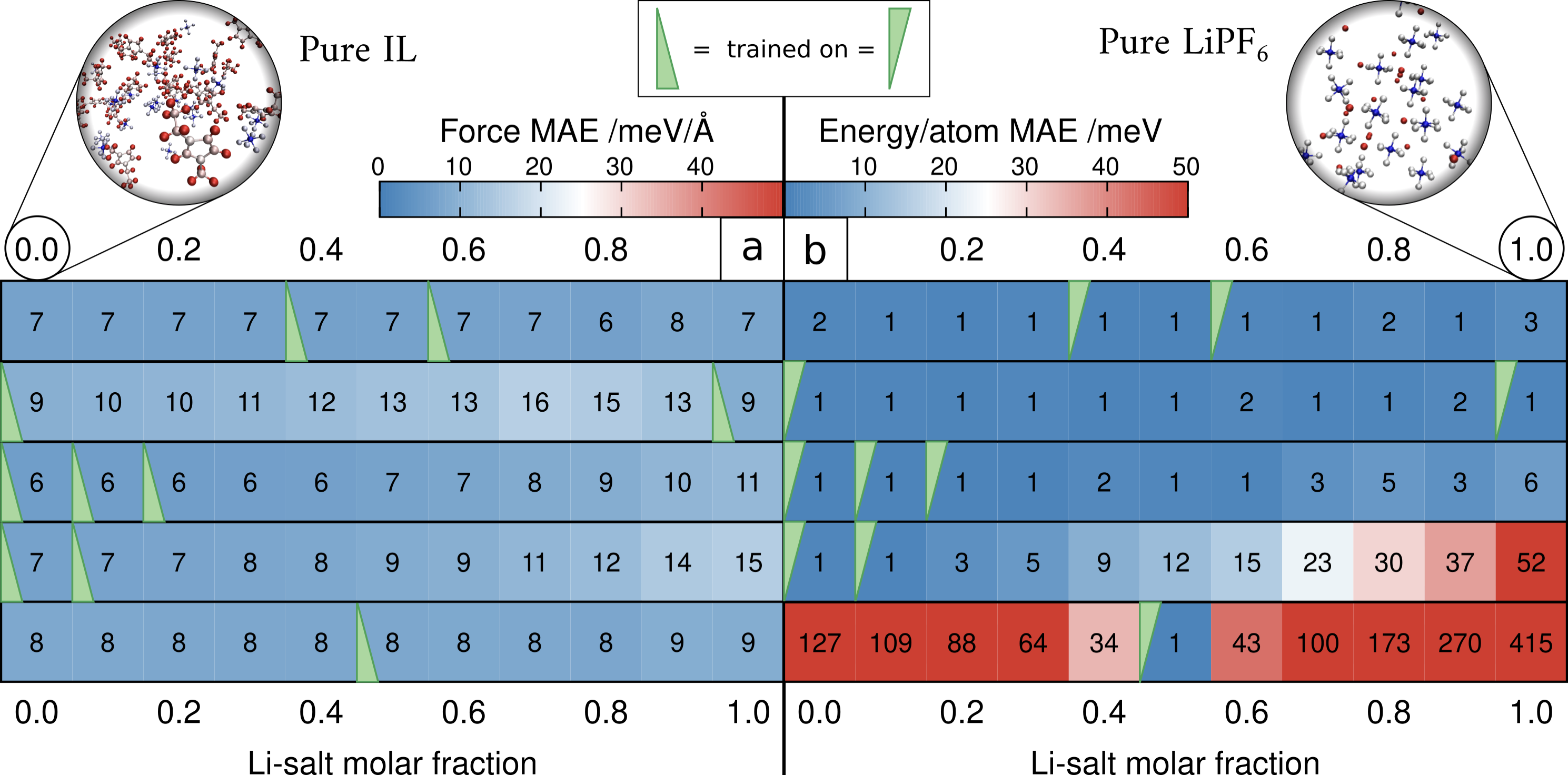}
\caption{Transferability investigation of \np{} for a typical salt-in-IL, \lipf{} in \oldil{}, trained on classical data. Each row corresponds to a different \np{} model trained on the molar fractions shown by the green triangles. Each number at each molar fraction is the mean absolute error (MAE) on an independent test set for forces, as shown on the left (a) in units of meV\angstrom$^{-1}$, and energy, on the right (b) in units of meV/atom.}
\label{fig:CT}
\end{figure*}

With this question in mind, we choose to study the transferability of a well-known salt-in-IL (SiIL), \lipf{} in \oldil{}, the results of which are shown in Fig.~\ref{fig:CT}. Each row represents a different \np{} model trained on classical data of specific molar fractions, as indicated by the green triangles, with the values at each molar fraction (in each box) being the test error on an independent dataset. On the left is the force mean absolute error (MAE) in meV\angstrom$^{-1}$ and on the right is the energy MAE in meV/atom. In what follows, we describe the results and implications of Fig.~\ref{fig:CT}.

If a \np{} model is trained on a single composition at a Li-salt molar fraction of 0.5, we find the force MAE to be 8~meV\angstrom$^{-1}$ and the energy MAE to be 1~meV/atom at the same mole fraction. As we are using a large dataset from a classical force field, these error values are considered as the baseline, and we will not consider their values to be significant. For compositions away from the 0.5 molar fraction trained on the force MAE remains relatively constant, but the energy MAE increases to $100-400$~meV/atom at the ends of the composition range. Therefore, training a \np{} model only on one composition works for the composition trained on, as is often done for MLIP, but the energy transferability of the \np{} model becomes worse the further out-of-distribution the composition becomes. 

%
%
%
%
%


Alternatively, multiple compositions could be trained on, which hopefully provides better transferability. The two rows above are examples of this strategy for adjacent compositions, trained on 2 and 3 compositions, comprising of the 0,0.1 and 0,0.1,0.2 Li-salt molar fraction compositions, respectively. Again, the force MAEs are relatively constant across the entire composition range, albeit increasing to values larger than 10~meV\angstrom$^{-1}$ (approximately $\times 2$ those trained on) at 0.9-1. The increase in force MAEs can be traced back to the increase in force errors of \lipf{}, with the forces of Li$^+$ becoming significantly worse. This is perhaps not surprising as the interactions of \lipf{} have not been sampled well. For the \np{} model trained on 2 compositions (0 and 0.1), the energy MAE increases to 30-50~meV/atom at 0.8-1 molar fraction, which is better than the model trained on a single composition, but still corresponds to a $\times 50$ increase. The model trained on 3 compositions (0-0.2), has similar trends, but the errors are slightly reduced in magnitude, and only corresponds to a $\times 6$ increase in energy MAE at 1.0 Li-salt molar fraction.

Finally, if we train on 2 compositions, but not at one end of the composition range, we can obtain energy and force MAEs that are typically within $\times 2$ of the MAEs of the compositions trained on. Training on the Li-salt molar fractions of 0 and 1 gives energy MAEs within $\times 2-3$ over the entire composition range, and the force errors also remain within this limit. The increase in the force error for this model at intermediate Li-salt molar fractions can be traced back to an increase in the error of all elements, which could be rooted in the fact that no Li$^{+}$-[EMIM]$^+$ interactions are included in the dataset. Overall, training on molar fractions of 0.4 and 0.6 results in a model with force and energy MAEs which are roughly constant across the entire composition range, i.e., within $\times 2$ (apart from 1.0 Li-salt molar fraction compositions), which appears to be the best approach for training a transferable \np{} model for this system. 



%
%
%
%
%


In the SI Section 3, we show this same transferability test performed with Allegro, where we find qualitatively exactly the same results. To further ensure these conclusions are robust, we performed a similar analysis where we took only 50 frames from the training/validation set and 50 frames from the test set and calculated the \textit{ab initio} energies and forces to train an Allegro model (see SI Section 3). We find the conclusions drawn from the previous tests with classical force field data are robust, with the transferability investigation on only a small amount of DFT data to be qualitatively identical. \textit{Therefore, such transferability tests can be performed with classical force field data first, before proceeding with expensive DFT calculations}. In addition, in the SI Section 3 we show a transferability test for an example class of Li-P-S-Cl solid electrolyte, and find the conclusions are again robust, further supporting our findings.


\np{} and Allegro learn directly on the force and energy values, with forces being atom-based quantities and the energy being the total energy of the simulation box. We find that the forces predicted by \np{} and Allegro generalize well, even if only 1 composition is sampled, but only provided interactions between species have been well sampled. This can be achieved from sampling from the middle of the composition range (0.4-0.6), or ``high-entropy'' configurations. Whereas, if we only sample from ends of the composition, the training data does not include enough cation-cation interactions, for example. The sampling of frames and compositions to learn forces more accurately could be achieved through employing uncertainty quantification~\cite{Zhu2023}, as is typically used when performing active learning~\cite{Vandermause2020,Xie2023}.



In contrast, the generalization of the energy appears to be significantly more dependent on how many compositions are selected, which could be rooted in how MLIPs decompose the total energy into a sum of atomic energies, which is an uncontrolled approximation, at least in terms of the MLIP~\cite{Behler2019,Yoo19E,Chong2023}. To understand these observations, and generalize our results, we perform a simple analysis. Let's assume that there is just a single training frame for each composition. For the model trained on the Li-salt molar fraction of 0.5, let's assume the total energy can be written 
\begin{equation}
    E_{0.5} = \mu_{c1}N_{c1,0.5} + \mu_{c2}N_{c2,0.5} + \mu_{a}N_{a,0.5},
\end{equation}

\noindent where $\mu_j$ and $N_j$ are the energy per ion and the number of ions, respectively. Note this simple, linear equation assumes no ions interact (i.e., the electrolyte is an ideal mixture), but this can still provide us insight into the energy generalization. The $N_j$ values are known from the training data, but we are using the MLIP (we used \np{}/Allegro here, but the results are expected to apply to other methods~\cite{Batatia2023FM}) to find the $\mu_j$'s. As we only have 1 equation but 3 unknowns, we are under-determined, and no unique solution exits (as an infinite set of solutions are valid). Therefore, the MLIP solves for a linearly dependent set of values for $\mu_j$ which satisfy $E_{0.5}$, but the MLIP lacks constraints to reproduce $E$ for other compositions. Thus, the energy generalizes poorly for this case (as clearly seen in the SI Section 3).





Instead, for example, we train on the best case of Li-salt molar fractions of $0.4+0.6$, and because $N_a = 20$ in all frames from the salts sharing the same anion, we can write the equations as 
\begin{equation}
    E_{0.4} - \mu_{a}N_a = \mu_{c1}N_{c1,0.4} + \mu_{c2}N_{c2,0.4},
\end{equation}

\noindent and 
\begin{equation}
    E_{0.6} - \mu_{a}N_a = \mu_{c1}N_{c1,0.6} + \mu_{c2}N_{c2,0.6}.
\end{equation}

\noindent From these equations, we can determine $\mu_{c1}$ and $\mu_{c2}$ up to some arbitrary constant incorporated in $\mu_a$. If fact, for this SiIL system, the only quantity which is well defined is the energy difference between the cations, and it should not be possible to reliably determine all $\mu$'s without arbitrary constants. Therefore, with only 2 compositions to train on, the MLIP can learn how the energy should change with composition reasonably well. This logic explains why training on 2 compositions instead of 1 significantly helps, but training on 3 instead of 2 does not substantially improve learning the energy to be generalizable, as the 3rd equation does not allow all the $\mu$'s to be determined uniquely. 


If this is the case, we should find that the predictions of the energy of Li$^{+}$, for example, should not necessarily be the same across different models trained, but that the energy difference between Li$^+$ and [EMIM]$^+$ should be model independent (if trained on 2 or more compositions). We confirm with Allegro on DFT data that this is the case (at least approximately), as shown in the SI Section 3. 

%
%
%
%
%


%
%
%
%
%

%
%
Overall, it appears that the MLIPs struggle most with finding the linear, constant shift contributions to the energy from different species, but the methods can typically capture non-linearities arising from local interactions. These observations allow us to generalize the results obtained here to have some simple guidelines for training complex (ionic) mixtures. For an $N$-component, single phase, bulk system with $M$ constraints (such as electroneutrality), we are required to sample $C = N-M$ unique compositions to ensure that the MLIP can learn how the total energy should (at least approximately) change with composition. Note that this condition sets the minimum number of compositions to sample to ensure the energy can be correctly learnt, and using more compositions should mainly contribute improvements through there being more data, provided all interactions/environments have been well sampled in those compositions. On top of this condition, it is preferred to use compositions which thoroughly sample all interactions, i.e., compositions with ``high entropy''. To further sample frames/compositions for the energy, uncertainty quantification could again be used~\cite{Zhu2023}, or the local predicted rigidity of compositions could be quantified~\cite{Chong2023}.

Here, we studied a 3-component system, but we had the constraint that the number of anions remained the same (to enforce electroneutrality), which meant we only needed 2 unique compositions to ensure the energy is transferable, even though we could only learn the energies of each ion up to some constant. If, for example, we had two ILs with different cations and anions which could be mixed together, we would again only need 2 compositions as we have 2 constraints of the electroneutrality of each salt, which means we can determine the energy of each salt, but not necessarily the ions within each salt. Again, sampling compositions near 0.5, the high entropy state, will allow interactions between ions to be well sampled. Consider again a case of a system with 2 types of cations and 2 types of anions, but there are 4 salts from the mixtures of cations-anions that we have access to. Even though we still require electroneutrality, we would require 4 compositions to determine the energy of each ion, as there are multiple ways of satisfying electroneutrality. This case should allow the energy of each ion to be determined without a constant, and should allow the MLIP to be even more transferable. This logic can be extended to other complex mixtures, not just electrolytes, and the results are expected to be robust and useful for anyone training bespoke models on complex mixtures using these methods. 
%
%
%




These results are not the first time transferable MLIPs have been investigated, but, as far as the authors are aware, it is the first time some general guidelines have been clearly outlined. In Ref.~\cite{Yoo19E}, Yoo \textit{et al.} noted that when developing a transferable model for mixtures of Ge-Te, training on a single mixture near a 1:1 ratio was not accurate, but when training on all the compositions studied (5 in total), satisfactory generalization was obtained. From the guidelines outlined here, we expect that only 2 compositions would be needed. In Ref.~\citenum{Rajni2022}, Chahal \textit{et al.} investigated a molten salt mixture, LiF-NaF-ZrF$_4$. They also found that training on a single composition did not produce adequate transferability, and reported training on 42-29-29 and 26-37-37 mole \% resulted in a satisfactory model. This model was tested on 38-51-11, 40-46-14 and 28-32-40, and it was demonstrated that their model could capture various properties, such as coordination numbers and diffusion coefficients. We expect their developed model has learnt forces to be transferable, owing to ``high-entropy'' mixtures being trained on, but as LiF-NaF-ZrF$_4$ could be considered as a 3-component salt mixture, we expect 3 compositions would be needed to learn the energy to be generalizable for different salt ratios. It is expected, however, that their model has learnt how the energy changes when changing the ratio of LiF to NaF-ZrF$_4$, owing to different LiF to NaF-ZrF$_4$ being sampled. As compositions relatively close to those trained on were tested, the errors incurred for the erroneous energy generalization would not be serious, and should not significantly alter their predictions for coordination numbers or diffusion coefficients, for example.

There has recently been significant efforts to develop ``general'' or ``foundational'' MLIPs~\cite{Chen2022,Deng2023,Merchant2023,Batatia2023FM,Musaelian2023bio,Kovacs2023,Yang24MatterSim} trained on large datasets, such as the Materials Project~\cite{Jain2013} or the SPICE dataset~\cite{Eastman2023}. We hope the quantitative approach taken here can be further developed and applied to these larger models, to reveal their limitations and guide improvements. In the context of electrolytes, Gong \textit{et al.}~\cite{Gong24Bamboo} have developed a large dataset, for carbonate-based Li-ion electrolytes, and shown some initial results for chemical transferability, but further tests should be performed on this dataset with the guidelines outlined here to understand the limitations of this dataset. 

As we have demonstrated that we can train a \np{}/Allegro model to be transferable, and some simple guidelines for how this can be achieved, we next move onto assessing the accuracy of Allegro when being applied to a novel system. 







As summarized previously, many ILs are excellent ``green solvents'' for reactions~\cite{Welton1999,Hallett2011,Hermann2008}. One property in this area that requires extremely accurate energy predictions is the solubility of solutes, as any error in the energy change becomes exponentially worse for the solubility prediction~\cite{Liu2014,Khawaja2017,Wenny2022}. For example, calculating the solubility of O$_2$ or CO$_2$ in ILs requires that these species are inserted into the IL and the energy change calculated. Classical force fields have typically been used to compute solubilities, but the energy of insertion of these molecular species is not at the same level of accuracy as DFT, which means the solubility computed from classical force fields is sometimes erroneous~\cite{Liu2014,Khawaja2017,Shi2014}. MLIPs promise a way to bridge this gap: they are DFT-accurate, but are still computationally cheap enough to perform such calculations~\cite{Fukushima2019,Jinnouchi2020}. 

\begin{figure*}
    \centering
    \includegraphics[width= 1\textwidth]{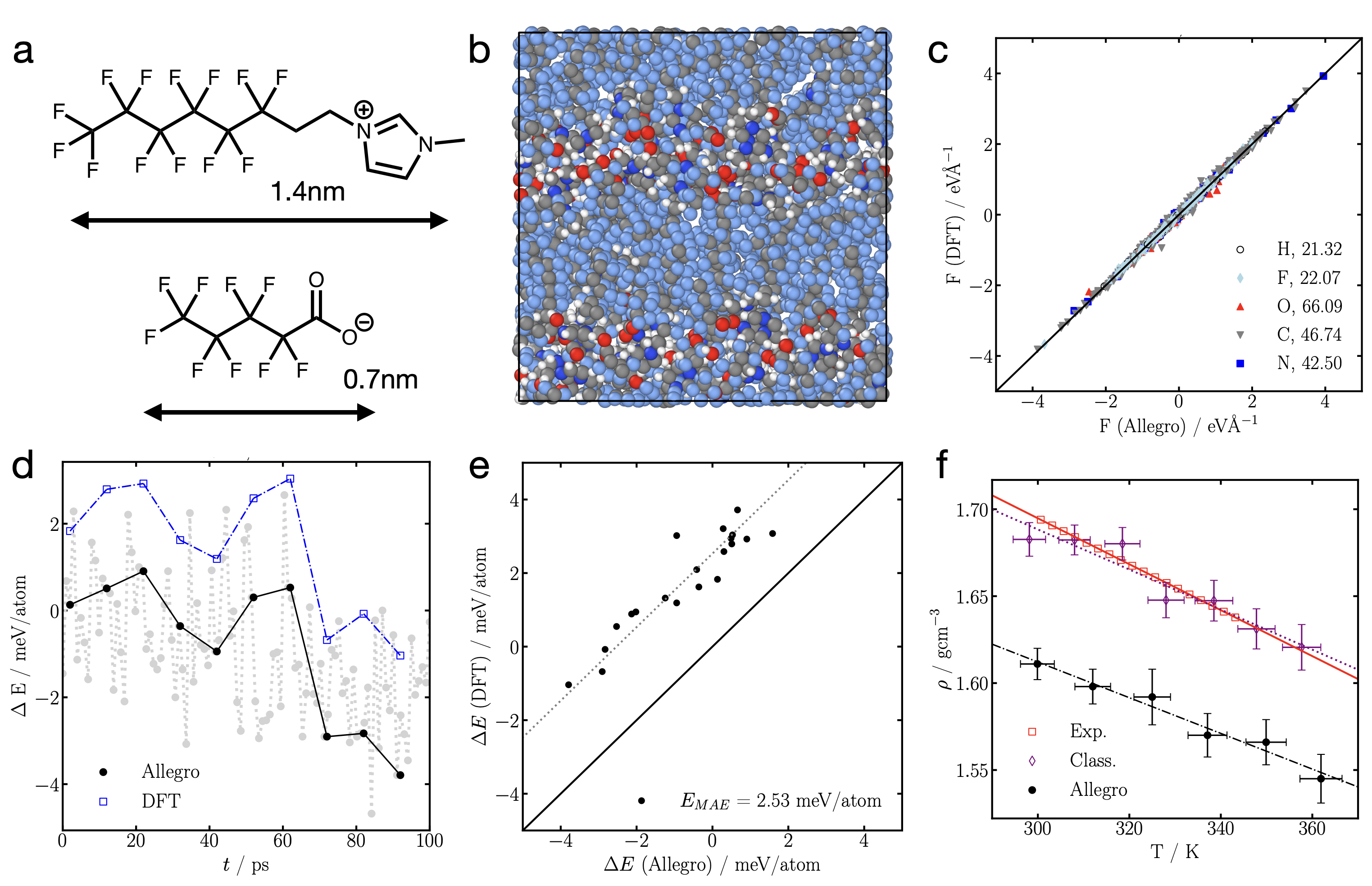}
    \caption{Allegro model for \newil{}. (a) - Schematic of \newil{}, with approximate scale of each ion. (b) - Structure of \newil{} showing its heterogeneous nature, generated using Ovito~\cite{Stukowski2010}. (c) - Force parity plot between Allegro and DFT for all the elements and components from a single frame at 50~ps through an Allegro simulation at 300~K. Force MAE of each element is shown in the legend in units of meV\angstrom$^{-1}$ (d) - Energy comparison between Allegro, and that calculated by DFT using the Allegro structure, as a function of time. The grey values are also Allegro, but were not used to calculate energy/forces in DFT. (e) - Energy parity plot between Allegro and DFT, half of the frames of which are shown in (d). (f) - Comparison of density as a function of temperature for the experiments, classical force field, and the Allegro model (see \sinfo{} Section 5 for details).}
    \label{fig:NIL}
\end{figure*}

To work towards this goal, we synthesize and characterize (details in SI Section 1) a novel IL \newil{}, the structure of which is shown in Fig.~\ref{fig:NIL}(a), that is of interest for its O$_2$ solubility~\cite{Wenny2022,Wenny2023}. As can be seen in Fig.~\ref{fig:NIL}(b), this IL has a highly heterogeneous, lamellar-like structure, which creates voids for O$_2$ to dissolve into~\cite{Wenny2022}. In Ref.~\citenum{Wenny2022}, the solubility of O$_2$ was correlated with thermal expansion and isothermal compressibility of the ILs, both of which are related to the formation of voids in the IL~\cite{Wenny2022}. Therefore, we compare the experimental density, thermal expansion and isothermal compressibility against the MLIP we develop for this novel IL.

We developed and tested the accuracy of an Allegro model for \newil{}, whilst only using a small amount of data and only performing 1 iterative training round at most, as it is expensive to keep (re-)training deep learning models. While developing this model, we investigated the role of how the dataset was generated, and hyperparameter optimization for the errors and stability of the model, full details of which can be found in the SI Sections 4-5. Here we only show results for one model. Specifically a model with $r_{max} = 7$, $l_{max} = 2$, $n_{layers} = 2$ (referred to as $722$), which proved to be reliable, with a scale factor of 100 on the stress loss (referred to as $11100$, as the scale of the energy and force are 1). Overall, we found using large $r_{max}$ was required for accurate predictions of the density, which is perhaps not surprising, as large $r_{max}$ captures more long-ranged electrostatics and dispersion interactions, and that $l_{max} =2$ was essential for a chemically stable model. We also found using the dataset of 180 frames generated from running short AIMD simulations, starting from the uncorrelated OPLS structures (see SI Sections 4-5), performed significantly better than training directly on the OPLS structures. Performing the short AIMD runs (only 50 steps) from the OPLS structures didn't dramatically alter the structures, but did lower the energy by $\sim$10~eV (with 530 atoms) and significantly reduced the pressure of the box, which suggests the quality of the data to be extremely important when in this light data regime. In addition, we included 40 frames of volume scans generated from a previous Allegro model. We validate the predictions of this model against DFT and also our experiments. 





To test this Allegro model, we run several short 100~ps NPT simulations with 530 atoms to collect 20 structure predictions (separated by $\sim$10~ps), and their corresponding energy and forces. In Fig.~\ref{fig:NIL}(c) we show a parity plot for the forces predicted by Allegro in comparison to the DFT forces computed from these structures. Overall, excellent agreement can be found, with force MAEs in the range of 20-70 meV\angstrom$^{-1}$. These errors are practically identical to the MAEs obtained validating the potential. Therefore, the Allegro model for \newil{} appears to have \textit{ab initio} accurate forces for this IL, with only a small amount of DFT data. In Fig.~\ref{fig:NIL}(d) we show how the energy evolves with the simulation from Allegro, and the frames picked out to test the energy and forces on. In Fig.~\ref{fig:NIL}(d), we also show the corresponding DFT energy calculated from those structures. Clearly, the DFT energy is higher than the Allegro model for all tested structures, by $\sim$2-4 meV/atom, which is several times the error obtained from the dataset that it is tested on, albeit still a respectable value for the amount of data. Importantly, the difference in energy does not get significantly worse with time. In Fig.~\ref{fig:NIL}(e) we further show this with an energy parity plot, where the DFT values are clearly larger than the Allegro energies. It is likely that this difference between Allegro and DFT originates from long-ranged interactions, as Allegro is a strictly local model~\cite{Musaelian2023}. Further development of these methods to include long-ranged interactions should improve the accuracy of the models~\cite{Behler2019,Yue2021LR,Zhang2022LR}.

In Fig.~\ref{fig:NIL}(f) we show the density predictions of our Allegro model for \newil{} at various temperatures, calculated from the last half of a 1~ns NPT run with 4240 atoms at 1~bar, and we also show the predictions from the classical force field and the experimental values. The experiment finds the density of \newil{} to be slightly less than 1.7 gcm$^{-3}$ at 300~K, decreasing with temperature to 1.64 gcm$^{-3}$ at 340~K. The values from the classical force field are extremely close to the experimental values because the charges were rescaled such that the density matched the experimental density (see SI Section 4). The density prediction from the Allegro model for \newil{} is 1.61 gcm$^{-3}$ at 300~K. The difference between the Allegro model and experiments could be attributed to the employed PBE exchange-correlation functional, as found by others~\cite{Liang2020,Li2021CP}. Despite this, the thermal expansion from Allegro, -1.03$\times 10^{-3}$gcm$^{-3}$K$^{-1}$, agrees reasonably well with that from experiments, -1.32$\times 10^{-3}$gcm$^{-3}$K$^{-1}$. Moreover, the isothermal compressibility of \newil{} is determined to be 8.6$\times 10^{-10}$~Pa$^{-1}$ at 298~K from experiments, and from fitting the Tait equation of state to the results from our Allegro model, we find a value of 6.46$\times 10^{-10}$~Pa$^{-1}$ at 300~K. While the thermal expansion and isothermal compression are not in perfect agreement with experiments, they are within what can be expected from the employed exchange-correlation functional, as others have found for different ILs and molten salts~\cite{Liang2020,Li2021CP}, which appears to be the main limitation. 


Overall, we have demonstrated that the equivariant MLIPs, NequIP and Allegro, can be used for data-efficient, transferable and accurate simulations of ILs and their mixtures. Specifically, we first showed that there exists a minimum number of compositions that need to be trained on to learn the energy to be transferable, using a salt-doped-IL as the test case. In contrast, forces could be learnt to be transferable with any number of compositions, but sampling ``high-entropy'' compositions often worked better than sampling pure compositions. Secondly, we demonstrated that these methods could be used for accurate simulations of a novel IL. Moreover, we showed that only $\sim$200 DFT frames with 1 iterative training loop were needed for reasonable robustness, demonstrating the data-efficient nature of the method. We did, however, find that how the data was generated was important, with frames generated from short AIMD runs being more reliable than frames generated from classical force fields. The main limitation of the method is the employed DFT settings, but now a method to accurately train them has been outlined, more accurate and expensive methods can be employed. 

\section*{Conflict of Interests}
The authors declare no competing financial interests.
%

\begin{suppinfo}
The \sinfo{} contains: Experimental synthesis and characterization, and additional simulation methods and results
\end{suppinfo}

\section{Acknowledgements}

We thank Rajni Chahal and Stephen Lam for discussions and confirming our predictions of the energy generalisation of their work is accurate. This work used beamline 12-ID-B at the Advanced Photon Source, a U.S. Department of Energy (DOE) Office of Science User Facility operated for the DOE Office of Science by Argonne National Laboratory under contract no. DE-AC02-06CH11357. This research was supported under a Multidisciplinary University Research Initiative, sponsored by the Department of the Navy, Office of Naval Research, under Grant N00014-20-1-2418. Work at Harvard by Z.A.H.G., J.H.Y., A.C., J.D., K.B., B.R.D, A.J., L.S., S.B., A.M., B.K., and N.M. was supported by Robert Bosch Research North America. J.H.Y. acknowledges funding from the Harvard University Center for the Environment. J.A.M. acknowledges support from the Arnold and Mabel Beckman Foundation through a Beckman Young Investigator grant. B. R. D. was supported by a NASA Space Technology Graduate Research Opportunity, under grant number 80NSSC20K1189. A.J. was supported by a Aker Scholarship.

\onecolumn

\section{Supporting Information}

\renewcommand{\theequation}{S\arabic{equation}}
\setcounter{equation}{0}

\renewcommand{\thetable}{S\arabic{table}}
\setcounter{table}{0}

\renewcommand{\thefigure}{S\arabic{figure}}
\setcounter{figure}{0}

\subsection{1. Experimental Synthesis and Characterisation}
\label{sec:EXP}

\subsubsection{Synthesis}

\begin{figure}[h!]
    \centering
    \includegraphics[scale=1]{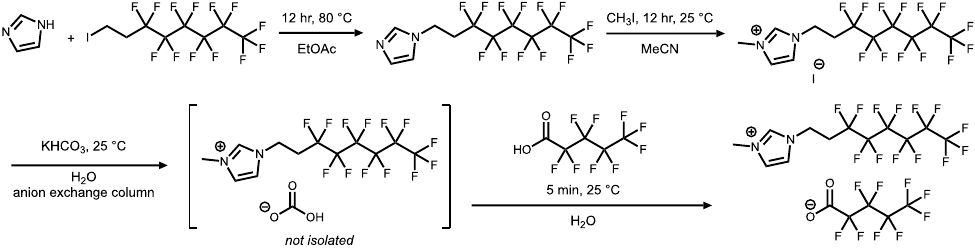}
    \caption{Experimental synthesis schematic.}
    \label{fig:EXP_SYN}
\end{figure}

We synthesized \newil{} according to Ref.~\citenum{Wenny2023}, which relies on a bicarbonate intermediate reacting with perfluoropentanoic acid to form the desired ionic liquid (IL), with concomitant formation of CO$_2$ and H$_2$O (Fig.~\ref{fig:EXP_SYN}). Briefly, 1,1,1,2,2,3,3,4,4,5,5,6,6-tridecafluoro-8-iodooctane and imidazole were reacted to form 1-(3,3,4,4,5,5,6,6,7,7,8,8,8-tridecafluorooctyl)~imidazole. This precursor was methylated with methyl iodide to form 1-(3,3,4,4,5,5,6,6,7,7,8,8,8-~tridecafluorooctyl)-3-methylimidazolium iodide. A bicarbonate-loaded anion exchange column was then used to generate 1-(3,3,4,4,5,5,6,6,7,7,8,8,8-tridecafluorooctyl)-3-methylimidazolium bicarbonate as an un-isolated intermediate. Reaction of this intermediate with perfluoropentanoic acid in water resulted in the formation of 1-(3,3,4,4,5,5,6,6,7,7,8,8,8-tridecafluorooctyl)-3-methylimidazolium perfluoropentanoate (\newil{}) as a pale-yellow liquid. Complete experimental details, including 1H and 19F NMR data consistent with the formation of the target IL, are included in the supporting information of Ref.~\citenum{Wenny2023}.

A residual iodide content of 1492~ppm was measured via ion chromatography by Galbraith Laboratories. A residual water content of 3800~ppm was measured via Karl Fisher titration on a Mettler Toledo C20 Coulometer. Note that the Karl Fisher titration was performed on a sample that was dried under vacuum for $\sim$5 h at ambient temperature, and the sample had to be exposed to air prior to performing the titration. Prior to further physical property measurements, the IL was dried at 50~$^{\circ}$C under vacuum for at least 12 h and was subsequently stored in a dry glovebox.

\subsubsection{Variable-Temperature Density Measurement}

The density of \newil{} was measured using an Anton Paar DMA 4100 variable temperature density meter, which has an accuracy of $\pm$0.0001~gcm$^{-3}$. Prior to each measurement, the instrument calibration was verified at 20 $^{\circ}$C and ambient pressure. Briefly, the density of ambient lab air and the density of nanopure water were recorded and compared to the expected values. A verification was deemed successful if the density of air was 0.0012 $\pm$0.0002~gcm$^{-3}$ and the density of nanopure water was 0.9982 $\pm$0.0002~gcm$^{-3}$ at 20 $^{\circ}$C. After calibration verification, the sample cell was briefly dried using a built-in air pump to prepare for sample injection. For each measurement, a pre-dried sample (1.1~mL) was loaded into a 3~mL syringe in a dry glovebox and temporarily protected with a N$_2$ headspace. The syringe was brought out of the dry glovebox, and the sample was immediately injected into the sample cell, with the syringe left in the injection port to prevent movement of the sample or the entry of gas bubbles. The density of the sample was recorded at 2.5~$^{\circ}$C steps between 2.5~$^{\circ}$C and 70~$^{\circ}$C. The sample was then carefully removed using the syringe, and the sample cell was washed with at least 20~mL of nanopure water and dried with an air pump to prepare for the next measurement.

\subsubsection{Isothermal Compressibility Measurement}

\begin{figure}[h!]
    \centering
    \includegraphics[width= 0.75\textwidth]{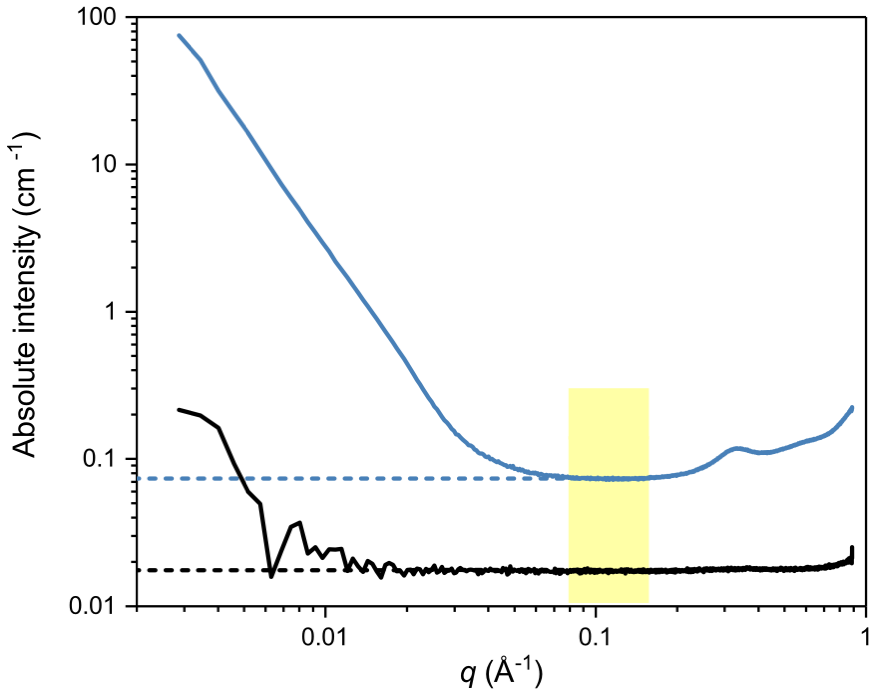}
    \caption{Small-angle X-ray scattering patterns of water (black) and \newil{} (blue). Dashed lines represent linear extrapolations to $q$ $\rightarrow$ 0 from the region of $q$ highlighted in yellow.}
    \label{fig:SAXS}
\end{figure}

As described in our previous work~\cite{Wenny2022}, isothermal compressibility values were obtained using small-angle X-ray scattering (SAXS) measurements performed at beamline 12-ID-B at the Advanced Photon Source at Argonne National Laboratory. Samples were probed using 13.3~keV X-rays, and the sample-to-detector distance was calibrated using a silver behenate standard. Scattered radiation was detected using a Pilatus 2~M detector. Beam defocusing and short exposure times (0.1~s) were used to reduce the chance of sample damage by the beam. The sample holder described previously, which consisted of a quartz capillary glued into a metal ring and connected to tubes to allow injection of the liquid sample, was used to obtain high-quality background and calibration data. Specifically, after measurement of the empty capillary as a background, a measurement of the SAXS pattern of nanopure water was used to obtain an absolute intensity calibration. Subsequently, \newil{} was injected into the sample holder and its SAXS pattern was recorded (Fig.~\ref{fig:SAXS}).

The zero-angle scattering intensity of the SAXS pattern was obtained as the y-intercept of a linear fit to the low-angle, featureless region between 0.08 and 0.16 \AA$^{-1}$. The known isothermal compressibility of water (4.524~$\times$ 10$^{-10}$ Pa$^{-1}$)~\cite{CRCRumble} was used to obtain a scaling factor, $a$, using an expression that relates the zero-angle scattering intensity, $I(0)$, to the isothermal compressibility, $\beta_T = aI(0)/(\eta^2k_BT)$, where $\eta$ is the scattering length density obtained using the molecular formula and density of the sample at 25~$^{\circ}$C, $k_B$ is the Boltzmann constant, and $T$ is temperature in K. This expression, with the value of obtained from water, can be then used to provide the isothermal compressibility of an unknown liquid such as \newil{}. 

\subsection{2. Initial Structure Generation and Equilibration Procedure}
\label{sec:EP}

To generate an initial configuration, we start by creating a low-density, non-overlapping structures by placing the ions randomly on the vertices of a three-dimensional cubic grid~\cite{fadel2019role, molinari2019general, molinari2020chelation, Molinari2019anomalies}. To overcome local energy barriers, in search of a lower energy minima, we then use an equilibration routine that comprises of a set of energy minimizations, compression/decompression and annealing stages, broadly based on previous works~\cite{molinari2016molecular, molinari2019general}. The equilibration procedure comprises the following energy minimization, compression and decompression, and annealing stages shown in Tab.~\ref{tab:equilibrationroutine}.


\begin{table}[!htb]
\centering
\begin{tabular}{l c c c c} \hline \hline
Stage											&	Duration / \si{\nano\second} 	&  Total  / \si{\nano\second}	&	T / \si{\kelvin}	&	P / \si{\mega\pascal} \\ \hline
\tikzmark{e}Energy min. 	&	-			&	-			&	-			&	-			\\
NPT (comp.)		&	\num{0.5}	&	\num{0.5}	&	\num{1}	&	\num{0.01} $\rightarrow$ \num{0.2} \\
Energy min.		&	-			&	-			&	-			&	-			\\
NPT (comp.$+$heating)			&	\num{0.5}	&	\num{1.0}	&	\num{1} $\rightarrow$ \num{300}	&	\num{0.2} $\rightarrow$ \num{10} \\
NPT (decomp.) &	\num{0.4}	&	\num{1.4}	&	\num{300}	&	\num{10} $\rightarrow$ \num{0.1} \\
NPT (heating)					&	\num{0.8}	&	\num{2.2}	&	\num{300} $\rightarrow$ \num{400}	&	\num{0.1} $\rightarrow$ \num{1} \\
NPT (cooling)					&	\num{0.8}	&	\num{3.0}	&	\num{400} $\rightarrow$ \num{300}	&	\num{1} $\rightarrow$ \num{0.1} \\
NPT (comp.)			&	\num{1.0}	&	\num{4.0}	&	\num{300}	&	\num{0.1} $\rightarrow$ \num{100} \\
NPT (decomp.)		&	\num{1.0}	&	\num{5.0}	&	\num{300}	&	\num{100} $\rightarrow$ \num{0.1} \\
\tikzmark{f}NPT										&	\num{1.0}	&	\num{6.0}	&	\num{300} &	\num{0.1} \\ \hline\hline
\end{tabular}
\caption{\textmd{Stages used to equilibrate all the initial structures. The stages are sequentially run. During heating/cooling and compression/decompression stages, the initial/final temperature/pressure is indicated in the table as ``$\text{Initial}\rightarrow\text{Final}$''. The change in temperature/pressure is performed at a constant rate over the entire duration of the stage.}}
\begin{tikzpicture}[overlay, remember picture, yshift=.25\baselineskip, shorten >=.5pt, shorten <=.5pt]
    \draw [->] ([xshift=-2mm, yshift=2mm]{pic cs:e}) -- ([xshift=-2mm]{pic cs:f});
\end{tikzpicture}
\label{tab:equilibrationroutine}
\end{table}

\newpage
\subsection{3. Compositional Transferability with Allegro}
\label{sec:ttl}



\subsubsection{Salt-in-Ionic Liquid - Transferability with Classical Data}

In Tab.~\ref{tab:E_CFF} and \ref{tab:F_CFF} we show the energy and force mean absolute error (MAE), respectively, for the transferability test with classical data using Allegro. It is shown in a similar format to the main text, where the top row is the Li-salt molar fraction of each column shown, and all other rows represent different Allegro models trained on the Li-salt molar fractions, indicated in bold. As can be seen, we find qualitatively exactly the same results as NequIP, as shown in the main text, albeit with different absolute values. We find that the energy generalization appears to be better for Allegro than NequIP, which is presumably because of the utilized hyperparameters.


\begin{table}[]
    \centering
    \begin{tabular}{ccccccccccc}
    \hline \hline
    0.0 & 0.1 & 0.2 & 0.3 & 0.4 & 0.5 & 0.6 & 0.7 & 0.8 & 0.9 & 1.0 \\
     \hline
     1.1 & 1.9 & 1.1 & 1.1 & \textbf{0.8} & 0.9 & \textbf{1.0} & 1.5 & 2.4 & 2.6 & 3.7 \\
     \textbf{0.8} & 0.9 & 0.8 & 0.9 & 1.5 & 0.9 & 1.4 & 1.3 & 2.0 & 1.0 & \textbf{1.2} \\
     \textbf{0.8} & \textbf{0.9} & \textbf{0.7} & 0.8 & 0.9 & 0.8 & 1.0 & 1.3 & 2.5 & 2.1 & 2.9 \\
     \textbf{0.9} & \textbf{0.9} & 0.8 & 0.9 & 0.9 & 1.3 & 1.4 & 1.5 & 1.0 & 1.3 & 1.7 \\
     86.6 & 75.6 & 60.6 & 44.1 & 23.6 & \textbf{0.8} & 29.9 & 68.0 & 117.8 & 184.7 & 282.1 \\
    \hline\hline
    \end{tabular}
    \caption{\textmd{Energy MAE (meV/atom) transferability test with classical data.}}
    \label{tab:E_CFF}
\end{table}
\begin{table}[]
    \centering
    \begin{tabular}{ccccccccccc}
    \hline \hline
    0.0 & 0.1 & 0.2 & 0.3 & 0.4 & 0.5 & 0.6 & 0.7 & 0.8 & 0.9 & 1.0 \\
     \hline
     6.55 & 6.42 & 6.48 & 6.43 & \textbf{6.69} & 6.53 & \textbf{6.65} & 6.63 & 6.96 & 6.88 & 6.29 \\
     \textbf{7.12} & 7.22 & 7.69 & 8.03 & 8.67 & 8.47 & 8.99 & 9.97 & 9.77 & 9.52 & \textbf{4.91} \\
     \textbf{6.58} & \textbf{6.51} & \textbf{6.76} & 6.91 & 7.48 & 7.93 & 8.47 & 9.12 & 10.48 & 11.34 & 10.93 \\
     \textbf{6.64} & \textbf{6.61} & 6.96 & 7.16 & 7.84 & 8.63 & 9.23 & 10.10 & 11.78 & 12.42 & 13.10 \\
     8.04 & 7.82 & 7.90 & 7.83 & 8.08 & \textbf{7.78} & 7.96 & 8.01 & 8.09 & 8.63 & 7.12 \\
    \hline\hline
    \end{tabular}
    \caption{\textmd{Force MAE (meV/\AA$^{-1}$) for transferability test with classical data.}}
    \label{tab:F_CFF}
\end{table}

\subsubsection{Salt-in-Ionic Liquid - Transferability with DFT Data}

In Tab.~\ref{tab:E} and \ref{tab:F} we show the energy and force MAE, respectively, for the transferability test with DFT data using Allegro. It is shown in a similar format to the main text, where the top row is the Li-salt molar fraction of each column shown, and all other rows represent different Allegro models trained on the Li-salt molar fractions, indicated in bold. 

In comparison to the test errors for the classical force field trained with NequIP/Allegro, we find qualitatively exactly the same result for Allegro trained on significantly less DFT data, albeit with different absolute values for the errors. When training on a single composition with DFT data from Quantum Espresso, it is not essential to utilize the single atom energies in Allegro, but it does make the energy errors $\times$100 worse for compositions not trained on. For models trained on multiple compositions, we find it is essential to use single atom energies to train Allegro to a similar standard as an Allegro model trained on a single composition. This is because of the large disparity in total energies of the boxes for DFT data, and without the single atom energies, Allegro assumes all atoms have the same energy, set to the mean energy per atom, which can be a bad approximation.

\begin{table}[]
    \centering
    \begin{tabular}{ccccccccccc}
    \hline \hline
    0.0 & 0.1 & 0.2 & 0.3 & 0.4 & 0.5 & 0.6 & 0.7 & 0.8 & 0.9 & 1.0 \\
     \hline
      0.3 & 0.5 & 0.3 & 0.4 & \textbf{0.3} & 0.3 & \textbf{0.5} & 0.3 & 0.3 & 0.4 & 2.7 \\
      \textbf{0.3} & 0.6 & 2.5 & 4.4 & 6.6 & 6.2 & 9.8 & 11.5 & 10.7 & 7.8 & \textbf{0.3} \\
      \textbf{0.3} & \textbf{0.2} & \textbf{0.3} & 0.5 & 0.7 & 0.8 & 1.0 & 1.2 & 5.2 & 11.4 & 13.6 \\
      \textbf{0.3} & \textbf{0.2} & 0.4 & 0.4 & 0.5 & 2.4 & 2.4 & 2.9 & 11.1 & 22.7 & 22.6 \\
      1$\cdot10^{4}$ & 8$\cdot10^{3}$ & 7$\cdot10^{3}$ & 5$\cdot10^{3}$ & 3$\cdot10^{3}$ & \textbf{0.3} & 3$\cdot10^{3}$ & 7$\cdot10^{3}$ & 1$\cdot10^{4}$ & 2$\cdot10^{4}$ & 3$\cdot10^{4}$ \\
      

    \hline\hline
    \end{tabular}
    \caption{\textmd{Energy MAE (meV/atom) transferability test with DFT data.}}
    \label{tab:E}
\end{table}
\begin{table}[]
    \centering
    \begin{tabular}{ccccccccccc}
    \hline \hline
    0.0 & 0.1 & 0.2 & 0.3 & 0.4 & 0.5 & 0.6 & 0.7 & 0.8 & 0.9 & 1.0 \\
     \hline
     22.96 & 22.32 & 22.94 & 22.26 & \textbf{23.26} & 22.97 & \textbf{23.22} & 24.01 & 25.03 & 26.31 & 24.48 \\
     \textbf{27.31} & 28.31 & 30.87 & 32.13 & 35.75 & 35.16 & 38.20 & 39.36 & 40.89 & 35.75 & \textbf{19.87} \\
     \textbf{17.70} & \textbf{17.61} & \textbf{19.75} & 19.74 & 21.29 & 28.66 & 33.14 & 44.58 & 47.46 & 53.63 & 76.59 \\
     \textbf{19.56} & \textbf{20.10} & 24.47 & 25.15 & 28.27 & 42.44 & 52.46 & 77.10 & 76.65 & 90.89 & 162.43 \\
     28.64 & 27.69 & 28.25 & 27.11 & 28.35 & \textbf{28.58} & 28.37 & 29.56 & 30.15 & 31.80 & 29.09 \\
    \hline\hline
    \end{tabular}
    \caption{\textmd{Force MAE (meV\AA$^{-1}$) for transferability test with DFT data.}}
    \label{tab:F}
\end{table}

In addition to the models trained on compositions in the main text, we further investigate training on different compositions for Allegro-DFT, to ensure the observed trends are robust. In Tab.~\ref{tab:E_M} and \ref{tab:F_M}, respectively, we show the energy and force MAE. The bottom set of rows represents models trained on a single composition. We find that the energy MAE consistently gets worse the further from the single composition trained on, almost independent of the composition that the model is trained on. The force MAE is reasonable across the entire composition range for the model trained on $0.5$, but training models on $0.1$ or $0.9$ results in force errors which progressively get worse for compositions away from that trained on, which occurs because of poor sampling of some interactions.

The next set of models are ones trained on 2 non-adjacent compositions. We find good energy and force MAEs across all compositions, indicating the model has learnt to be transferable, for all composition combinations apart from training on the $0.0+1.0$ data, which occurs because of not sampling Li-EMIM interactions. Training on 2 adjacent compositions is often worse than training on 2 non-adjacent compositions, but it also becomes highly dependent on the composition sampling. The $0.3+0.4$ and $0.6+0.7$ models have good energy and force MAEs for all compositions, but the $0.0+0.1$ and $0.9+1.0$ force and energy MAEs get worse the further from these compositions. Very similar observations are found for the next set of models upon training on 3 compositions, albeit with slightly better MAEs. 

Therefore, the observations drawn from Tab.~\ref{tab:E} and \ref{tab:F} are robust, with the optimal training procedure on 2 (or more) intermediate, non-adjacent compositions, such as the $0.4+0.6$ model. We also find consistent indications that there is asymmetry in the obtained errors when training on 2 or 3 adjacent compositions. Generally, training on compositions at the 1.0 end results in smaller force and energy MAEs at the 0.0 end, in comparison to models trained on the 0.0 end for the errors at the 1.0 end. This asymmetry could arise from the strong Li-PF$_6$ interactions, whereas the EMIM-PF$_6$ interactions are weaker~\cite{McEldrew2021}. Moreover, learning the energy of 1.0 composition appears to be the most difficult, as energy errors are consistently larger.

\begin{table}[]
    \centering
    \begin{tabular}{ccccccccccc}
    \hline \hline
    0.0 & 0.1 & 0.2 & 0.3 & 0.4 & 0.5 & 0.6 & 0.7 & 0.8 & 0.9 & 1.0 \\
     \hline
      \textbf{0.3} & 0.3 & 0.3 & 0.3 & 0.3 & \textbf{0.3} & 0.5 & 0.3 & 0.3 & 0.2 & \textbf{0.3} \\
    0.6 & 0.3 & 0.4 & 0.5 & 0.7 & 0.5 & 0.3 & 0.8 & \textbf{0.7} & \textbf{0.3} & \textbf{0.3} \\
    0.4 & 0.7 & 0.4 & 0.4 & \textbf{0.3} & \textbf{0.3} & \textbf{0.5} & 0.2 & 0.2 & 0.3 & 2.1 \\
    \textbf{0.3} & \textbf{0.2} & \textbf{0.3} & 0.5 & 0.7 & 0.8 & 1.0 & 1.2 & 5.2 & 11.4 & 13.6 \\
    \hline 
      \textbf{0.3} & \textbf{0.2} & 0.4 & 0.4 & 0.5 & 2.4 & 2.4 & 2.9 & 11.1 & 22.7 & 22.6 \\
      0.3 & 0.2 & 0.2 & \textbf{0.3} & \textbf{0.3} & 0.4 & 0.3 & 0.8 & 0.9 & 0.4 & 0.7 \\
      1.2 & 0.8 & 0.8 & 0.6 & 0.6 & 0.3 & \textbf{0.4} & \textbf{0.5} & 0.8 & 0.9 & 0.2 \\
      2.4 & 1.8 & 2.2 & 2.1 & 2.1 & 1.4 & 2.2 & 1.5 & 0.7 & \textbf{0.6} & \textbf{0.2} \\
     \hline 
     \textbf{0.3} & 0.6 & 2.5 & 4.4 & 6.6 & 6.2 & 9.8 & 11.5 & 10.7 & 7.8 & \textbf{0.3} \\
      0.3 & \textbf{0.3} & 0.3 & 0.3 & 0.3 & 0.3 & 0.3 & 0.4 & 0.3 & \textbf{0.4} & 1.7 \\
      0.3 & 0.3 & \textbf{0.3} & 0.5 & 0.5 & 0.5 & 0.3 & 0.8 & \textbf{0.4} & 0.3 & 1.9 \\
      0.4 & 0.2 & 0.3 & \textbf{0.3} & 0.3 & 0.4 & 0.2 & \textbf{0.6} & 0.4 & 0.4 & 1.1 \\
      0.3 & 0.5 & 0.3 & 0.4 & \textbf{0.3} & 0.3 & \textbf{0.5} & 0.3 & 0.3 & 0.4 & 2.7 \\
      \hline 
      2$\cdot10^{3}$ & \textbf{0.3} & 2$\cdot10^{3}$ & 4$\cdot10^{3}$ & 6$\cdot10^{3}$ & 1$\cdot10^{4}$ & 1$\cdot10^{4}$ & 2$\cdot10^{4}$ & 2$\cdot10^{4}$ & 3$\cdot10^{4}$ & 5$\cdot10^{4}$ \\
      1$\cdot10^{4}$ & 8$\cdot10^{3}$ & 7$\cdot10^{3}$ & 5$\cdot10^{3}$ & 3$\cdot10^{3}$ & \textbf{0.3} & 3$\cdot10^{3}$ & 7$\cdot10^{3}$ & 1$\cdot10^{4}$ & 2$\cdot10^{4}$ & 3$\cdot10^{4}$ \\
      5$\cdot10^{3}$ & 5$\cdot10^{3}$ & 5$\cdot10^{3}$ & 4$\cdot10^{3}$ & 4$\cdot10^{3}$ & 4$\cdot10^{3}$ & 3$\cdot10^{3}$ & 2$\cdot10^{3}$ & 1$\cdot10^{3}$ & \textbf{0.7} & 2$\cdot10^{3}$ \\

    \hline\hline
    \end{tabular}
    \caption{\textmd{Energy MAE (meV/atom) transferability test with DFT data for additional composition mixtures trained on.}}
    \label{tab:E_M}
\end{table}
\begin{table}[]
    \centering
    \begin{tabular}{ccccccccccc}
    \hline \hline
    0.0 & 0.1 & 0.2 & 0.3 & 0.4 & 0.5 & 0.6 & 0.7 & 0.8 & 0.9 & 1.0 \\
     \hline
      \textbf{23.61} & 23.23 & 23.91 & 23.36 & 24.48 & \textbf{24.25} & 24.59 & 25.47 & 25.70 & 24.43 & \textbf{19.01} \\
    34.64 & 32.89 & 33.39 & 31.64 & 31.82 & 30.32 & 29.15 & 27.31 & \textbf{25.57} & \textbf{22.14} & \textbf{17.50} \\
     19.92 & 19.48 & 19.93 & 19.32 & \textbf{20.08} & \textbf{19.91} & \textbf{20.46} & 20.82 & 22.54 & 23.57 & 22.04 \\
     \textbf{17.70} & \textbf{17.61} & \textbf{19.75} & 19.74 & 21.29 & 28.66 & 33.14 & 44.58 & 47.46 & 53.63 & 76.59 \\
    \hline 
      \textbf{19.56} & \textbf{20.10} & 24.47 & 25.15 & 28.27 & 42.44 & 52.46 & 77.10 & 76.65 & 90.89 & 162.43 \\
      20.51 & 20.23 & 20.84 & \textbf{20.35} & \textbf{21.60} & 22.14 & 22.77 & 24.01 & 26.16 & 27.26 & 28.14 \\
      24.68 & 23.62 & 24.19 & 23.33 & 24.18 & 22.97 & \textbf{23.39} & \textbf{23.38} & 24.07 & 23.91 & 23.15 \\
      66.95 & 58.05 & 60.36 & 56.14 & 56.84 & 51.16 & 48.47 & 44.13 & 39.34 & \textbf{29.57} & \textbf{19.23} \\
     \hline 
       \textbf{27.31} & 28.31 & 30.87 & 32.13 & 35.75 & 35.16 & 38.20 & 39.36 & 40.89 & 35.75 & \textbf{19.87} \\
       24.50 & \textbf{24.16} & 24.99 & 24.53 & 25.74 & 25.61 & 25.72 & 26.80 & 26.18 & \textbf{23.92} & 22.77 \\
       23.20 & 22.93 & \textbf{23.41} & 22.85 & 23.71 & 23.70 & 23.80 & 23.54 & \textbf{23.73} & 23.45 & 22.04 \\
       22.60 & 21.97 & 22.66 & \textbf{22.00} & 23.03 & 22.52 & 22.98 & \textbf{23.30} & 22.53 & 25.12 & 24.83 \\
       22.96 & 22.32 & 22.94 & 22.26 & \textbf{23.26} & 22.97 & \textbf{23.22} & 24.01 & 25.03 & 26.31 & 24.48 \\
     \hline
     23.23 & \textbf{23.38} & 25.24 & 25.21 & 28.52 & 32.98 & 35.24 & 42.67 & 50.80 & 64.70 & 75.00 \\
     28.64 & 27.69 & 28.25 & 27.11 & 28.35 & \textbf{28.58} & 28.37 & 29.56 & 30.15 & 31.80 & 29.09 \\
     69.69 & 61.20 & 63.62 & 58.78 & 60.19 & 54.20 & 50.88 & 47.14 & 41.29 & \textbf{32.07} & 25.30 \\
    \hline\hline
    \end{tabular}
    \caption{\textmd{Force MAE (meV\AA$^{-1}$) for transferability test with DFT data for additional composition mixtures trained on.}}
    \label{tab:F_M}
\end{table}

As outlined in the main text, a NequIP/Allegro model trained on a single composition should not learn the energy of each species correctly, and will only be able to reproduce the energy at the composition trained on. A model trained on two compositions should be able to learn the energy difference between the cations (Li$^+$ and [EMIM]$^+$), but might only know the energy of each ion up to some arbitrary constant. In Tab.~\ref{tab:EA_salts}, we show the energy of each ion, obtained from summing up the pair energies in the Allegro test set, with the same order as Tab.~\ref{tab:E_M}. The energy difference between EMIM-Li was obtained to be $-1463.4$~eV from DFT.


For models trained on a single composition, the energy difference between the cations varies by over 300~eV and is 150-500~eV out from the DFT value, which highlights that this difference in the ion energies is not meaningful in these models and that this is the source of the large energy errors obtained for these models. As can be seen in Tab.~\ref{tab:EA_salts}, all the models trained on 2 or 3 compositions have a very close energy differences between the cations, and they are all close to the DFT value of $-1463.4$~eV, as the simple analysis in the main text indicated. Generally, models trained on non-adjacent compositions find the energy difference between cations closer to the DFT values than models trained on adjacent compositions. This can be understood from the fact that obtaining gradients from noisy data is more reliable when using larger intervals. Moreover, it is clear in Tab.~\ref{tab:EA_salts} that training on 3 compositions instead of 2 does not generally improve the value for the energy difference between cations, and therefore, the generalization of the energy predictions. This further verifies that there is a minimum number of compositions required to learn the chemical potential difference between cations.

\begin{figure}
\centering
\includegraphics[width= 0.95\textwidth]{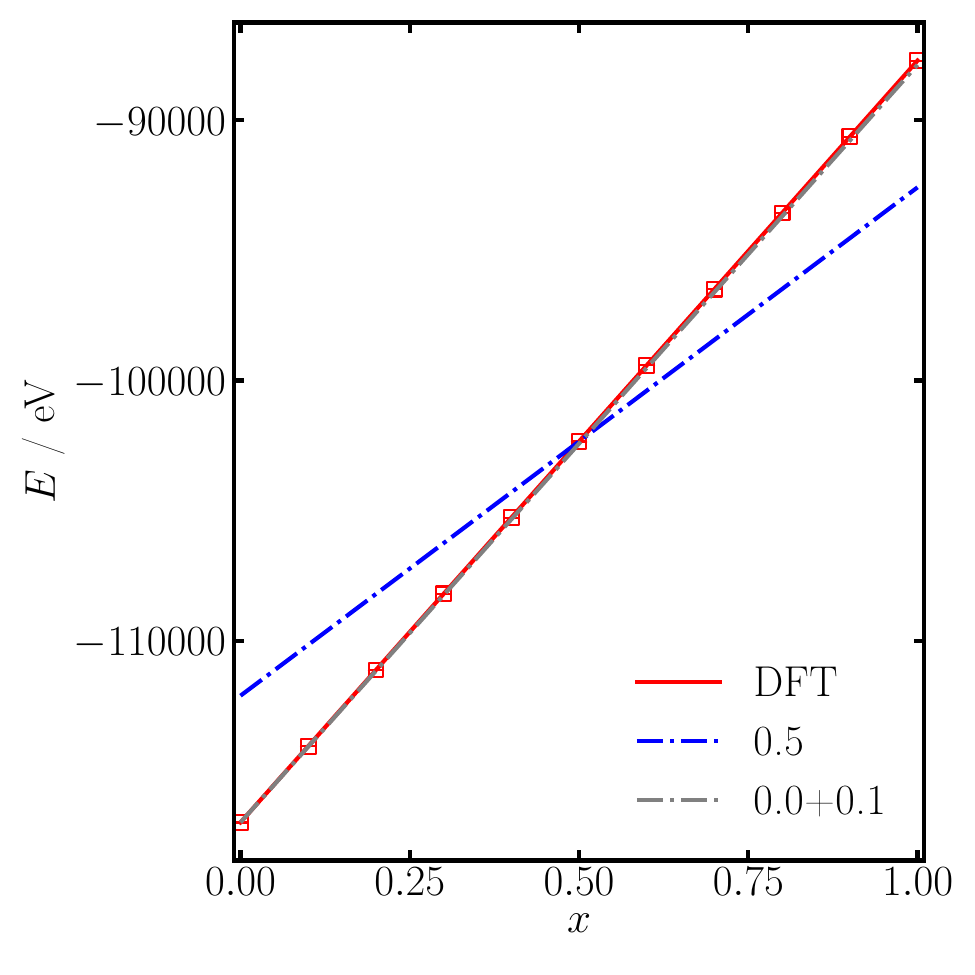}
\caption{Total energy as a function of mole fraction of Li-salt. Symbols represent the DFT test set, with the standard deviation of the energies at each composition being smaller than the symbol size. The solid red line is fitted to the DFT values, and the gradient yields a value of $-1463.4$~eV for the energy difference between EMIM and Li. The dotted-dashed blue line represents the model trained on a single composition of 0.5, where the gradient of the line is determined from the energy difference of EMIM and Li, which can be seen to agree poorly with the DFT values. The grey dotted-dashed line represents a model trained on 0.0+0.1 compositions, which can be seen to agree well at low Li-salt mole fractions, but the deviations with the DFT values become visible for high Li-salt mole fractions, which is the source of the energy error of this model.}
\label{fig:EA_tot}
\end{figure}

In Fig.~\ref{fig:EA_tot} we show the total energies from DFT for each composition, after averaging over the 50 frames for each composition. The energy is linear with the mole fraction of Li-salt, which was used to obtain the energy difference between EMIM-Li of $-1463.4$~eV. We also visually show the above results for the Allegro models trained on $0.5$ and $0.0+0.1$. As can be seen clearly, the $0.5$ model can reproduce the energy of $0.5$, but because the gradient of the line is not correct, there are large energy errors for other compositions, which get progressively worse the further from the composition that is trained on. For the $0.0+0.1$ model, which is probably the worst 2 composition model, it can be seen that the energy prediction only slightly deviates from the DFT value, which gives rise to the energy errors reported in Tab.~\ref{tab:E}. 



In Tab.~\ref{tab:EA_salts} for models trained on 2 or 3 compositions, it is apparent that a lot of models are reporting the same values for the energy of each ion, which should only be known up to some constant. The $0.0+0.1$ and $0.0+0.1+0.2$ models are clear examples of this, which predict very different energies for ions than all other models. Whereas all other 2-3 composition models are finding values within several eV, which might suggest that the models are finding meaningful values for the energy of each ion. This observation is difficult to verify with ionic systems, however, but could be easily checked with mixtures of neutral molecules. 


\begin{table}[]
    \centering
    \begin{tabular}{ccccc}
    \hline
    \hline
    Model & Li & EMIM & PF$_6$ & EMIM - Li \\
    \hline
    $0.0+0.5+1.0$ & -219.0 & -1682.9 & -4166.1 & -1463.9 \\
    $0.8+0.9+1.0$ & -216.2 & -1680.3 & -4168.8 & -1464.0 \\
    $0.4+0.5+0.6$ & -220.5 & -1684.0 & -4164.8 & -1463.6 \\
    $0.0+0.1+0.2$ & -166.5 & -1625.8 & -4220.9 & -1459.3 \\
    \hline
    $0.0+0.1$     & -131.3 & -1586.7 & -4258.0 & -1455.5 \\
    $0.3+0.4$     & -207.3 & -1664.5 & -4180.9 & -1457.1 \\
    $0.6+0.7$     & -215.9 & -1678.0 & -4170.0 & -1462.1 \\
    $0.9+1.0$     & -213.7 & -1678.2 & -4171.1 & -1464.5 \\
    \hline
    $0.0+1.0$     & -217.9 & -1682.3 & -4166.3 & -1464.3 \\
    $0.1+0.9$     & -220.0 & -1684.0 & -4165.0 & -1464.0 \\
    $0.2+0.8$     & -219.9 & -1683.0 & -4165.6 & -1463.0 \\
    $0.3+0.7$     & -218.8 & -1682.3 & -4166.5 & -1463.6 \\
    $0.4+0.6$     & -219.5 & -1682.3 & -4166.3 & -1462.8 \\
    \hline
    $0.1$          & -218.9 & -1276.8 & -4533.4 & -1058.0 \\
    $0.5$          & -215.0 & -1191.5 & -4413.9 & -976.6 \\
    $0.9$          & -212.6 & -1528.1 & -4186.8 & -1315.6 \\
    \hline
    \hline
    \end{tabular}
    \caption{\textmd{Energy predictions of Allegro for each ion, averaged over all compositions, in units of eV. The energy difference between EMIM and Li was determined from the DFT data to be $-1463.4$~eV from a linear fit to the 50 averaged energies at each composition, which had an MAE of $0.3$~eV. }}
    \label{tab:EA_salts}
\end{table}

\subsubsection{Solid Electrolyte - Transferability with DFT Data}

In addition to performing a transferability test on the salt-in-ionic liquid, we further tested the robustness of these results on a solid electrolyte LPSCl. We studied the solid electrolyte chemistries Li$_{6.5}$PS$_{5.5}$Cl$_{0.5}$, Li$_{6}$PS$_{5}$Cl and Li$_{5.5}$PS$_{4.5}$Cl$_{1.5}$, which are referred to as LPSCl$|0.5$,  LPSCl$|1.0$ and LPSCl$|1.5$, respectively. While these compositions are not as diverse as those investigated with the liquid electrolyte, they should still provide an indication if the same rules apply in solids.

To generate the data for this test, we employed the active learning workflow FLARE~\cite{Vandermause2020,Xie2023}, using the Vienna Ab initio Simulation Package (VASP 6)~\cite{Kresse1993M,Kresse1996E,Kresse1996E2} to perform the DFT calculations. The PBE exchange-correlation functional was used~\cite{Perdew1981,Perdew1996}. Cubic cells with experimental lattice parameters were used for each composition~\cite{Deiseroth2008,Zhou2022,Adeli2019}. Starting from the fully ordered (Cl$^-$/S$^{2-}$ sublattice) and fully occupied (48 Li$^+$) unit cell, disordering of Cl$^-$/S$^{2-}$ was randomly generated and the corresponding number of Li$^+$ was removed to maintain charge neutrality. A $4\times4\times4$ k-point mesh with a plane-wave cutoff energy of 700~eV was used. The electronic self-consistent loop convergence was set to 10$^{-5}$~eV. The MD survey during the on-the-fly training was performed with LAMMPS~\cite{LAMMPS2022} using NVT-ensemble with Nos\'e-Hoover thermostat at 300~K for 1~ns with a timestep of 1~fs for all compositions.

The LPSCl$|0.5$ simulations contained 54 atoms, with a total of 64 frames collected; 52 of which were used to train the model, 6 for validation and another 6 for testing. The LPSCl$|1.0$ simulations contained 52 atoms, with a total of 41 frames collected; 33 of which were used to train the model, 4 for validation and another 4 for testing. The LPSCl$|1.5$ simulations contained 50 atoms, with a total of 57 frames collected; 47 of which were used to train the model, 5 for validation and another 5 for testing. In all cases, the test set was taken from the last frames collected from the active learning, and the training-validation frames were drawn randomly from the remaining (initial) frames.

In Tab.~\ref{tab:SE_E} and Tab.~\ref{tab:SE_F} we show, respectively, the energy and force transferability for models trained on all possible combinations of the compositions studied. Similar to our observations reported in the main text and above, we find that training on a single composition does not result in a model with good energy generalization, but using 2 compositions (especially if they are not adjacent compositions) results in good energy generalization, not significantly worse than training on all compositions. 

\begin{table}[]
    \centering
    \begin{tabular}{cccc}
    \hline
    \hline
    Model & LPSCl$|0.5$ & LPSCl$|1.0$ & LPSCl$|1.5$ \\
    \hline
    $0.5+1+1.5$ & \textbf{1.7} & \textbf{0.5} & \textbf{0.7} \\
    \hline
    $0.5+1.0$ & \textbf{1.6} & \textbf{2.7} & 9.0 \\
    $1.0+1.5$ & 10.1 & \textbf{2.1} & \textbf{0.7} \\
    $0.5+1.5$ & \textbf{2.0} & 0.7 & \textbf{1.1} \\
    \hline
    $0.5$ & \textbf{2.1} & 107.3 &  221.9 \\
    $1.0$ & 113.7 & \textbf{2.0} & 111.7  \\
    $1.5$ & 81.7 & 41.6 & \textbf{0.8}  \\
    \hline
    \hline
    \end{tabular}
    \caption{\textmd{Energy MAE (meV/atom) for transferability test with DFT data for the studied solid electrolytes. }}
    \label{tab:SE_E}
\end{table}

\begin{table}[]
    \centering
    \begin{tabular}{cccc}
    \hline
    \hline
    Model & LPSCl$|0.5$ & LPSCl$|1.0$ & LPSCl$|1.5$ \\
    \hline
    $0.5+1+1.5$ & \textbf{28.8} & \textbf{41.5} & \textbf{32.9} \\
    \hline
    $0.5+1.0$ & \textbf{31.3} & \textbf{48.6} & 41.0  \\
    $1.0+1.5$ & 40.5 & \textbf{49.0} & \textbf{34.6} \\
    $0.5+1.5$ & \textbf{34.4} & 56.9 & \textbf{35.4} \\
    \hline
    $0.5$ & \textbf{38.2} & 70.5 & 47.1 \\
    $1.0$ & 49.2 & \textbf{70.7} & 56.0  \\
    $1.5$ & 61.2 & 106.8 & \textbf{47.5} \\
    \hline
    \hline
    \end{tabular}
    \caption{\textmd{Force MAE (meV\AA$^{-1}$) for transferability test with DFT data for the studied solid electrolytes.}}
    \label{tab:SE_F}
\end{table}

LPSCl has Cl$^-$/S$^{2-}$ disorder, so we can tune the composition of S and Cl, and the amount of Li is changed to compensate the charge. Therefore, we essentially only have 2 variables, and we should be able to learn the energy to be generalizable with 2 compositions, which we found to be true. In this example, we expect the energy of Li+S-Cl to be learnt with 2 compositions, and to be model independent, but we do not expect the per species energies to be model independent. 

In Tab.~\ref{tab:EA_LPSCl} we display the predicted values of the energy of each species, which was determined from the average of the per atom energies of each species over all the compositions studied, for each model as indicated. As before, it can clearly be seen there are large variations in the per species energy, over several eV in some cases, which is to be expected from the previous paragraph. We expect, however, that for models trained on 2 compositions the energy of Li+S-Cl should be approximately independent of the model. The value of the energy of Li+S-Cl was determined from DFT to be $-4.73$~eV, from the energy difference between the LPSCl|$0.5$ and LPSCl|$1.5$ averaged over the test set. It can be seen all models trained on 2 compositions are within 0.7~eV of this value obtained from DFT, whereas the models trained on a single composition are over 1~eV out of this value. Moreover, the model trained on all 3 compositions is not significantly better than the models trained on 2 compositions, which further supports the fact that there is a minimum number of compositions required to learn the energy.

\begin{table}[]
    \centering
    \begin{tabular}{cccccc}
    \hline
    \hline
    Model & Li & S & P & Cl & Li+S-Cl \\
    \hline
    $0.5+1+1.5$ & -3.58 & -4.09 & -6.68 & -3.55 & -4.12 \\
    \hline
    $0.5+1.0$ & -3.62 & -4.22 & -5.81 & -3.58 & -4.26 \\
    $1.0+1.5$ & -3.16 & -4.93 & -5.82 & -2.79 & -5.30 \\
    $0.5+1.5$ & -2.73 & -5.13 & -6.18 & -3.93 & -3.93 \\
    \hline
    $0.5$ & -3.85 & -4.18 & -4.82 & -1.91 & -6.12 \\
    $1.0$ & -3.26 & -4.99 & -6.00 & -1.62 & -6.63 \\
    $1.5$ & -2.51 & -5.88 & -2.35 & -5.74 & -2.65 \\
    \hline
    \hline
    \end{tabular}
    \caption{\textmd{Energy predictions of Allegro for each species, averaged over all compositions, in units of eV. The energy of Li+S-Cl was determined from the DFT data to be $-4.73$~eV, from the energy difference between the LPSCl|$0.5$ and LPSCl|$1.5$ averaged over the test set. }}
    \label{tab:EA_LPSCl}
\end{table}

\clearpage
\newpage
\subsection{4. Structure Generation for \newil{}}
\label{sec:SGDFT}

\begin{figure}[h!]
    \centering
    \includegraphics[scale=1.5]{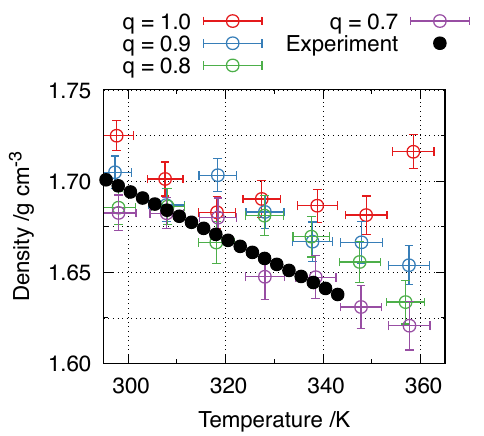}
    \caption{Density versus temperature of OPLS-based MD at different point charge rescaling and experiments.}
    \label{fig:OPLS_Malia_density_vs_temperature}
\end{figure}

With the aim of sampling configurations for \newil{} as close as possible to experiments, we rescale the nominal point charges assigned to atomic species by the classical force field. This is commonly done in the IL literature to mimic the average charge screening due to polarization, and charge transfer effects~\cite{doherty2017revisiting, mogurampelly2017structure, pal2017effects}. Figure~\ref{fig:OPLS_Malia_density_vs_temperature} shows density as a function of temperature for experimental measurements and computed values with different charge rescaling. We chose a rescaling of the points charges of \SI{70}{\percent}.

We selected a handful (60 at each temperature) of the configurations generated from this classical force field with rescaled charges, and used them to calculate the energy, force and stress from DFT. The force and energy values from DFT for these OPLS sampled structures appeared reasonable, but the diagonal components of the stress tensor were of the order of $\sim$1~GPa. This suggested there could be an issue with the DFT settings, such as the employed PBE exchange-correlation functional not localizing charges on the ions correctly. To test the origin of this systematic stress, we performed several checks. First, we turned off the D3 correction and performed a variable-cell BFGS relaxation to ensure the system remained condensed. We found the system reached a density of $\sim$1.8~gcm$^{-3}$ after 50 steps, which is slightly higher than the OPLS density and the experimental value. Therefore, the ions appeared to still bind without a D3 correction, which suggested they remained at least partially charged.


%
%

During the variable cell relaxation test, it was noticed that the stress dropped rapidly, even without the cell changing size dramatically. This suggested the origin was not from the cell size, but some structural features of the ions, perhaps from the inconsistency between the potential energy surface of OPLS and PBE+D3. With this in mind, we performed fixed-cell AIMD on each structure for 50 steps (with a $\sim$1~fs timestep) at the corresponding temperatures (using the Verlet algorithm and a tolerance for rescaling velocities of 10~K) and also 10 steps of fixed-cell relaxation for each structure.

%
%

\begin{figure}
\centering
\includegraphics[width= 1\textwidth]{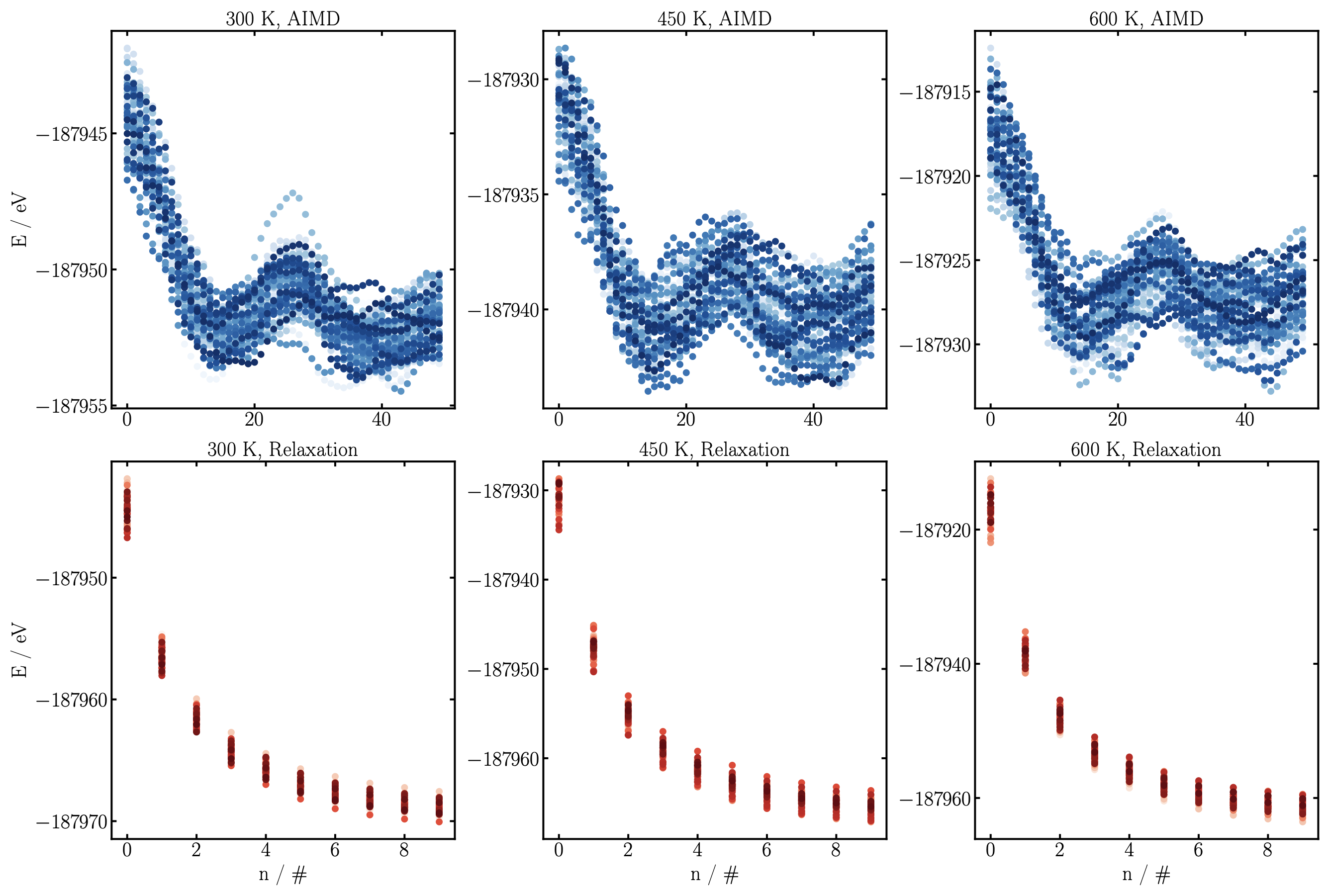}
\caption{Energy as a function of step number for AIMD trajectories (top panels, with a timestep of $\sim1$~fs) and relaxations (bottom panels), for the 3 temperatures used to generate the structures (300~K - left; 450~K - middle; 600~K - right).}
\label{fig:E_AIMD_R}
\end{figure}

In Fig.~\ref{fig:E_AIMD_R} we display how the energy changes for a short AIMD run and relaxation, starting from the OPLS generated structures. At 300~K, the AIMD trajectory lowers its energy by 5-10~eV in 10-20 steps, and fluctuates about this value for the remaining 40-30 steps. Similar observations are found for 450~K and 600~K, with the energy drop being more substantial. As expected, the relaxations starting from the OPLS structures at different temperatures drop significantly, by over 20~eV in all cases.

\begin{figure}
\centering
\includegraphics[width= 1\textwidth]{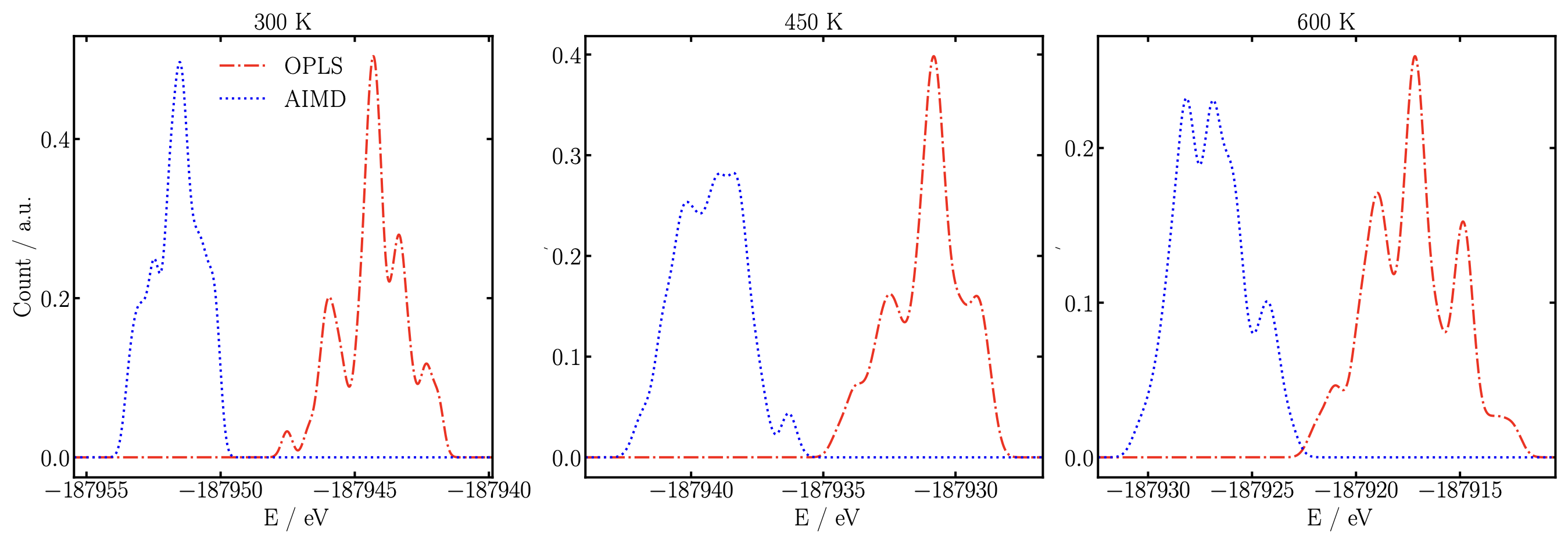}
\caption{Distribution of energy values for each temperature, as indicated, from the initial OPLS structures and 50th frame from the AIMD trajectory. A Gaussian smearing scheme has been employed, with smearing values of 0.2, 0.3, 0.4~eV at 300, 450, 600~K, respectively.}
\label{fig:E_dist}
\end{figure}

In Fig.~\ref{fig:E_dist} we show the energy distributions for OPLS and AIMD at each temperature, obtained from a Gaussian smearing scheme. The energy distributions of the OPLS structures are highly peaked and not very normally distributed. The energy distributions obtained from the 50 AIMD steps, starting from these OPLS structures, are less peaked, more normally distributed, and generally less spread out in energy. The energy shifts between the OPLS-AIMD structures are almost comparable to changing the temperature by 150~K.

\begin{figure}
\centering
\includegraphics[width= 1\textwidth]{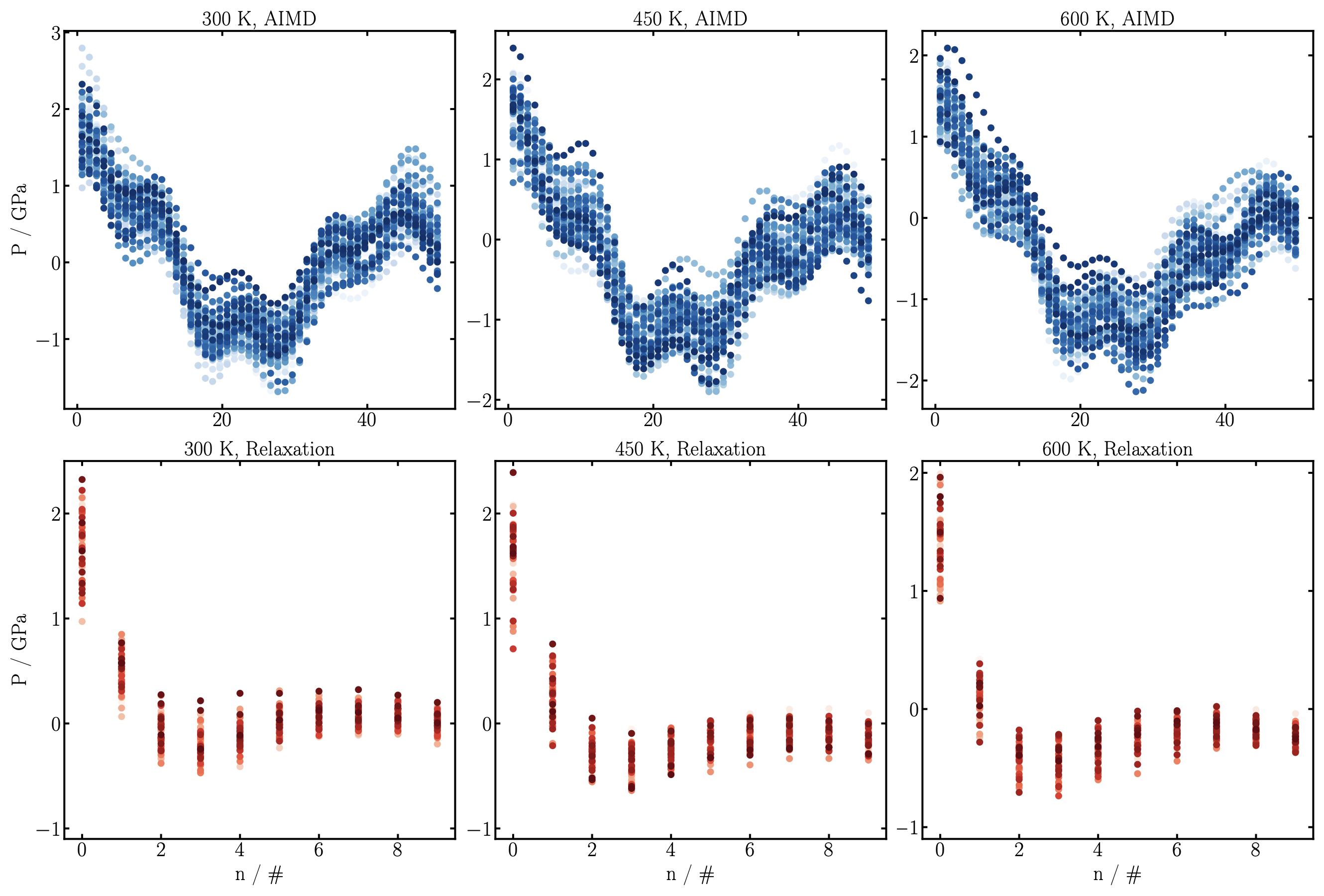}
\caption{Pressure as a function of step number for AIMD trajectories (top panels) and relaxations (bottom panels), for the 3 temperatures use to generate the structures (300~K - left; 450~K - middle; 600~K - right).}
\label{fig:P_AIMD_R}
\end{figure}

In Fig.~\ref{fig:P_AIMD_R} we display the corresponding pressures. Naturally, the relaxations dramatically reduce the pressure, with values close to, or just less than, 0 at the end of 10 steps. The pressure from the short AIMD trajectories oscillates significantly, but after 40 steps the pressure is close to 0.

Finally, for a single AIMD trajectory and relaxation, we display all of the forces on the atoms as a function of the number of steps. The relaxations reduce the forces to within approximately a factor of $1/2$ in 2 steps, and are close to 0 by 10 steps. The short AIMD trajectory has force values that are not significantly decreasing/increasing in overall magnitude throughout the short run.

\begin{figure}
\centering
\includegraphics[width= 1\textwidth]{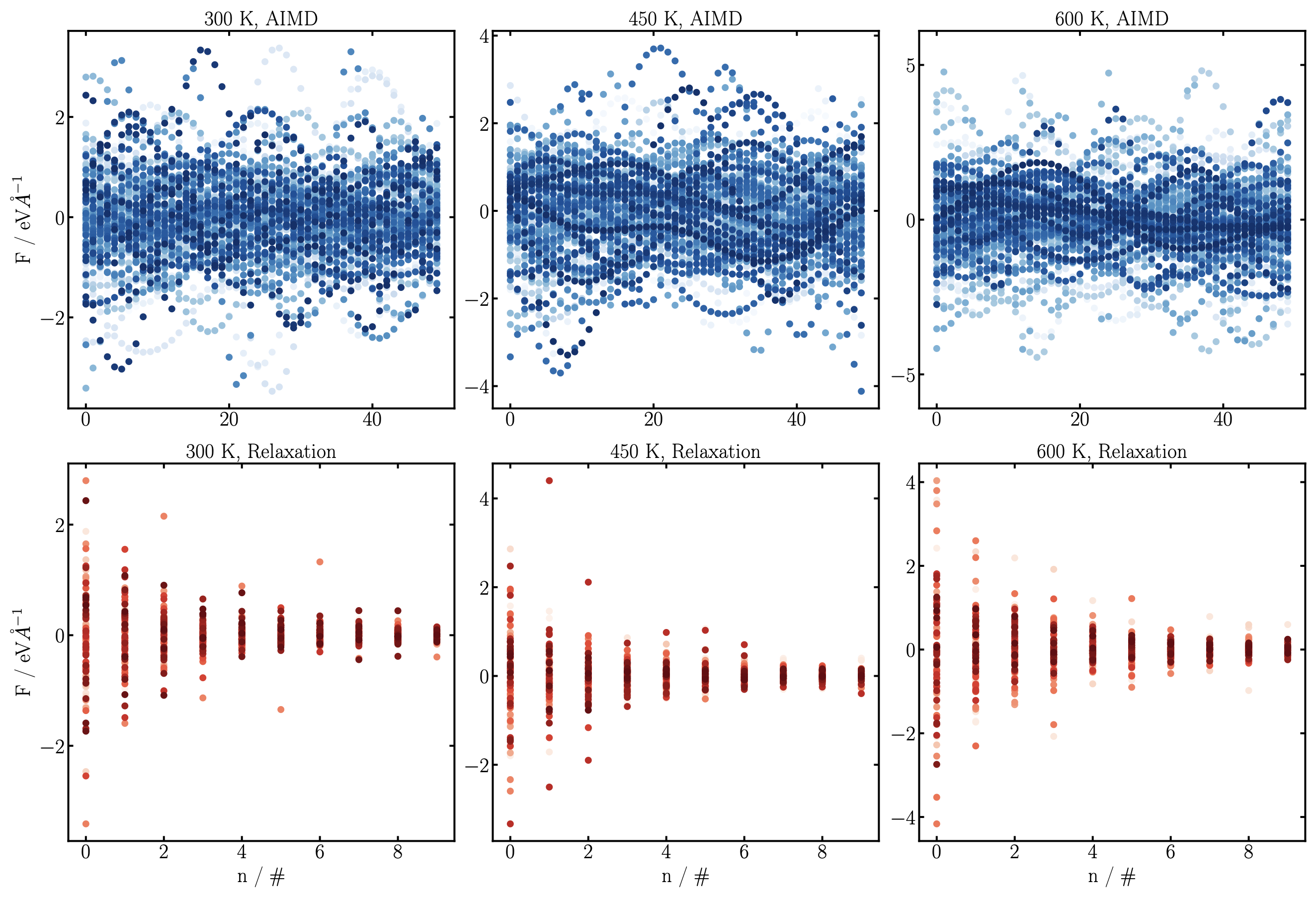}
\caption{Force as a function of step number for single AIMD trajectory (top panels) and a single relaxation (bottom panels), for the 3 temperatures use to generate the structures (300~K - left; 450~K - middle; 600~K - right).}
\label{fig:F_AIMD_R}
\end{figure}

\begin{figure}
\centering
\includegraphics[width= 1\textwidth]{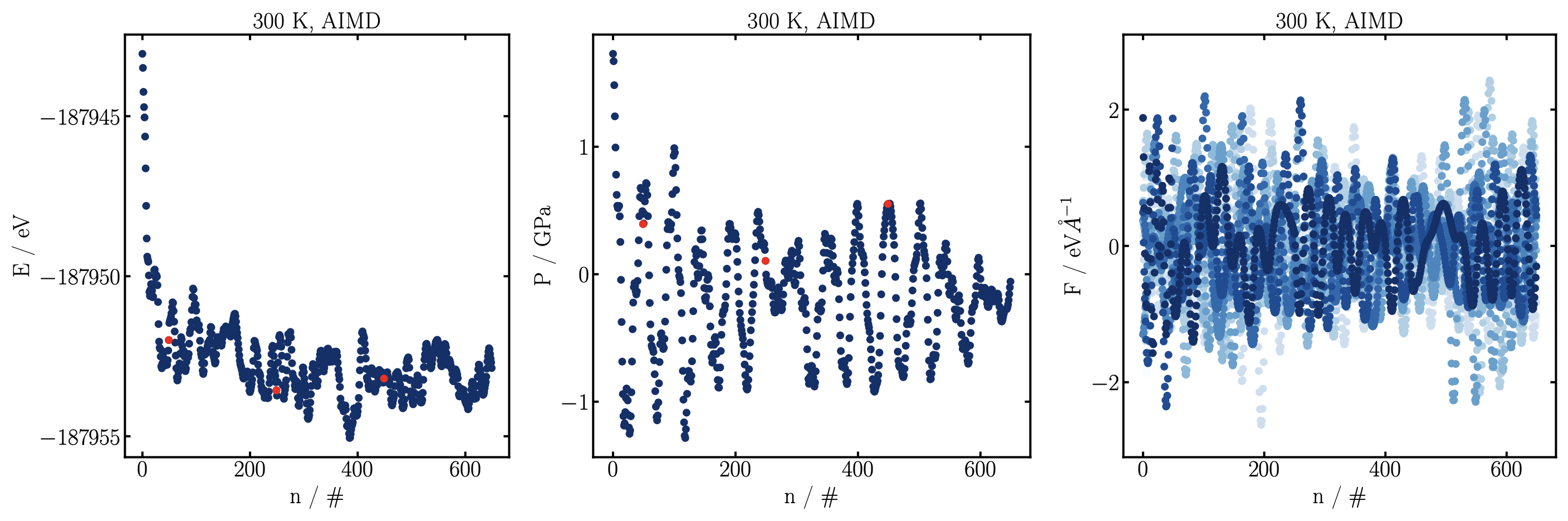}
\caption{Energy, pressure and force as a function of step number for a single, longer AIMD trajectory at 300~K.}
\label{fig:LONG_AIMD}
\end{figure}

In Fig.~\ref{fig:LONG_AIMD} we show the energy, pressure and forces for all atoms for a longer trajectory, slightly over 0.6~ps. We find that the energy reduces a little more after the 50th step, but only by 1-2~eV, the pressure remains to oscillate around 0, and the force values remain within $\pm$3~eV\AA$^{-1}$. 

Overall, it appears the OPLS structures are systematically higher in energy, which could potentially cause problems upon training a model on this data. Therefore, we train models on the OPLS generated structures, from the 50th frames of the AIMD trajectories (starting from these OPLS structures), and the 2nd frames from the relaxations (starting from these OPLS structures). Later, we also investigate the role of choosing different frames from the AIMD trajectories.

\newpage
\subsection{5. Allegro Details for \newil{}}
\label{sec:AE}

In this section, we compare MLIPs for \newil{} trained on three datasets: (1) the dataset of structures generated directly from the OPLS force field, referred to as the ``OPLS'' dataset (2) the dataset from propagating these structures in AIMD at the corresponding temperatures for 50 steps, referred to as the ``AIMD'' dataset (3) the dataset from the relaxations, using step 2, as this reduces the energy and pressure, but keeps the forces non-zero, which is referred to as the ``Rel.'' dataset. 

First, we performed hyperparameter tests to see what settings work best. Note this is not an exhaustive test, but these investigations allow us to see what helps the Allegro architecture perform well. For each hyperparameter and dataset, in addition to showing errors of the model, we also test for chemical stability and density at 300~K, obtained from a 100~ps NPT simulation with 10 ion pairs with a timestep of 0.5~fs. The model hyperparameter notation is as follows: $r_{max}l_{max}n_{layer}$ $L_FL_EL_S$, where $r_{max}$ is the maximum range of interactions taken into account, $l_{max}$ determines the angular resolution of the model, $n_{layer}$ is the number of layers in the network (which is related to the body-order of the interactions accounted for), and $L_F$,$L_E$,$L_S$ is, respectively, the loss coefficient of the force, energy and stress. The values of the force MAE in meV\AA$^{-1}$ are shown for each atom type. The energy MAE in meV/atom and stress MAE in 10$^{-5}$eV\AA$^{-3}$. 





\begin{table}[]
    \centering
    \begin{tabular}{llccccccccc}
    \hline \hline
    \textbf{Model}    && C & O & N & F & H & $E$ & $\sigma$ & CS & $\rho$ \\
    \hline 
    \textbf{422 111}  && 63.15  & 70.32 & 47.46 & 35.52 & 25.06 & 1.89 & 60.53 & Y & 0.2-0.5 \\
    \textbf{522 111}  && 53.30 & 63.66 & 43.51 & 29.47 & 21.53 & 0.84 & 27.88 & Y & 0.3-1.2 \\
    \textbf{622 111}  && 47.80 & 51.97 & 42.79 & 25.60 & 19.47 & 0.45 & 14.50 & Y & 1.4-1.7 \\
    \textbf{722 111}  && \textbf{42.34} & \textbf{46.68} & \textbf{39.71} & \textbf{22.01} & \textbf{18.06} & \textbf{0.42} & \textbf{10.65} & Y & 1.5-1.7\\
    \hline
    \textbf{532 111} && 51.33 & 59.16 & \textbf{38.54} & 28.66 & \textbf{19.85} & 0.65 & 25.50 & Y & 1.0-1.5\\
    \textbf{512 111}  && 59.09 & 60.91 & 53.38 & 32.82 & 26.67 & 0.89 & 30.08 & N & - \\
    \textbf{612 111} && 52.15 & 52.10 & 50.76 & 28.43 & 24.10 & 0.51 & 16.29 & Y & 1.4-1.7 \\
    \textbf{712 111} && \textbf{47.35} & \textbf{46.09} & 49.62 & \textbf{25.34} & 22.59 & \textbf{0.40} & \textbf{13.24} & N & - \\
    \hline
    \textbf{622 1101}  && \textbf{49.60} & \textbf{55.29} & \textbf{42.12} & \textbf{26.53} & \textbf{19.98} & 0.40 & 12.14 & Y/N & 1.5-1.7/-\\
    \textbf{622 110$^{2}$1}  && 54.69 & 59.82 & 47.31 & 28.71 & 22.90 & \textbf{0.30} & \textbf{9.67} & N & -\\
    \textbf{622 110$^{3}$1}  && 89.30 & 86.12 & 85.02 & 47.75 & 48.52 & 0.47 & 20.03 & N & - \\
    \hline
    \textbf{622 1110}  && 47.68 & 51.77 & 42.75 & 25.66 & 19.50 & 0.42 & 10.44 & Y & 1.5-1.7 \\
    \textbf{622 1110$^{2}$}  && \textbf{46.08} & \textbf{51.64} & \textbf{41.20} & \textbf{25.09} & \textbf{18.81} & \textbf{0.42} & 8.20 & Y & 1.5-1.7 \\
    \textbf{622 1110$^{3}$}  && 47.80 & 51.68 & 41.54 & 25.90 & 19.69 & 0.43 & \textbf{7.68} & N & - \\
    \hline \hline
    \end{tabular}
    \caption{\textmd{Hyperparameter scan for Allegro potentials trained on OPLS snapshots.}}
    \label{tab:OPLS_HPS}
\end{table}

\begin{figure}
    \centering
    \includegraphics[width= 0.7\textwidth]{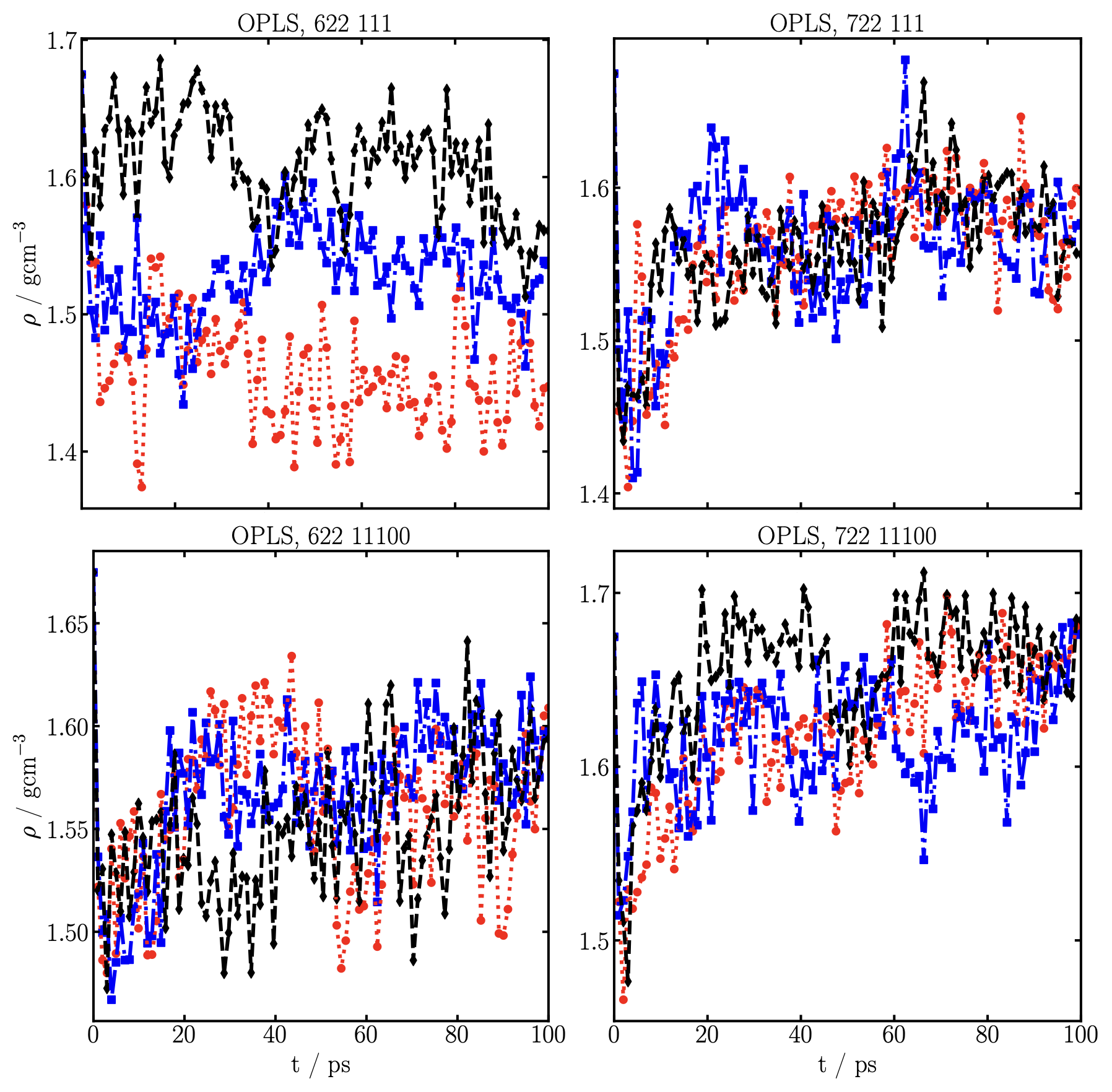}
    \caption{Several NPT simulations for several Allegro potentials over 100~ps, demonstrating the predicted densities and their variability between different runs. Each point corresponds to a 1~ps step, with lines drawn between each point for guidance.}
    \label{fig:HPT}
\end{figure}

In Tab.~\ref{tab:OPLS_HPS} we show the results for different hyperparameters, only being tested on the ``OPLS'' dataset. The real-space cut-off, $r_{max}$, to include pair-wise interactions was varied between $4-7$~\AA. We found the errors decreased with increasing $r_{max}$, and the density increases to values closer to the classical force field. Values of $r_{max} = 4,5$ are not promising, but $r_{max} = 6,7$ appear to have good errors and reasonable density values, as also seen in Fig.~\ref{fig:HPT}. Increasing the angular resolution of the Allegro model, through varying $l_{max}$ (the maximum order of the spherical harmonic used), generally made the model more accurate in terms of errors, but larger $l_{max}$ makes the Allegro model significantly more computationally expensive. The $l_{max} = 1$ models were not chemically stable, which discounted these models from being further tested. Using $l_{max} = 2$ is generally considered to be a good trade-off between computational expense and accuracy, and we proceed with this value. Similarly, using 2 layers in the Allegro model is considered to be a good compromise, and we do not vary this hyperparameter.

In Tab.~\ref{tab:OPLS_HPS} we also tested increasing the scale factors of energy and stress, as it was noticed that the force loss dominated the loss function by orders of magnitude. Increasing the factor in front of the energy loss reduced the errors on the energy, but made the model chemically unstable. Whereas increasing the factor in front of the stress initially decreased errors of force and stress and improved the densities, as seen in Fig.~\ref{fig:HPT}, but values that are too large made the Allegro model chemically unstable.

Overall, we found Allegro models with $r_{max} = 7$, $l_{max} = 2$ and 2 layers, referred to as a $722$ model, with either scaling the loss of the stress by 100 (referred to a $11100$, as the factor multiplying the standard Allegro loss function is 1 for force, 1 for energy, and 100 for stress) or not (scale factor of 1 for forces, energy and stress),  as a reasonable starting point. We further tested these hyperparameters for the 3 datasets, and show the results in Tab.~\ref{tab:OPLS_AIMD_R}. While we cannot directly compare the errors of the 3 different datasets, we can compare their densities and chemical stability, and also assess hyperparameter choices for each dataset. In most cases, there is very little difference in the force and energy error between using a stress scale factor of 1 and 100, but the scale of 100 does improve the stress error significantly. Moreover, the model trained on relaxation data becomes chemically more stable with the stress scale factor of 100.


\begin{table}[]
    \centering
    \begin{tabular}{llccccccccc}
    \hline \hline
    \textbf{Data} & \textbf{Model}    & C & O & N & F & H & $E$ & $\sigma$ & CS & $\rho$ \\
    \hline 
     OPLS & \textbf{722 111}  & \textbf{42.34} & \textbf{46.68} & 39.71 & \textbf{22.01} & \textbf{18.06} & \textbf{0.42} & 10.65 & Y & 1.5-1.7 \\
     & \textbf{722 1110$^{2}$}  & 42.91 & 46.71 & \textbf{39.67} & 22.39 & 18.37 & 0.49 & \textbf{8.56} & Y & 1.5-1.7 \\
    \hline
    AIMD & \textbf{722 111}  & 45.21 & 53.52 & 43.33 & 22.90 & 20.23 & \textbf{0.46} & 11.10 & Y & 1.5-1.7\\
     & \textbf{722 1110$^{2}$}  & \textbf{44.96} & \textbf{52.94} & \textbf{41.37} & \textbf{22.79} & \textbf{19.75} & 0.48 & \textbf{8.17} & Y & 1.5-1.7 \\
    \hline
    Rel. & \textbf{722 111}  & \textbf{36.25} & 43.12 & 32.18 & \textbf{19.20} & \textbf{15.71} & \textbf{0.42} & 11.12 & N & - \\
     & \textbf{722 1110$^{2}$}  & 36.31 & \textbf{42.36} & \textbf{31.69} & 19.21 & 15.87 & 0.49 & \textbf{7.82} & Y & 1.5-1.7 \\
    \hline \hline
    \end{tabular}
    \caption{\textmd{Errors for different datasets with different scales on the stress.}}
    \label{tab:OPLS_AIMD_R}
\end{table}

\begin{figure}
\centering
\includegraphics[width=1\textwidth]{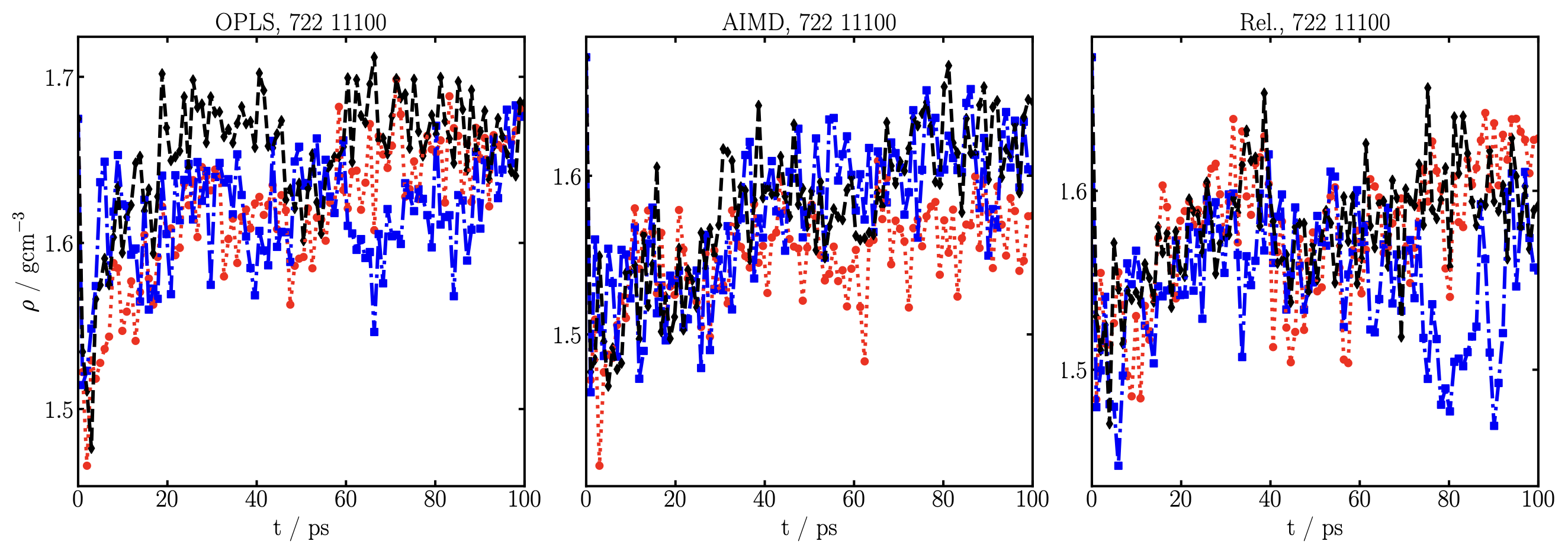}
\caption{Several NPT simulations for several Allegro potentials over 100~ps, demonstrating the predicted densities and their variability between different runs. Each point corresponds to a 1~ps step, with lines drawn between each point for guidance.}
\label{fig:QDC}
\end{figure}



\begin{figure}
\centering
\includegraphics[width=1\textwidth]{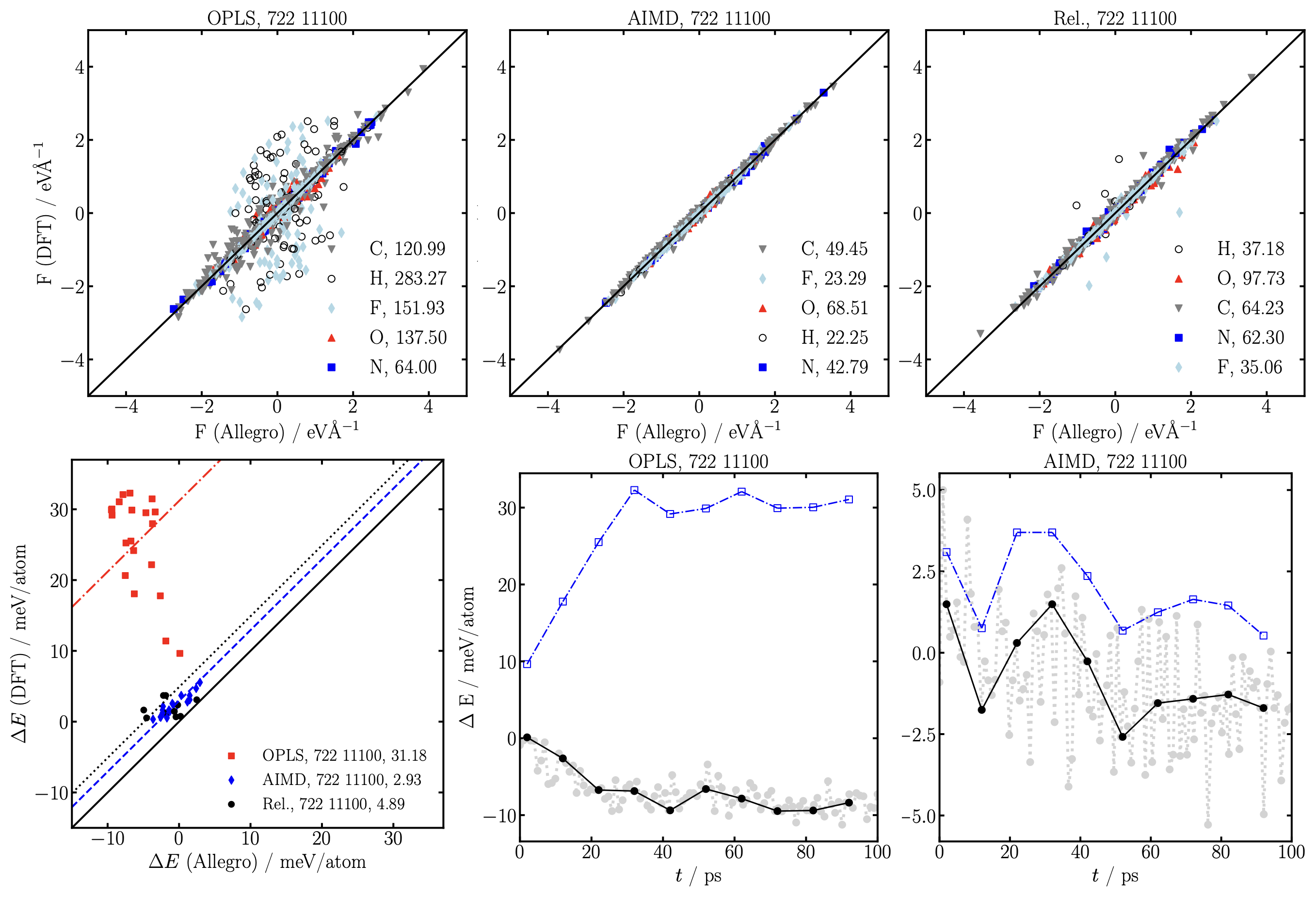}
\caption{Force parity plots for various Allegro models, taken from a single frame at 300~K approximately 50~ps through the NPT simulation. All force components of all atoms are shown, with the force MAE in meV\AA$^{-1}$ being indicated in the plot. Energy parity plot between DFT and Allegro from 10-20 frames selected from 1-2 100~ps NPT simulations. The corresponding energy MAE is given in the legend, and a diagonal indicating the average energy MAE for each dataset is provided. The frames selected for the OPLS and AIMD datasets are time resolved (middle-right), showing that the OPLS $722$ model gets worse over 30~ps and then stabilizes, while the AIMD potential does not significantly drift over time.}
\label{fig:PP_F_E_722}
\end{figure}

We further test the $722$ $11100$ models for each dataset. For the 100~ps NPT simulations of these models, shown in Fig.~\ref{fig:QDC}, we select 10 frames from 1-2 runs and compute the energy and forces from DFT. In Fig.~\ref{fig:PP_F_E_722}(top panels) we compare the force parity plots for one of these frames, taken at approximately 50~ps through the simulation, for each model. The force parity plot for the OPLS model is much worse than the errors indicated when validating, with hydrogen and fluorine being particularly bad. The Rel. model is also worse than indicated upon validating, but it is significantly better than the OPLS model. In contrast, the AIMD model has errors comparable to the errors indicated in Tab.~\ref{tab:OPLS_AIMD_R}. Overall, the forces of the AIMD model appear to be accurate, but the OPLS and Rel. models have significant failures. We also plot the energy parity plot in Fig.~\ref{fig:PP_F_E_722} for each of these datasets, where $\Delta E = 0$ corresponds to $E = -187955$~eV, which is approximately the equilibrium values found from long AIMD run (see Fig.~\ref{fig:LONG_AIMD}). The energy error of the AIMD and Rel. models are 2.93 and 4.89~meV/atom, respectively, which is approximately an order of magnitude larger than the errors indicated when validating, but still a respectable value. In contrast, the OPLS model is significantly worse, with deviations of up to 40~meV/atom, which approximately corresponds to the energy difference between the OPLS-AIMD frames, as seen in Fig.~\ref{fig:E_dist}. We time resolve the energy plot for one run, as seen in Fig.~\ref{fig:PP_F_E_722}, and it is clear the DFT energies for the AIMD model tracks the predicts from Allegro, albeit with a roughly constant shift, but the OPLS model gets worse over the course of the first 30~ps of the simulation before plateauing. 


\begin{figure}
\centering
\includegraphics[width=1\textwidth]{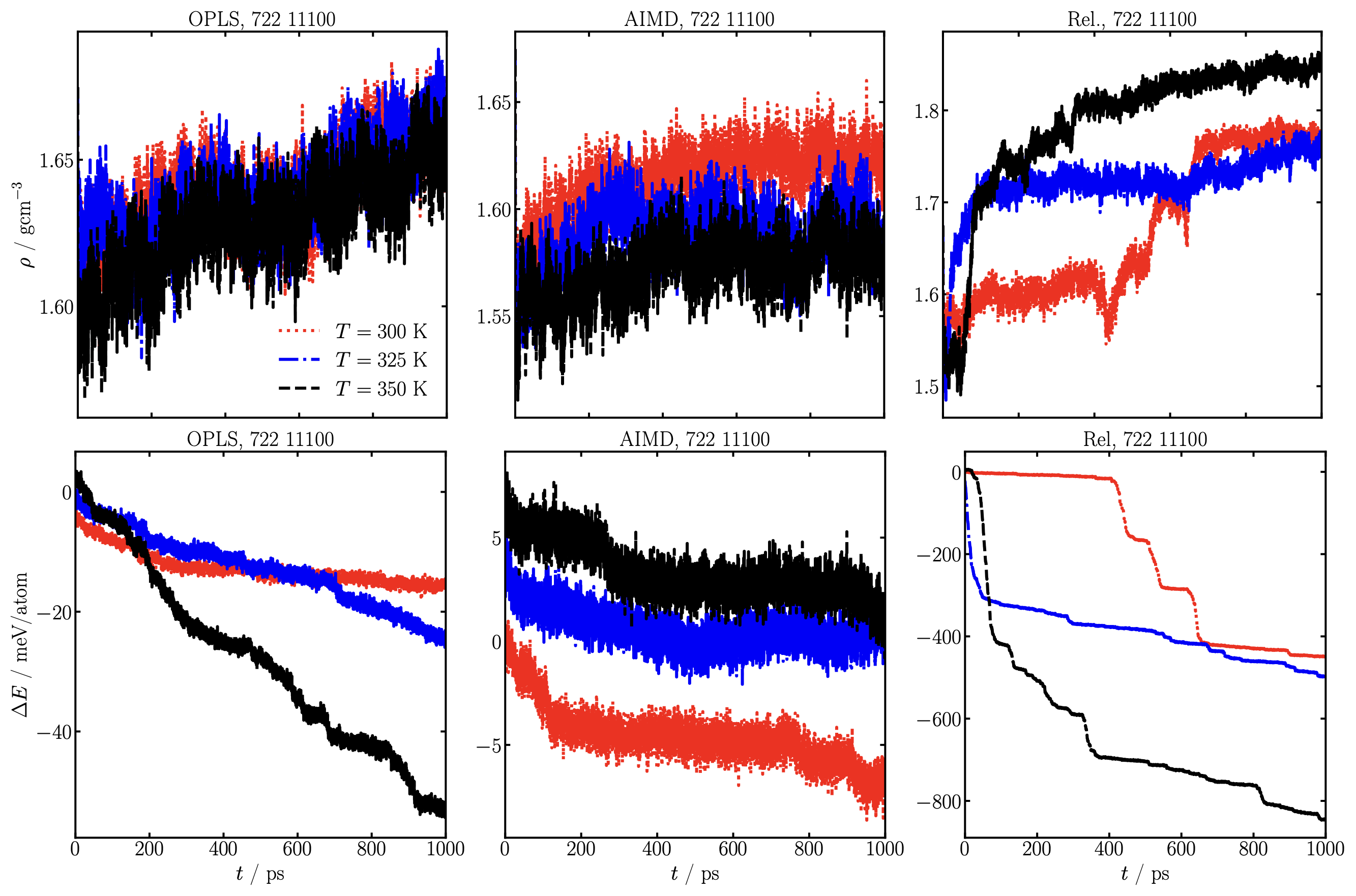}
\caption{Density and energy as a function of time for 3 temperatures, as indicated, for various Allegro potentials.}
\label{fig:NPT_722_E}
\end{figure}

To further test these models, we computed the densities of these models over 1~ns at 300-350~K with 4240 atoms ($2\times2\times2$ replicate of the unit cell trained on), the results of which are shown in Fig.~\ref{fig:NPT_722_E}, with the density values being given in Tab.~\ref{tab:rho_DT}. The AIMD model performs reasonably well, with the densities equilibrating to values expected, with the density decreasing with increasing temperature. Whereas the OPLS model densities continue to increase with time, and are relatively independent of temperature. The Rel. model density at 300~K initially appears reasonable, but after 400~ps there is a dramatic increase in the density. The Rel. model at higher temperatures fails at even shorter time scales, and the density typically increases with temperature, indicating a failure of this model to generalize to higher temperatures.

In Fig.~\ref{fig:NPT_722_E}, we also show the corresponding energies to the density plots. The value $\Delta E = 0$ corresponds to $E = -1503640$~eV, which is the approximate potential energy from the long AIMD simulation at 300~K. For the OPLS model, we find that the energy decreases by up to 70~meV/atom over the course of 1~ns. Moreover, the energy of higher temperatures are lower, which suggests a pathological problem with the OPLS model. In contrast, the AIMD model lowers the energy by approximately 5 meV/atom, with much more constant values and higher energies for higher temperatures. Finally, the Rel. model has significant failures in the energy predictions, which can be correlated with the unphysical jumps in the density. Therefore, the AIMD model predicts stable energies and densities in NPT, but the OPLS and Rel. models again have significant failures. We find that plotting the energy is a good way to identify problems for system sizes where explicit DFT checks cannot be performed.

\begin{figure}
\centering
\includegraphics[width=0.7\textwidth]{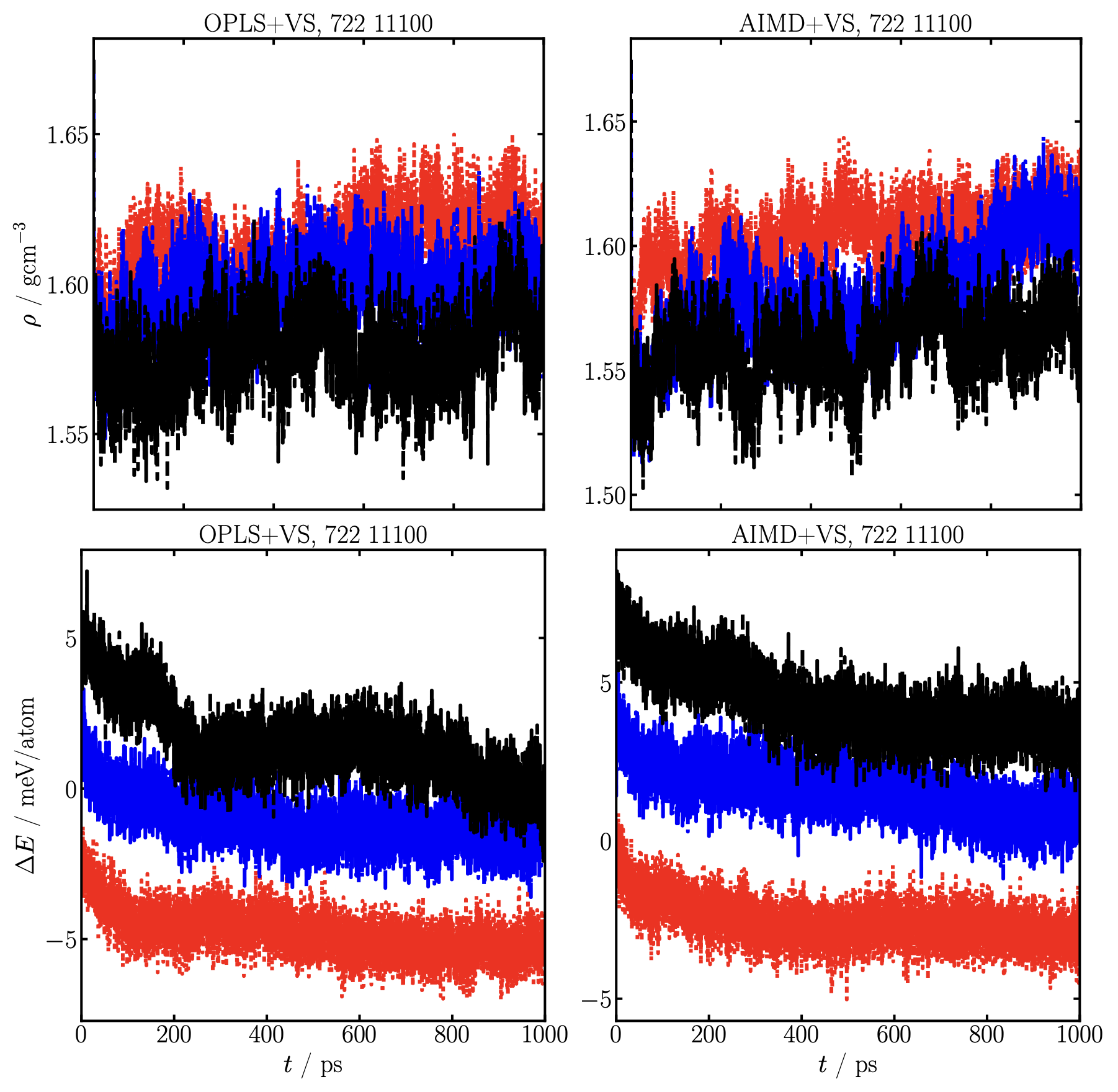}
\caption{Density as a function of time for 3 temperatures for various Allegro potentials. Energy as a function of time for 3 temperatures for various Allegro potentials.}
\label{fig:VS_NPT_722}
\end{figure}

\begin{figure}
\centering
\includegraphics[width=1\textwidth]{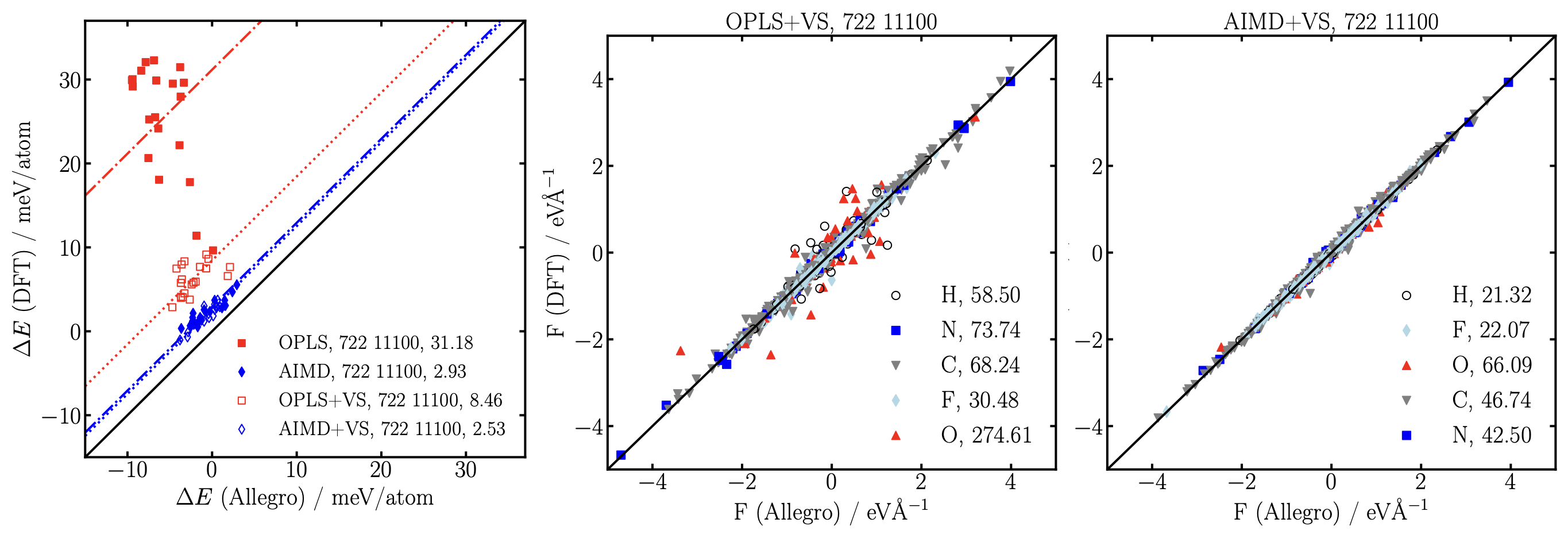}
\caption{Energy parity plot (left) between DFT and Allegro from 10-20 frames selected from 1-2 100~ps NPT simulations. Force parity plots (middle-right), taken from a single frame at 300~K approximately 50~ps through the NPT simulation. All force components of all atoms are shown, with the Force MAE in meV\AA$^{-1}$ being indicated in the plot.}
\label{fig:EF_PP_VS_722}
\end{figure}

To see if we can improve these density and energy values, we collected 40 frames of volume scans, over a density range of 1.67-0.87 gcm$^{-3}$, and added these to each dataset. Naturally, the force, energy and stress errors reduce as shown in Tab.~\ref{tab:VS}, as expected upon adding more data, especially in this light data regime. The Rel. model became chemically unstable, but the AIMD and OPLS models remained stable within 100~ps and were tested further. 

In Fig.~\ref{fig:VS_NPT_722}, we plot the predicted density and energy values over 1~ns for the AIMD+VS and OPLS+VS models. For the OPLS+VS model, we find that the densities equilibrate, with larger densities at lower temperatures. For the AIMD+VS model, the densities are again stable, albeit with a slightly lower density than without the volume scans, but both OPLS+VS and AIMD+VS give similar densities. Moreover, the energy of the OPLS+VS model is now stable, and agrees reasonably well with the AIMD+VS model. 

These improvements in the predictions for the OPLS+VS model can be understood from the parity plots in Fig.~\ref{fig:EF_PP_VS_722}. As can be seen, the energy errors significantly improve, and no longer get progressively worse with time. Moreover, the force errors are better than without the additional volume scans, although the error on the oxygen atoms remain to be much larger than expected. With continued iterative training (as is often done with these methods) of the OPLS+VS model, it is expected that these errors eventually converge to those found for the AIMD models. The AIMD+VS model has a similar accuracy to the AIMD model. Overall, it appears that the quality of the data (structures) used to the train the Allegro models can significantly impact the predictions of the model. If the structures are sampled from the DFT potential energy surface consistently, it is clear that Allegro can learn a more accurate model than if it is attempting to generalise from OPLS structures. 



\begin{table}[]
    \centering
    \begin{tabular}{lllccccccccc}
    \hline \hline
    \textbf{Data} & \textbf{Model}  && C & O & N & F & H & $E$ & $\sigma$ & CS & $\rho$ \\
    \hline 
    OPLS & \textbf{722 1110$^{2}$}  && 42.91 & 46.71 & 39.67 & 22.39 & 18.37 & 0.49 & 8.56 & Y & 1.5-1.7 \\
    +VS & \textbf{722 1110$^{2}$}  && \textbf{36.48} & \textbf{39.97} & \textbf{33.84} & \textbf{19.76} & \textbf{15.98} & \textbf{0.42} & \textbf{7.80} & Y & 1.5-1.7 \\
    \hline
    AIMD & \textbf{722 1110$^{2}$}  && 44.96 & 52.94 & 41.37 & 22.79 & 19.75 & 0.48 & 8.17 & Y & 1.5-1.7 \\
    +VS & \textbf{722 1110$^{2}$}  && \textbf{38.23} & \textbf{42.87} & \textbf{34.07} & \textbf{20.18} & \textbf{17.38} & \textbf{0.38} & \textbf{8.09} & Y & 1.5-1.7 \\
    \hline
    Rel.  & \textbf{722 1110$^{2}$}  &&  36.31 & 42.36 & 31.69 & 19.21 & 15.87 & 0.49 & 7.82 & Y & 1.5-1.7 \\
    +VS & \textbf{722 1110$^{2}$} && \textbf{30.96} & \textbf{34.44} & \textbf{26.63} & \textbf{17.08} & \textbf{13.54} & \textbf{0.41} & \textbf{6.97} & N & - \\
    \hline \hline
    \end{tabular}
    \caption{\textmd{Effect of volume scans being added to dataset for different hyperparameters and datasets.}}
    \label{tab:VS}
\end{table}

\begin{table}[]
    \centering
    \begin{tabular}{lllccccccccc}
    \hline \hline
    \textbf{Data} & \textbf{Model}  && $\rho$ / gcm$^{-3}$ & $\rho$ / gcm$^{-3}$  & $\rho$ / gcm$^{-3}$ \\
    \hline
    OPLS & \textbf{722 1110$^{2}$}  && 1.648 $\pm$ 0.013 & 1.647 $\pm$ 0.014 & 1.639 $\pm$ 0.012 \\
    +VS & \textbf{722 1110$^{2}$}   && 1.619 $\pm$ 0.010 & 1.600 $\pm$ 0.010 & 1.579 $\pm$ 0.012 \\
    \hline
    AIMD & \textbf{722 1110$^{2}$}  && 1.623 $\pm$ 0.008 & 1.590 $\pm$ 0.012 & 1.578 $\pm$ 0.010 \\
    +VS & \textbf{722 1110$^{2}$}   && 1.611 $\pm$ 0.009 & 1.592 $\pm$ 0.016 & 1.566 $\pm$ 0.013 \\
        & \textbf{722 1110$^{2}$}   && 1.598 $\pm$ 0.010 & 1.584 $\pm$ 0.011 & 1.545 $\pm$ 0.014 \\
    \hline
    Rel. & \textbf{722 1110$^{2}$}  && 1.740 $\pm$ 0.046 & 1.736 $\pm$ 0.017 & 1.835 $\pm$ 0.012 \\
    \hline \hline
    \textbf{Data} & \textbf{Model}  && T / K & T / K  & T / K \\
    \hline
    OPLS & \textbf{722 1110$^{2}$}  && 299.9 $\pm$ 3.8 & 324.9 $\pm$ 4.0 & 350.0 $\pm$ 4.3 \\
    +VS & \textbf{722 1110$^{2}$}   && 300.0 $\pm$ 3.8 & 324.9 $\pm$ 4.1 & 350.1 $\pm$ 4.3 \\
    \hline
    AIMD & \textbf{722 1110$^{2}$}  && 300.1 $\pm$ 3.8 & 325.0 $\pm$ 4.0 & 349.9 $\pm$ 4.4 \\
    +VS & \textbf{722 1110$^{2}$}   && 299.9 $\pm$ 3.7 & 325.0 $\pm$ 4.0 & 349.9 $\pm$ 4.4 \\
        & \textbf{722 1110$^{2}$}   && 312.0 $\pm$ 3.9 & 331.9 $\pm$ 4.1 & 361.9 $\pm$ 4.6 \\
    \hline
    Rel. & \textbf{722 1110$^{2}$}  && 300.1 $\pm$ 3.8 & 325.0 $\pm$ 4.1 & 350.1 $\pm$ 4.4 \\
    \hline \hline
    \end{tabular}
    \caption{\textmd{Density values for each NPT shown in Figs.~\ref{fig:NPT_722_E}-~\ref{fig:VS_NPT_722}.}}
    \label{tab:rho_DT}
\end{table}

   
    
    

\begin{figure}
\centering
\includegraphics[width=0.7\textwidth]{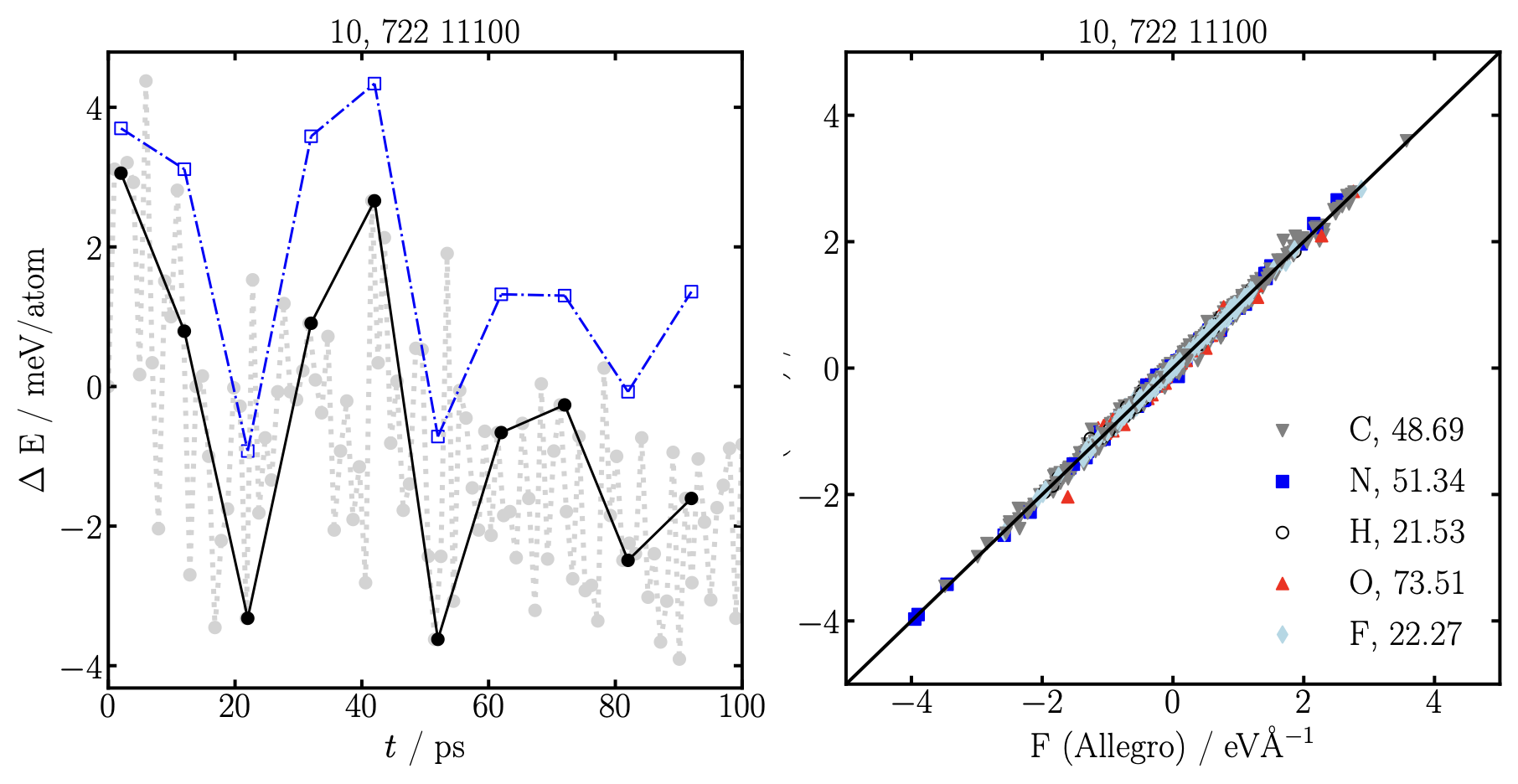}
\caption{Comparison of energy as a function of time (left) between DFT and Allegro (trained on the 10th AIMD frames) for 10-20 frames selected from a 100~ps NPT simulations. Force parity plots (right), taken from a single frame at 300~K approximately 50~ps through the NPT simulation. All force components of all atoms are shown, with the Force MAE in meV\AA$^{-1}$ being indicated in the plot.}
\label{fig:EF_10}
\end{figure}


\begin{figure}
\centering
\includegraphics[width=0.7\textwidth]{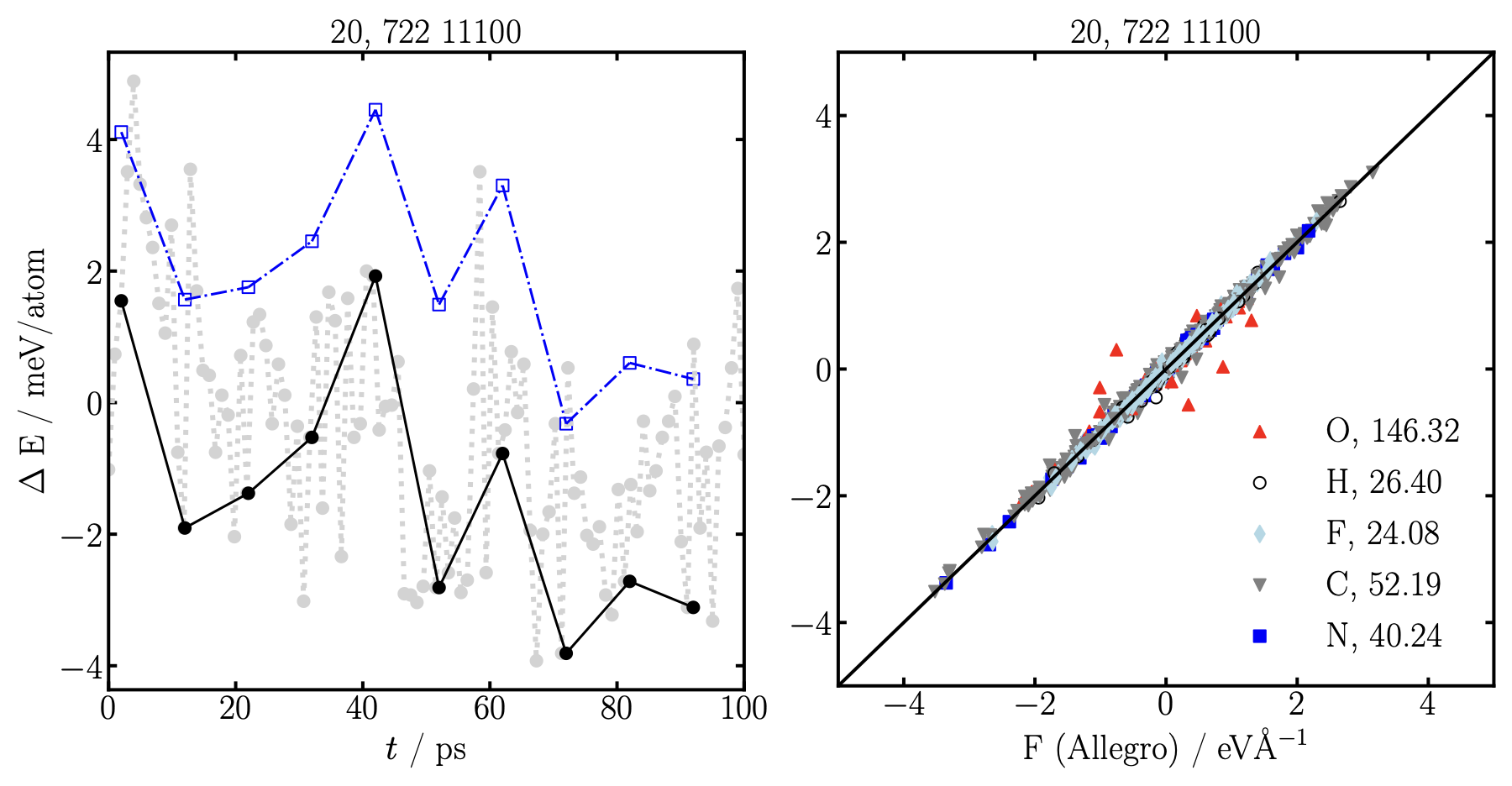}
\caption{Comparison of energy as a function of time (left) between DFT and Allegro (trained on the 20th AIMD frames) for 10-20 frames selected from a 100~ps NPT simulations. Force parity plots (right), taken from a single frame at 300~K approximately 50~ps through the NPT simulation. All force components of all atoms are shown, with the Force MAE in meV\AA$^{-1}$ being indicated in the plot.}
\label{fig:EF_20}
\end{figure}

Finally, we aim to answer if Allegro models can be trained on a dataset comprising of a smaller number of AIMD steps. So far, we have focused on a dataset solely composed of the 50th frames, approximately 50~fs long, but it would be computationally cheaper if we would only need to do 10-20 steps instead. In Fig.~\ref{fig:P_AIMD_R}, we can see that after 10 steps the pressure is close to 0 and the energy has dropped to roughly its equilibrium value. Therefore, we trained a potential where we only retain the 10th frames. In contrast, at the 20th step, the pressure is now large and negative, whilst the energy is approximately at its equilibrium value. We also train a potential on the 20th frames to see if this impacts the quality of the potential.

In Fig.~\ref{fig:EF_10} we display the comparison between the Allegro model trained on the 10th AIMD frames and DFT, on frames collected over a 100~ps NPT simulation. Overall, the results for the force parity plots and time-resolved energy plots are similar to the 50th frames previously shown. In contrast, the Allegro model trained on the 20th frames, shown in Fig.~\ref{fig:EF_20}, appears to perform more similar to the OPLS dataset than the AIMD dataset. Therefore, even the step number in the AIMD trajectory that the frames are collected from appears to matter, with those with pressures close to 0 performing better than those with large pressures.

\bibliography{literature.bib}

\providecommand{\latin}[1]{#1}
\makeatletter
\providecommand{\doi}
  {\begingroup\let\do\@makeother\dospecials
  \catcode`\{=1 \catcode`\}=2 \doi@aux}
\providecommand{\doi@aux}[1]{\endgroup\texttt{#1}}
\makeatother
\providecommand*\mcitethebibliography{\thebibliography}
\csname @ifundefined\endcsname{endmcitethebibliography}
  {\let\endmcitethebibliography\endthebibliography}{}
\begin{mcitethebibliography}{86}
\providecommand*\natexlab[1]{#1}
\providecommand*\mciteSetBstSublistMode[1]{}
\providecommand*\mciteSetBstMaxWidthForm[2]{}
\providecommand*\mciteBstWouldAddEndPuncttrue
  {\def\EndOfBibitem{\unskip.}}
\providecommand*\mciteBstWouldAddEndPunctfalse
  {\let\EndOfBibitem\relax}
\providecommand*\mciteSetBstMidEndSepPunct[3]{}
\providecommand*\mciteSetBstSublistLabelBeginEnd[3]{}
\providecommand*\EndOfBibitem{}
\mciteSetBstSublistMode{f}
\mciteSetBstMaxWidthForm{subitem}{(\alph{mcitesubitemcount})}
\mciteSetBstSublistLabelBeginEnd
  {\mcitemaxwidthsubitemform\space}
  {\relax}
  {\relax}

\bibitem[Welton(1999)]{Welton1999}
Welton,~T. Room-temperature ionic liquids: solvents for synthesis and
  catalysis. \emph{Chem. Rev.} \textbf{1999}, \emph{99}, 2071--2084\relax
\mciteBstWouldAddEndPuncttrue
\mciteSetBstMidEndSepPunct{\mcitedefaultmidpunct}
{\mcitedefaultendpunct}{\mcitedefaultseppunct}\relax
\EndOfBibitem
\bibitem[Weing\"artner(2008)]{Hermann2008}
Weing\"artner,~H. Understanding ionic liquids at the molecular level: facts,
  problems, and controversies. \emph{Angew. Chem. Int. Ed.} \textbf{2008},
  \emph{47}, 654--670\relax
\mciteBstWouldAddEndPuncttrue
\mciteSetBstMidEndSepPunct{\mcitedefaultmidpunct}
{\mcitedefaultendpunct}{\mcitedefaultseppunct}\relax
\EndOfBibitem
\bibitem[Hallett and Welton(2011)Hallett, and Welton]{Hallett2011}
Hallett,~J.~P.; Welton,~T. Room-temperature ionic liquids: solvents for
  synthesis and catalysis. 2. \emph{Chem. Rev.} \textbf{2011}, \emph{111},
  3508--3576\relax
\mciteBstWouldAddEndPuncttrue
\mciteSetBstMidEndSepPunct{\mcitedefaultmidpunct}
{\mcitedefaultendpunct}{\mcitedefaultseppunct}\relax
\EndOfBibitem
\bibitem[Fedorov and Kornyshev(2014)Fedorov, and Kornyshev]{Fedorov2014}
Fedorov,~M.~V.; Kornyshev,~A.~A. Ionic liquids at electrified interfaces.
  \emph{Chem. Rev.} \textbf{2014}, \emph{114}, 2978--3036\relax
\mciteBstWouldAddEndPuncttrue
\mciteSetBstMidEndSepPunct{\mcitedefaultmidpunct}
{\mcitedefaultendpunct}{\mcitedefaultseppunct}\relax
\EndOfBibitem
\bibitem[Watanabe \latin{et~al.}(2017)Watanabe, Thomas, Zhang, Ueno, Yasuda,
  and Dokko]{Watanabe2017}
Watanabe,~M.; Thomas,~M.~L.; Zhang,~S.; Ueno,~K.; Yasuda,~T.; Dokko,~K.
  Application of ionic liquids to energy storage and conversion materials and
  devices. \emph{Chem. Rev.} \textbf{2017}, \emph{117}, 7190--7239\relax
\mciteBstWouldAddEndPuncttrue
\mciteSetBstMidEndSepPunct{\mcitedefaultmidpunct}
{\mcitedefaultendpunct}{\mcitedefaultseppunct}\relax
\EndOfBibitem
\bibitem[Yao \latin{et~al.}(2022)Yao, Chen, Fu, and Zhang]{Yao2022}
Yao,~N.; Chen,~X.; Fu,~Z.-H.; Zhang,~Q. Applying classical, ab initio, and
  machine-learning molecular dynamics simulations to the liquid electrolyte for
  rechargeable batteries. \emph{Chem. Rev.} \textbf{2022}, \emph{122},
  10970--11021\relax
\mciteBstWouldAddEndPuncttrue
\mciteSetBstMidEndSepPunct{\mcitedefaultmidpunct}
{\mcitedefaultendpunct}{\mcitedefaultseppunct}\relax
\EndOfBibitem
\bibitem[Hu \latin{et~al.}(2011)Hu, Liu, Xu, and Zhang]{Hu2011}
Hu,~Y.-F.; Liu,~Z.-C.; Xu,~C.-M.; Zhang,~X.-M. The molecular characteristics
  dominating the solubility of gases in ionic liquids. \emph{Chem. Soc. Rev.}
  \textbf{2011}, \emph{40}, 3802--3823\relax
\mciteBstWouldAddEndPuncttrue
\mciteSetBstMidEndSepPunct{\mcitedefaultmidpunct}
{\mcitedefaultendpunct}{\mcitedefaultseppunct}\relax
\EndOfBibitem
\bibitem[Wenny \latin{et~al.}(2022)Wenny, Molinari, Slavney, Thapa, Lee,
  Kozinsky, and Mason]{Wenny2022}
Wenny,~M.~B.; Molinari,~N.; Slavney,~A.~H.; Thapa,~S.; Lee,~B.; Kozinsky,~B.;
  Mason,~J.~A. Understanding relationships between free volume and oxygen
  absorption in ionic liquids. \emph{J. Phys. Chem. B} \textbf{2022},
  \emph{126}, 1268--1274\relax
\mciteBstWouldAddEndPuncttrue
\mciteSetBstMidEndSepPunct{\mcitedefaultmidpunct}
{\mcitedefaultendpunct}{\mcitedefaultseppunct}\relax
\EndOfBibitem
\bibitem[Niedermeyer \latin{et~al.}(2012)Niedermeyer, Hallett, Villar-Garcia,
  Hunt, and Welton]{Niedermeyer2012}
Niedermeyer,~H.; Hallett,~J.~P.; Villar-Garcia,~I.~J.; Hunt,~P.~A.; Welton,~T.
  Mixtures of ionic liquids. \emph{Chem. Soc. Rev.} \textbf{2012}, \emph{41},
  7780--7802\relax
\mciteBstWouldAddEndPuncttrue
\mciteSetBstMidEndSepPunct{\mcitedefaultmidpunct}
{\mcitedefaultendpunct}{\mcitedefaultseppunct}\relax
\EndOfBibitem
\bibitem[Izgorodina \latin{et~al.}(2017)Izgorodina, Seeger, Scarborough, and
  Tan]{Izgorodina2017}
Izgorodina,~E.~I.; Seeger,~Z.~L.; Scarborough,~D. L.~A.; Tan,~S. Y.~S. Quantum
  chemical methods for the prediction of energetic, physical, and spectroscopic
  properties of ionic liquids. \emph{Chem. Rev.} \textbf{2017}, \emph{117},
  6696--6754\relax
\mciteBstWouldAddEndPuncttrue
\mciteSetBstMidEndSepPunct{\mcitedefaultmidpunct}
{\mcitedefaultendpunct}{\mcitedefaultseppunct}\relax
\EndOfBibitem
\bibitem[Dong \latin{et~al.}(2017)Dong, Liu, Dong, Zhang, and Zhang]{Dong2017}
Dong,~K.; Liu,~X.; Dong,~H.; Zhang,~X.; Zhang,~S. Multiscale studies on ionic
  liquids. \emph{Chem. Rev.} \textbf{2017}, \emph{117}, 6636--6695\relax
\mciteBstWouldAddEndPuncttrue
\mciteSetBstMidEndSepPunct{\mcitedefaultmidpunct}
{\mcitedefaultendpunct}{\mcitedefaultseppunct}\relax
\EndOfBibitem
\bibitem[Jeanmairet \latin{et~al.}(2022)Jeanmairet, Rotenberg, and
  Salanne]{Jeanmairet2022}
Jeanmairet,~G.; Rotenberg,~B.; Salanne,~M. Microscopic simulations of
  electrochemical double-layer capacitors. \emph{Chem. Rev.} \textbf{2022},
  \emph{122}, 10860--10898\relax
\mciteBstWouldAddEndPuncttrue
\mciteSetBstMidEndSepPunct{\mcitedefaultmidpunct}
{\mcitedefaultendpunct}{\mcitedefaultseppunct}\relax
\EndOfBibitem
\bibitem[McEldrew \latin{et~al.}(2021)McEldrew, Goodwin, Zhao, Bazant, and
  Kornyshev]{McEldrew2021corr}
McEldrew,~M.; Goodwin,~Z. A.~H.; Zhao,~H.; Bazant,~M.~Z.; Kornyshev,~A.~A.
  Correlated ion transport and the gel phase in room temperature ionic liquids.
  \emph{J. Phys. Chem. B} \textbf{2021}, \emph{125}, 2677--2689\relax
\mciteBstWouldAddEndPuncttrue
\mciteSetBstMidEndSepPunct{\mcitedefaultmidpunct}
{\mcitedefaultendpunct}{\mcitedefaultseppunct}\relax
\EndOfBibitem
\bibitem[Dajnowicz \latin{et~al.}(2022)Dajnowicz, Agarwal, Stevenson, Jacobson,
  Ramezanghorbani, Leswing, Friesner, Halls, and Abel]{Dajnowicz2022}
Dajnowicz,~S.; Agarwal,~G.; Stevenson,~J.~M.; Jacobson,~L.~D.;
  Ramezanghorbani,~F.; Leswing,~K.; Friesner,~R.~A.; Halls,~M.~D.; Abel,~R.
  High-dimensional neural network potential for liquid electrolyte simulations.
  \emph{J. Phys. Chem. B} \textbf{2022}, \emph{126}, 6271--6280\relax
\mciteBstWouldAddEndPuncttrue
\mciteSetBstMidEndSepPunct{\mcitedefaultmidpunct}
{\mcitedefaultendpunct}{\mcitedefaultseppunct}\relax
\EndOfBibitem
\bibitem[Behler(2021)]{Behler2019}
Behler,~J. Four generations of high-dimensional neural network potentials.
  \emph{Chem. Rev.} \textbf{2021}, \emph{121}, 10037--10072\relax
\mciteBstWouldAddEndPuncttrue
\mciteSetBstMidEndSepPunct{\mcitedefaultmidpunct}
{\mcitedefaultendpunct}{\mcitedefaultseppunct}\relax
\EndOfBibitem
\bibitem[Deringer \latin{et~al.}(2019)Deringer, Caro, and
  Cs\'{a}nyi]{Deringer2019}
Deringer,~V.~L.; Caro,~M.~A.; Cs\'{a}nyi,~G. Machine learning interatomic
  potentials as emerging tools for materials science. \emph{Adv.Mater.}
  \textbf{2019}, \emph{31}, 1902765\relax
\mciteBstWouldAddEndPuncttrue
\mciteSetBstMidEndSepPunct{\mcitedefaultmidpunct}
{\mcitedefaultendpunct}{\mcitedefaultseppunct}\relax
\EndOfBibitem
\bibitem[Unke \latin{et~al.}(2021)Unke, Chmiela, Sauceda, Gastegger, Poltavsky,
  Sch{\"u}tt, Tkatchenko, and M\"{u}ller]{Unke2021}
Unke,~O.~T.; Chmiela,~S.; Sauceda,~H.~E.; Gastegger,~M.; Poltavsky,~I.;
  Sch{\"u}tt,~K.~T.; Tkatchenko,~A.; M\"{u}ller,~K.-R. Machine learning force
  fields. \emph{Chem. Rev.} \textbf{2021}, \emph{121}, 10142--10186\relax
\mciteBstWouldAddEndPuncttrue
\mciteSetBstMidEndSepPunct{\mcitedefaultmidpunct}
{\mcitedefaultendpunct}{\mcitedefaultseppunct}\relax
\EndOfBibitem
\bibitem[Deringer \latin{et~al.}(2021)Deringer, Bart\'{o}k, Bernstein, Wilkins,
  Ceriotti, and Cs\'{a}nyi]{Deringer2021}
Deringer,~V.~L.; Bart\'{o}k,~A.~P.; Bernstein,~N.; Wilkins,~D.~M.;
  Ceriotti,~M.; Cs\'{a}nyi,~G. Gaussian process regression for materials and
  molecules. \emph{Chem. Rev.} \textbf{2021}, \emph{121}, 10073--10141\relax
\mciteBstWouldAddEndPuncttrue
\mciteSetBstMidEndSepPunct{\mcitedefaultmidpunct}
{\mcitedefaultendpunct}{\mcitedefaultseppunct}\relax
\EndOfBibitem
\bibitem[Batzner \latin{et~al.}(2023)Batzner, Musaelian, and
  Kozinsky]{Batzner2023}
Batzner,~S.; Musaelian,~A.; Kozinsky,~B. Advancing molecular simulation with
  equivariant interatomic potentials. \emph{Nat. Rev. Phys.} \textbf{2023},
  \emph{5}, 437--438\relax
\mciteBstWouldAddEndPuncttrue
\mciteSetBstMidEndSepPunct{\mcitedefaultmidpunct}
{\mcitedefaultendpunct}{\mcitedefaultseppunct}\relax
\EndOfBibitem
\bibitem[Montes-Campos \latin{et~al.}(2022)Montes-Campos, Carrete, Bichelmaier,
  Varela, and Madsen]{Campos2022}
Montes-Campos,~H.; Carrete,~J.; Bichelmaier,~S.; Varela,~L.~M.; Madsen,~G.
  K.~H. A differentiable neural-network force field for ionic liquids. \emph{J.
  Chem. Inf. Model.} \textbf{2022}, \emph{62}, 88--101\relax
\mciteBstWouldAddEndPuncttrue
\mciteSetBstMidEndSepPunct{\mcitedefaultmidpunct}
{\mcitedefaultendpunct}{\mcitedefaultseppunct}\relax
\EndOfBibitem
\bibitem[Ling \latin{et~al.}(2023)Ling, Li, Wang, Lu, Wang, Wang, and
  He]{Ling2023}
Ling,~Y.; Li,~K.; Wang,~M.; Lu,~J.; Wang,~C.; Wang,~Y.; He,~H. Revisiting the
  structure, interaction, and dynamical property of ionic liquid from the deep
  learning force field. \emph{J Power Sources} \textbf{2023}, \emph{555},
  232350\relax
\mciteBstWouldAddEndPuncttrue
\mciteSetBstMidEndSepPunct{\mcitedefaultmidpunct}
{\mcitedefaultendpunct}{\mcitedefaultseppunct}\relax
\EndOfBibitem
\bibitem[Tovey \latin{et~al.}(2020)Tovey, Krishnamoorthy, Sivaraman, Guo,
  Benmore, Heuer, and Holm]{Tovey2020}
Tovey,~S.; Krishnamoorthy,~A.~N.; Sivaraman,~G.; Guo,~J.; Benmore,~C.;
  Heuer,~A.; Holm,~C. DFT accurate interatomic potential for molten NaCl from
  machine learning. \emph{J. Phys. Chem. C} \textbf{2020}, \emph{124},
  25760--25768\relax
\mciteBstWouldAddEndPuncttrue
\mciteSetBstMidEndSepPunct{\mcitedefaultmidpunct}
{\mcitedefaultendpunct}{\mcitedefaultseppunct}\relax
\EndOfBibitem
\bibitem[Li \latin{et~al.}(2021)Li, K\"{u}\c{c}\"{u}kbenli, Lam, Khaykovich,
  Kaxiras, and Li]{Li2021CP}
Li,~Q.-J.; K\"{u}\c{c}\"{u}kbenli,~E.; Lam,~S.; Khaykovich,~B.; Kaxiras,~E.;
  Li,~J. Development of robust neural-network interatomic potential for molten
  salt. \emph{Cell Rep.} \textbf{2021}, \emph{2}, 100359\relax
\mciteBstWouldAddEndPuncttrue
\mciteSetBstMidEndSepPunct{\mcitedefaultmidpunct}
{\mcitedefaultendpunct}{\mcitedefaultseppunct}\relax
\EndOfBibitem
\bibitem[Mondal \latin{et~al.}(2023)Mondal, Kussainova, Yue, and
  Panagiotopoulos]{Mondal2023}
Mondal,~A.; Kussainova,~D.; Yue,~S.; Panagiotopoulos,~A.~Z. Modeling chemical
  reactions in alkali carbonate-hydroxide electrolytes with deep learning
  potentials. \emph{J. Chem. Theory Comput.} \textbf{2023}, \emph{19},
  4584--4595\relax
\mciteBstWouldAddEndPuncttrue
\mciteSetBstMidEndSepPunct{\mcitedefaultmidpunct}
{\mcitedefaultendpunct}{\mcitedefaultseppunct}\relax
\EndOfBibitem
\bibitem[Liang \latin{et~al.}(2020)Liang, Lu, and Yu]{Liang2020}
Liang,~W.; Lu,~G.; Yu,~J. Molecular dynamics simulations of molten magnesium
  chloride using machine-learning-based deep potential. \emph{Adv. Theory
  Simul.} \textbf{2020}, \emph{3}, 2000180\relax
\mciteBstWouldAddEndPuncttrue
\mciteSetBstMidEndSepPunct{\mcitedefaultmidpunct}
{\mcitedefaultendpunct}{\mcitedefaultseppunct}\relax
\EndOfBibitem
\bibitem[Batzner \latin{et~al.}(2021)Batzner, Musaelian, Sun, Geiger, Mailoa,
  Kornbluth, Molinari, Smidt, and Kozinsky]{batzner2021se}
Batzner,~S.; Musaelian,~A.; Sun,~L.; Geiger,~M.; Mailoa,~J.~P.; Kornbluth,~M.;
  Molinari,~N.; Smidt,~T.~E.; Kozinsky,~B. E(3)-equivariant graph neural
  networks for data-efficient and accurate interatomic potentials. \emph{Nat.
  Commun} \textbf{2021}, \emph{13}, 2453\relax
\mciteBstWouldAddEndPuncttrue
\mciteSetBstMidEndSepPunct{\mcitedefaultmidpunct}
{\mcitedefaultendpunct}{\mcitedefaultseppunct}\relax
\EndOfBibitem
\bibitem[Musaelian \latin{et~al.}(2023)Musaelian, Batzner, Johansson, Sun,
  Owen, Kornbluth, and Kozinsky]{Musaelian2023}
Musaelian,~A.; Batzner,~S.; Johansson,~A.; Sun,~L.; Owen,~C.~J.; Kornbluth,~M.;
  Kozinsky,~B. Learning local equivariant representations for large-scale
  atomistic dynamics. \emph{Nat. Commun} \textbf{2023}, \emph{14}, 579\relax
\mciteBstWouldAddEndPuncttrue
\mciteSetBstMidEndSepPunct{\mcitedefaultmidpunct}
{\mcitedefaultendpunct}{\mcitedefaultseppunct}\relax
\EndOfBibitem
\bibitem[Musaelian \latin{et~al.}(2023)Musaelian, Johansson, Batzner, and
  Kozinsky]{Musaelian2023bio}
Musaelian,~A.; Johansson,~A.; Batzner,~S.; Kozinsky,~B. Scaling the leading
  accuracy of deep equivariant models to biomolecular simulations of realistic
  size. \emph{arXiv:2304.10061 (accessed 2024-02-18)} \textbf{2023}, \relax
\mciteBstWouldAddEndPunctfalse
\mciteSetBstMidEndSepPunct{\mcitedefaultmidpunct}
{}{\mcitedefaultseppunct}\relax
\EndOfBibitem
\bibitem[Batatia \latin{et~al.}(2022)Batatia, Kov\'{a}cs, Simm, Ortner, and
  Cs\'{a}nyi]{Batatia2023}
Batatia,~I.; Kov\'{a}cs,~D.~P.; Simm,~G. N.~C.; Ortner,~C.; Cs\'{a}nyi,~G.
  MACE: higher order equivariant message passing neural networks for fast and
  accurate force fields. \emph{arXiv:2206.07697 (accessed 2024-02-04)}
  \textbf{2022}, \relax
\mciteBstWouldAddEndPunctfalse
\mciteSetBstMidEndSepPunct{\mcitedefaultmidpunct}
{}{\mcitedefaultseppunct}\relax
\EndOfBibitem
\bibitem[Molinari \latin{et~al.}(2019)Molinari, Mailoa, and
  Kozinsky]{molinari2019general}
Molinari,~N.; Mailoa,~J.~P.; Kozinsky,~B. General trend of a negative Li
  effective charge in ionic liquid electrolytes. \emph{J. Phys. Chem. Lett.}
  \textbf{2019}, \emph{10}, 2313--2319\relax
\mciteBstWouldAddEndPuncttrue
\mciteSetBstMidEndSepPunct{\mcitedefaultmidpunct}
{\mcitedefaultendpunct}{\mcitedefaultseppunct}\relax
\EndOfBibitem
\bibitem[Molinari \latin{et~al.}(2019)Molinari, Mailoa, Craig, Christensen, and
  Kozinsky]{Molinari2019anomalies}
Molinari,~N.; Mailoa,~J.~P.; Craig,~N.; Christensen,~J.; Kozinsky,~B. Transport
  anomalies emerging from strong correlation in ionic liquid electrolytes.
  \emph{J Power Sources} \textbf{2019}, \emph{428}, 27--36\relax
\mciteBstWouldAddEndPuncttrue
\mciteSetBstMidEndSepPunct{\mcitedefaultmidpunct}
{\mcitedefaultendpunct}{\mcitedefaultseppunct}\relax
\EndOfBibitem
\bibitem[McEldrew \latin{et~al.}(2021)McEldrew, Goodwin, Molinari, Kozinsky,
  Kornyshev, and Bazant]{McEldrew2021}
McEldrew,~M.; Goodwin,~Z. A.~H.; Molinari,~N.; Kozinsky,~B.; Kornyshev,~A.~A.;
  Bazant,~M.~Z. Salt-in-ionic-liquid electrolytes: Ion network formation and
  negative effective charges of alkali metal cations. \emph{J. Phys. Chem. B}
  \textbf{2021}, \emph{125}, 13752--13766\relax
\mciteBstWouldAddEndPuncttrue
\mciteSetBstMidEndSepPunct{\mcitedefaultmidpunct}
{\mcitedefaultendpunct}{\mcitedefaultseppunct}\relax
\EndOfBibitem
\bibitem[Goodwin \latin{et~al.}(2023)Goodwin, McEldrew, Kozinsky, and
  Bazant]{Goodwin23PRXE}
Goodwin,~Z.~A.; McEldrew,~M.; Kozinsky,~B.; Bazant,~M.~Z. Theory of cation
  solvation and ionic association in nonaqueous solvent mixtures. \emph{PRX
  Energy} \textbf{2023}, \emph{2}, 013007\relax
\mciteBstWouldAddEndPuncttrue
\mciteSetBstMidEndSepPunct{\mcitedefaultmidpunct}
{\mcitedefaultendpunct}{\mcitedefaultseppunct}\relax
\EndOfBibitem
\bibitem[Gouverneur \latin{et~al.}(2018)Gouverneur, Schmidt, and
  Sch{\"o}nhoff]{gouverneur2018negative}
Gouverneur,~M.; Schmidt,~F.; Sch{\"o}nhoff,~M. Negative effective Li
  transference numbers in Li salt/ionic liquid mixtures: does Li drift in the
  ``wrong'' direction? \emph{Phys. Chem. Chem. Phys.} \textbf{2018}, \emph{20},
  7470--7478\relax
\mciteBstWouldAddEndPuncttrue
\mciteSetBstMidEndSepPunct{\mcitedefaultmidpunct}
{\mcitedefaultendpunct}{\mcitedefaultseppunct}\relax
\EndOfBibitem
\bibitem[Brinkk\"otter \latin{et~al.}(2021)Brinkk\"otter, Mariani, Jeong,
  Passerini, and Sch\"onhoff]{Marc2021}
Brinkk\"otter,~M.; Mariani,~A.; Jeong,~S.; Passerini,~S.; Sch\"onhoff,~M. Ionic
  liquid in Li salt electrolyte: modifying the Li$^+$ transport mechanism by
  coordination to an asymmetric anion. \emph{Adv. Energy Sustainability Res.}
  \textbf{2021}, \emph{2}, 2000078\relax
\mciteBstWouldAddEndPuncttrue
\mciteSetBstMidEndSepPunct{\mcitedefaultmidpunct}
{\mcitedefaultendpunct}{\mcitedefaultseppunct}\relax
\EndOfBibitem
\bibitem[Wenny \latin{et~al.}(2023)Wenny, Walter, Slavney, and
  Mason]{Wenny2023}
Wenny,~M.~B.; Walter,~M.~V.; Slavney,~A.~H.; Mason,~J.~A. Generalizable
  synthesis of highly fluorinated ionic liquids. \emph{J. Phys. Chem. B}
  \textbf{2023}, \emph{127}, 2028--2033\relax
\mciteBstWouldAddEndPuncttrue
\mciteSetBstMidEndSepPunct{\mcitedefaultmidpunct}
{\mcitedefaultendpunct}{\mcitedefaultseppunct}\relax
\EndOfBibitem
\bibitem[Fadel \latin{et~al.}(2019)Fadel, Faglioni, Samsonidze, Molinari,
  Merinov, Goddard~III, Grossman, Mailoa, and Kozinsky]{fadel2019role}
Fadel,~E.~R.; Faglioni,~F.; Samsonidze,~G.; Molinari,~N.; Merinov,~B.~V.;
  Goddard~III,~W.~A.; Grossman,~J.~C.; Mailoa,~J.~P.; Kozinsky,~B. Role of
  solvent-anion charge transfer in oxidative degradation of battery
  electrolytes. \emph{Nat. Commun} \textbf{2019}, \emph{10}, 1--10\relax
\mciteBstWouldAddEndPuncttrue
\mciteSetBstMidEndSepPunct{\mcitedefaultmidpunct}
{\mcitedefaultendpunct}{\mcitedefaultseppunct}\relax
\EndOfBibitem
\bibitem[Molinari and Kozinsky(2020)Molinari, and
  Kozinsky]{molinari2020chelation}
Molinari,~N.; Kozinsky,~B. Chelation-induced reversal of negative cation
  transference number in ionic liquid electrolytes. \emph{J. Phys. Chem. B}
  \textbf{2020}, \emph{124}, 2676--2684\relax
\mciteBstWouldAddEndPuncttrue
\mciteSetBstMidEndSepPunct{\mcitedefaultmidpunct}
{\mcitedefaultendpunct}{\mcitedefaultseppunct}\relax
\EndOfBibitem
\bibitem[Molinari \latin{et~al.}(2016)Molinari, Khawaja, Sutton, and
  Mostofi]{molinari2016molecular}
Molinari,~N.; Khawaja,~M.; Sutton,~A.; Mostofi,~A. Molecular model for HNBR
  with tunable cross-link density. \emph{J Phys. Chem B} \textbf{2016},
  \emph{120}, 12700--12707\relax
\mciteBstWouldAddEndPuncttrue
\mciteSetBstMidEndSepPunct{\mcitedefaultmidpunct}
{\mcitedefaultendpunct}{\mcitedefaultseppunct}\relax
\EndOfBibitem
\bibitem[Hoover(1985)]{hoover1985canonical}
Hoover,~W.~G. Canonical dynamics: equilibrium phase-space distributions.
  \emph{Phys. Rev. A} \textbf{1985}, \emph{31}, 1695\relax
\mciteBstWouldAddEndPuncttrue
\mciteSetBstMidEndSepPunct{\mcitedefaultmidpunct}
{\mcitedefaultendpunct}{\mcitedefaultseppunct}\relax
\EndOfBibitem
\bibitem[Nos{\'e}(1984)]{nose1984unified}
Nos{\'e},~S. A unified formulation of the constant temperature molecular
  dynamics methods. \emph{J. Chem. Phys.} \textbf{1984}, \emph{81},
  511--519\relax
\mciteBstWouldAddEndPuncttrue
\mciteSetBstMidEndSepPunct{\mcitedefaultmidpunct}
{\mcitedefaultendpunct}{\mcitedefaultseppunct}\relax
\EndOfBibitem
\bibitem[Hoover(1986)]{hoover1986constant}
Hoover,~W.~G. Constant-pressure equations of motion. \emph{Phys. Rev. A}
  \textbf{1986}, \emph{34}, 2499\relax
\mciteBstWouldAddEndPuncttrue
\mciteSetBstMidEndSepPunct{\mcitedefaultmidpunct}
{\mcitedefaultendpunct}{\mcitedefaultseppunct}\relax
\EndOfBibitem
\bibitem[Plimpton(1995)]{plimpton1995fast}
Plimpton,~S. Fast Parallel Algorithms for Short-range Molecular Dynamics.
  \emph{J. Comput. Phys.} \textbf{1995}, \emph{117}, 1--19\relax
\mciteBstWouldAddEndPuncttrue
\mciteSetBstMidEndSepPunct{\mcitedefaultmidpunct}
{\mcitedefaultendpunct}{\mcitedefaultseppunct}\relax
\EndOfBibitem
\bibitem[Thompson \latin{et~al.}(2022)Thompson, Aktulga, Berger, Bolintineanu,
  Brown, Crozier, in~'t Veld, Kohlmeyer, Moore, Nguyen, Shan, Stevens,
  J.~Tranchida, and Plimpton]{LAMMPS2022}
Thompson,~A.~P.; Aktulga,~H.~M.; Berger,~R.; Bolintineanu,~D.~S.; Brown,~W.~M.;
  Crozier,~P.~S.; in~'t Veld,~P.~J.; Kohlmeyer,~A.; Moore,~S.~G.; Nguyen,~T.~D.
  \latin{et~al.}  LAMMPS - a flexible simulation tool for particle-based
  materials modeling at the atomic, meso, and continuum scales. \emph{Comp Phys
  Comm} \textbf{2022}, \emph{271}, 10817\relax
\mciteBstWouldAddEndPuncttrue
\mciteSetBstMidEndSepPunct{\mcitedefaultmidpunct}
{\mcitedefaultendpunct}{\mcitedefaultseppunct}\relax
\EndOfBibitem
\bibitem[Jorgensen and Tirado-Rives(1988)Jorgensen, and
  Tirado-Rives]{jorgensen1988opls}
Jorgensen,~W.~L.; Tirado-Rives,~J. The OPLS [optimized potentials for liquid
  simulations] potential functions for proteins, energy minimizations for
  crystals of cyclic peptides and crambin. \emph{JACS} \textbf{1988},
  \emph{110}, 1657--1666\relax
\mciteBstWouldAddEndPuncttrue
\mciteSetBstMidEndSepPunct{\mcitedefaultmidpunct}
{\mcitedefaultendpunct}{\mcitedefaultseppunct}\relax
\EndOfBibitem
\bibitem[Doherty \latin{et~al.}(2017)Doherty, Zhong, Gathiaka, Li, and
  Acevedo]{doherty2017revisiting}
Doherty,~B.; Zhong,~X.; Gathiaka,~S.; Li,~B.; Acevedo,~O. Revisiting OPLS force
  field parameters for ionic liquid simulations. \emph{J. Chem. Theory Comput.}
  \textbf{2017}, \emph{13}, 6131--6145\relax
\mciteBstWouldAddEndPuncttrue
\mciteSetBstMidEndSepPunct{\mcitedefaultmidpunct}
{\mcitedefaultendpunct}{\mcitedefaultseppunct}\relax
\EndOfBibitem
\bibitem[Mogurampelly and Ganesan(2017)Mogurampelly, and
  Ganesan]{mogurampelly2017structure}
Mogurampelly,~S.; Ganesan,~V. Structure and mechanisms underlying ion transport
  in ternary polymer electrolytes containing ionic liquids. \emph{J. Chem.
  Phys.} \textbf{2017}, \emph{146}, 074902\relax
\mciteBstWouldAddEndPuncttrue
\mciteSetBstMidEndSepPunct{\mcitedefaultmidpunct}
{\mcitedefaultendpunct}{\mcitedefaultseppunct}\relax
\EndOfBibitem
\bibitem[Pal \latin{et~al.}(2017)Pal, Beck, Lessnich, and
  Vogel]{pal2017effects}
Pal,~T.; Beck,~C.; Lessnich,~D.; Vogel,~M. Effects of silica surfaces on the
  structure and dynamics of room-temperature ionic liquids: a molecular
  dynamics simulation study. \emph{J. Phys. Chem. C} \textbf{2017}, \emph{122},
  624--634\relax
\mciteBstWouldAddEndPuncttrue
\mciteSetBstMidEndSepPunct{\mcitedefaultmidpunct}
{\mcitedefaultendpunct}{\mcitedefaultseppunct}\relax
\EndOfBibitem
\bibitem[Giannozzi \latin{et~al.}(2009)Giannozzi, Baroni, Bonini, Calandra,
  Car, Cavazzoni, Ceresoli, Chiarotti, Cococcioni, Dabo, Corso, de~Gironcoli,
  Fabris, Fratesi, Gebauer, Gerstmann, Gougoussis, Kokalj, Lazzeri,
  Martin-Samos, Marzari, Mauri, Mazzarello, Paolini, Pasquarello, Paulatto,
  Sbraccia, Scandolo, Sclauzero, Seitsonen, Smogunov, Umari, and
  Wentzcovitch]{Giannozzi2009}
Giannozzi,~P.; Baroni,~S.; Bonini,~N.; Calandra,~M.; Car,~R.; Cavazzoni,~C.;
  Ceresoli,~D.; Chiarotti,~G.~L.; Cococcioni,~M.; Dabo,~I. \latin{et~al.}
  QUANTUM ESPRESSO: a modular and open-source software project for quantum
  simulations of materials. \emph{J.Phys.: Condens.Matter} \textbf{2009},
  \emph{21}, 395502\relax
\mciteBstWouldAddEndPuncttrue
\mciteSetBstMidEndSepPunct{\mcitedefaultmidpunct}
{\mcitedefaultendpunct}{\mcitedefaultseppunct}\relax
\EndOfBibitem
\bibitem[Giannozzi \latin{et~al.}(2017)Giannozzi, Andreussi, Brumme, Bunau,
  Nardelli, Calandra, Car, Cavazzoni, Ceresoli, Cococcioni, Colonna, Carnimeo,
  Corso, de~Gironcoli, Delugas, Jr, Ferretti, Floris, Fratesi, Fugallo,
  Gebauer, Gerstmann, Giustino, Gorni, Jia, Kawamura, Ko, Kokalj,
  K\"u\c{c}\"ukbenli, Lazzeri, Marsili, Marzari, Mauri, Nguyen, Nguyen, de-la
  Roza, Paulatto, Ponc\'e, Rocca, Sabatini, Santra, Schlipf, Seitsonen,
  Smogunov, Timrov, Thonhauser, Umari, Vast, Wu, and Baroni]{Giannozzi2017}
Giannozzi,~P.; Andreussi,~O.; Brumme,~T.; Bunau,~O.; Nardelli,~M.~B.;
  Calandra,~M.; Car,~R.; Cavazzoni,~C.; Ceresoli,~D.; Cococcioni,~M.
  \latin{et~al.}  Advanced capabilities for materials modelling with Quantum
  ESPRESSO. \emph{J.Phys.: Condens.Matter} \textbf{2017}, \emph{29},
  465901\relax
\mciteBstWouldAddEndPuncttrue
\mciteSetBstMidEndSepPunct{\mcitedefaultmidpunct}
{\mcitedefaultendpunct}{\mcitedefaultseppunct}\relax
\EndOfBibitem
\bibitem[Perdew \latin{et~al.}(1996)Perdew, Burke, and Ernzerho]{Perdew1996}
Perdew,~J.; Burke,~K.; Ernzerho,~M. Generalized gradient approximation made
  simple. \emph{Phys. Rev. Lett.} \textbf{1996}, \emph{77}, 3865--3868\relax
\mciteBstWouldAddEndPuncttrue
\mciteSetBstMidEndSepPunct{\mcitedefaultmidpunct}
{\mcitedefaultendpunct}{\mcitedefaultseppunct}\relax
\EndOfBibitem
\bibitem[Grimme \latin{et~al.}(2010)Grimme, Antony, Ehrlich, and
  Krieg]{Grimme2010}
Grimme,~S.; Antony,~J.; Ehrlich,~S.; Krieg,~H. A consistent and accurate ab
  initio parametrization of density functional dispersion correction (DFT-D)
  for the 94 elements H-Pu. \emph{J. Chem. Phys} \textbf{2010}, \emph{132},
  154104\relax
\mciteBstWouldAddEndPuncttrue
\mciteSetBstMidEndSepPunct{\mcitedefaultmidpunct}
{\mcitedefaultendpunct}{\mcitedefaultseppunct}\relax
\EndOfBibitem
\bibitem[Garrity \latin{et~al.}(2014)Garrity, Bennett, Rabe, and
  Vanderbilt]{Garrity2014}
Garrity,~K.~F.; Bennett,~J.~W.; Rabe,~K.~M.; Vanderbilt,~D. Pseudopotentials
  for high-throughput DFT calculations. \emph{Comput. Mater. Sci.}
  \textbf{2014}, \emph{81}, 446--452\relax
\mciteBstWouldAddEndPuncttrue
\mciteSetBstMidEndSepPunct{\mcitedefaultmidpunct}
{\mcitedefaultendpunct}{\mcitedefaultseppunct}\relax
\EndOfBibitem
\bibitem[Zhang \latin{et~al.}(2018)Zhang, Han, Wang, Car, and E]{Zhang2018}
Zhang,~L.; Han,~J.; Wang,~H.; Car,~R.; E,~W. Deep potential molecular dynamics:
  a scalable model with the accuracy of quantum mechanics. \emph{Phys. Rev.
  Lett.} \textbf{2018}, \emph{120}, 143001\relax
\mciteBstWouldAddEndPuncttrue
\mciteSetBstMidEndSepPunct{\mcitedefaultmidpunct}
{\mcitedefaultendpunct}{\mcitedefaultseppunct}\relax
\EndOfBibitem
\bibitem[Batatia \latin{et~al.}(2023)Batatia, Benner, Chiang, Elena, Kov\'acs,
  Riebesell, Advincula, Asta, Avaylon, Baldwin, Berger, Bernstein, Bhowmik,
  Blau, C\u{a}rare, Darby, De, Pia, Deringer, Elijo\v{s}ius, El-Machachi,
  Falcioni, Fako, Ferrari, Genreith-Schriever, George, Goodall, Grey, Grigorev,
  Han, Handley, Heenen, Hermansson, Holm, Jaafar, Hofmann, Jakob, Jung, Kapil,
  Kaplan, Karimitari, Kermode, Kroupa, Kullgren, Kuner, Kuryla, Liepuoniute,
  Margraf, Magd\u{a}u, Michaelides, Moore, Naik, Niblett, Norwood, O'Neill,
  Ortner, Persson, Reuter, Rosen, Schaaf, Schran, Shi, Sivonxay, Stenczel,
  Svahn, Sutton, Swinburne, Tilly, van~der Oord, Varga-Umbrich, Vegge,
  Vondr\'ak, Wang, Witt, Zills, and Cs\'anyi]{Batatia2023FM}
Batatia,~I.; Benner,~P.; Chiang,~Y.; Elena,~A.~M.; Kov\'acs,~D.~P.;
  Riebesell,~J.; Advincula,~X.~R.; Asta,~M.; Avaylon,~M.; Baldwin,~W.~J.
  \latin{et~al.}  A foundation model for atomistic materials chemistry.
  \emph{arXiv:2401.00096 (accessed 2024-01-12)} \textbf{2023}, \relax
\mciteBstWouldAddEndPunctfalse
\mciteSetBstMidEndSepPunct{\mcitedefaultmidpunct}
{}{\mcitedefaultseppunct}\relax
\EndOfBibitem
\bibitem[Zhu \latin{et~al.}(2023)Zhu, Batzner, Musaelian, and
  Kozinsky]{Zhu2023}
Zhu,~A.; Batzner,~S.; Musaelian,~A.; Kozinsky,~B. Fast uncertainty estimates in
  deep learning interatomic potentials. \emph{J. Chem. Phys.} \textbf{2023},
  \emph{158}, 164111\relax
\mciteBstWouldAddEndPuncttrue
\mciteSetBstMidEndSepPunct{\mcitedefaultmidpunct}
{\mcitedefaultendpunct}{\mcitedefaultseppunct}\relax
\EndOfBibitem
\bibitem[Vandermause \latin{et~al.}(2020)Vandermause, Torrisi, Batzner, Xie,
  Sun, Kolpak, and Kozinsky]{Vandermause2020}
Vandermause,~J.; Torrisi,~S.~B.; Batzner,~S.; Xie,~Y.; Sun,~L.; Kolpak,~A.~M.;
  Kozinsky,~B. On-the-fly active learning of interpretable Bayesian force
  fields for atomistic rare events. \emph{npj Comput Mater} \textbf{2020},
  \emph{6}, 20\relax
\mciteBstWouldAddEndPuncttrue
\mciteSetBstMidEndSepPunct{\mcitedefaultmidpunct}
{\mcitedefaultendpunct}{\mcitedefaultseppunct}\relax
\EndOfBibitem
\bibitem[Xie \latin{et~al.}(2023)Xie, Vandermause, Ramakers, Protik, Johansson,
  and Kozinsky]{Xie2023}
Xie,~Y.; Vandermause,~J.; Ramakers,~S.; Protik,~N.~H.; Johansson,~A.;
  Kozinsky,~B. Uncertainty-aware molecular dynamics from bayesian active
  learning: phase transformations and thermal transport in SiC. \emph{npj
  Comput. Mater.} \textbf{2023}, \emph{9}, 36\relax
\mciteBstWouldAddEndPuncttrue
\mciteSetBstMidEndSepPunct{\mcitedefaultmidpunct}
{\mcitedefaultendpunct}{\mcitedefaultseppunct}\relax
\EndOfBibitem
\bibitem[Yoo \latin{et~al.}(2019)Yoo, Lee, Jeong, Lee, Watanabe, and
  Han]{Yoo19E}
Yoo,~D.; Lee,~K.; Jeong,~W.; Lee,~D.; Watanabe,~S.; Han,~S. Atomic energy
  mapping of neural network potential. \emph{Phys. Rev. Materials}
  \textbf{2019}, \emph{3}, 093802\relax
\mciteBstWouldAddEndPuncttrue
\mciteSetBstMidEndSepPunct{\mcitedefaultmidpunct}
{\mcitedefaultendpunct}{\mcitedefaultseppunct}\relax
\EndOfBibitem
\bibitem[Chong \latin{et~al.}(2023)Chong, Grasselli, Mahmoud, Morrow, Deringer,
  and Ceriotti]{Chong2023}
Chong,~S.; Grasselli,~F.; Mahmoud,~C.~B.; Morrow,~J.~D.; Deringer,~V.~L.;
  Ceriotti,~M. Robustness of local predictions in atomistic machine learning
  models. \emph{J. Chem. Theory Comput.} \textbf{2023}, \emph{19},
  8020--8031\relax
\mciteBstWouldAddEndPuncttrue
\mciteSetBstMidEndSepPunct{\mcitedefaultmidpunct}
{\mcitedefaultendpunct}{\mcitedefaultseppunct}\relax
\EndOfBibitem
\bibitem[Chahal \latin{et~al.}(2022)Chahal, Roy, Brehm, Banerjee, Bryantsev,
  and Lam]{Rajni2022}
Chahal,~R.; Roy,~S.; Brehm,~M.; Banerjee,~S.; Bryantsev,~V.; Lam,~S.~T.
  Transferable deep learning potential reveals intermediate-range ordering
  effects in LiF-NaF-ZrF$_4$ molten salt. \emph{JACS Au} \textbf{2022},
  \emph{2}, 2693--2702\relax
\mciteBstWouldAddEndPuncttrue
\mciteSetBstMidEndSepPunct{\mcitedefaultmidpunct}
{\mcitedefaultendpunct}{\mcitedefaultseppunct}\relax
\EndOfBibitem
\bibitem[Chen and Ong(2022)Chen, and Ong]{Chen2022}
Chen,~C.; Ong,~S.~P. A universal graph deep learning interatomic potential for
  the periodic table. \emph{Nat. Comput. Sci.} \textbf{2022}, \emph{2},
  718--728\relax
\mciteBstWouldAddEndPuncttrue
\mciteSetBstMidEndSepPunct{\mcitedefaultmidpunct}
{\mcitedefaultendpunct}{\mcitedefaultseppunct}\relax
\EndOfBibitem
\bibitem[Deng \latin{et~al.}(2023)Deng, Zhong, Jun, Riebesell, Han, Bartel, and
  Ceder]{Deng2023}
Deng,~B.; Zhong,~P.; Jun,~K.; Riebesell,~J.; Han,~K.; Bartel,~C.~J.; Ceder,~G.
  Chgnet as a pretrained universal neural network potential for charge-informed
  atomistic modelling. \emph{Nat. Mach. Intell.} \textbf{2023}, \emph{5},
  1031--1041\relax
\mciteBstWouldAddEndPuncttrue
\mciteSetBstMidEndSepPunct{\mcitedefaultmidpunct}
{\mcitedefaultendpunct}{\mcitedefaultseppunct}\relax
\EndOfBibitem
\bibitem[Merchant \latin{et~al.}(2023)Merchant, Batzner, Schoenholz, Aykol,
  Cheon, and Cubuk]{Merchant2023}
Merchant,~A.; Batzner,~S.; Schoenholz,~S.~S.; Aykol,~M.; Cheon,~G.;
  Cubuk,~E.~D. Scaling deep learning for materials discovery. \emph{Nature}
  \textbf{2023}, \emph{624}, 80--85\relax
\mciteBstWouldAddEndPuncttrue
\mciteSetBstMidEndSepPunct{\mcitedefaultmidpunct}
{\mcitedefaultendpunct}{\mcitedefaultseppunct}\relax
\EndOfBibitem
\bibitem[Kov\'{a}cs \latin{et~al.}(2023)Kov\'{a}cs, Moore, Browning, Batatia,
  Horton, Kapil, Witt, Cole, and Cs\'{a}nyi]{Kovacs2023}
Kov\'{a}cs,~D.~P.; Moore,~J.~H.; Browning,~N.~J.; Batatia,~I.; Horton,~J.~T.;
  Kapil,~V.; Witt,~W.~C.; Cole,~I.-B. M. D.~J.; Cs\'{a}nyi,~G. MACE-OFF23:
  transferable machine learning force fields for organic molecules.
  \emph{arXiv:2312.15211 (accessed 2014-01-12)} \textbf{2023}, \relax
\mciteBstWouldAddEndPunctfalse
\mciteSetBstMidEndSepPunct{\mcitedefaultmidpunct}
{}{\mcitedefaultseppunct}\relax
\EndOfBibitem
\bibitem[Yang \latin{et~al.}(2024)Yang, Hu, Zhou, Liu, Shi, Li, Li, Chen, Chen,
  Zeni, Horton, Pinsler, Fowler, Z{\"u}gner, Xie, Smith, Sun, Wang, Kong, Liu,
  Hao, and Lu]{Yang24MatterSim}
Yang,~H.; Hu,~C.; Zhou,~Y.; Liu,~X.; Shi,~Y.; Li,~J.; Li,~G.; Chen,~Z.;
  Chen,~S.; Zeni,~C. \latin{et~al.}  MatterSim: a deep learning atomistic model
  across elements, temperatures and pressures. \emph{arXiv:2405.04967
  (2024-06-22)} \textbf{2024}, \relax
\mciteBstWouldAddEndPunctfalse
\mciteSetBstMidEndSepPunct{\mcitedefaultmidpunct}
{}{\mcitedefaultseppunct}\relax
\EndOfBibitem
\bibitem[Jain \latin{et~al.}(2013)Jain, Ong, Hautier, Chen, Richards, Dacek,
  Cholia, Gunter, Skinner, Ceder, and Persson]{Jain2013}
Jain,~A.; Ong,~S.~P.; Hautier,~G.; Chen,~W.; Richards,~W.~D.; Dacek,~S.;
  Cholia,~S.; Gunter,~D.; Skinner,~D.; Ceder,~G. \latin{et~al.}  Commentary:
  The materials project: a materials genome approach to accelerating materials
  innovation. \emph{APL Mater.} \textbf{2013}, \emph{1}, 011002\relax
\mciteBstWouldAddEndPuncttrue
\mciteSetBstMidEndSepPunct{\mcitedefaultmidpunct}
{\mcitedefaultendpunct}{\mcitedefaultseppunct}\relax
\EndOfBibitem
\bibitem[Eastman \latin{et~al.}(2023)Eastman, Behara, Dotson, Galvelis, Herr,
  Horton, Mao, Chodera, Pritchard, Wang, Fabritiis, and Markland]{Eastman2023}
Eastman,~P.; Behara,~P.~K.; Dotson,~D.~L.; Galvelis,~R.; Herr,~J.~E.;
  Horton,~J.~T.; Mao,~Y.; Chodera,~J.~D.; Pritchard,~B.~P.; Wang,~Y.
  \latin{et~al.}  SPICE, a dataset of drug-like molecules and peptides for
  training machine learning potentials. \emph{Scientific Data} \textbf{2023},
  \emph{10}, 11\relax
\mciteBstWouldAddEndPuncttrue
\mciteSetBstMidEndSepPunct{\mcitedefaultmidpunct}
{\mcitedefaultendpunct}{\mcitedefaultseppunct}\relax
\EndOfBibitem
\bibitem[Gong \latin{et~al.}(2024)Gong, Zhang, Mu, Pu, Wang, Yu, Chen, Zheng,
  Wang, Chen, Wu, Shi, Gao, Yan, and Xiang]{Gong24Bamboo}
Gong,~S.; Zhang,~Y.; Mu,~Z.; Pu,~Z.; Wang,~H.; Yu,~Z.; Chen,~M.; Zheng,~T.;
  Wang,~Z.; Chen,~L. \latin{et~al.}  BAMBOO: a predictive and transferable
  machine learning force field framework for liquid electrolyte development.
  \emph{arXiv:2404.07181 (accessed 2024-06-22)} \textbf{2024}, \relax
\mciteBstWouldAddEndPunctfalse
\mciteSetBstMidEndSepPunct{\mcitedefaultmidpunct}
{}{\mcitedefaultseppunct}\relax
\EndOfBibitem
\bibitem[Liu \latin{et~al.}(2014)Liu, Dai, , and en~Jiang]{Liu2014}
Liu,~H.; Dai,~S.; ; en~Jiang,~D. Solubility of gases in a common ionic liquid
  from molecular dynamics based free energy calculations. \emph{J. Phys. Chem.
  B} \textbf{2014}, \emph{118}, 2719--2725\relax
\mciteBstWouldAddEndPuncttrue
\mciteSetBstMidEndSepPunct{\mcitedefaultmidpunct}
{\mcitedefaultendpunct}{\mcitedefaultseppunct}\relax
\EndOfBibitem
\bibitem[Khawaja \latin{et~al.}(2017)Khawaja, Sutton, and Mostofi]{Khawaja2017}
Khawaja,~M.; Sutton,~A.~P.; Mostofi,~A.~A. Molecular simulation of gas
  solubility in nitrile nutadiene rubber. \emph{J. Phys. Chem. B}
  \textbf{2017}, \emph{121}, 287--297\relax
\mciteBstWouldAddEndPuncttrue
\mciteSetBstMidEndSepPunct{\mcitedefaultmidpunct}
{\mcitedefaultendpunct}{\mcitedefaultseppunct}\relax
\EndOfBibitem
\bibitem[Shi \latin{et~al.}(2014)Shi, Thompson, Albenze, Steckel, Nulwala, and
  Luebke]{Shi2014}
Shi,~W.; Thompson,~R.~L.; Albenze,~E.; Steckel,~J.~A.; Nulwala,~H.~B.;
  Luebke,~D.~R. Contribution of the acetate anion to CO$_2$ solubility in ionic
  liquids: theoretical method development and experimental study. \emph{J.
  Phys. Chem. B} \textbf{2014}, \emph{118}, 7383--7394\relax
\mciteBstWouldAddEndPuncttrue
\mciteSetBstMidEndSepPunct{\mcitedefaultmidpunct}
{\mcitedefaultendpunct}{\mcitedefaultseppunct}\relax
\EndOfBibitem
\bibitem[Fukushima \latin{et~al.}(2019)Fukushima, Ushijima, Kumazoe, Koura,
  Shimojo, Shimamura, Misawa, Kalia, Nakano, and Vashishta]{Fukushima2019}
Fukushima,~S.; Ushijima,~E.; Kumazoe,~H.; Koura,~A.; Shimojo,~F.;
  Shimamura,~K.; Misawa,~M.; Kalia,~R.~K.; Nakano,~A.; Vashishta,~P.
  Thermodynamic integration by neural network potentials based on
  first-principles dynamic calculations. \emph{Phys. Rev. B} \textbf{2019},
  \emph{100}, 214108\relax
\mciteBstWouldAddEndPuncttrue
\mciteSetBstMidEndSepPunct{\mcitedefaultmidpunct}
{\mcitedefaultendpunct}{\mcitedefaultseppunct}\relax
\EndOfBibitem
\bibitem[Jinnouchi \latin{et~al.}(2020)Jinnouchi, Karsai, and
  Kresse]{Jinnouchi2020}
Jinnouchi,~R.; Karsai,~F.; Kresse,~G. Making free-energy calculations routine:
  Combining first principles with machine learning. \emph{Phys. Rev. B}
  \textbf{2020}, \emph{101}, 060201(R)\relax
\mciteBstWouldAddEndPuncttrue
\mciteSetBstMidEndSepPunct{\mcitedefaultmidpunct}
{\mcitedefaultendpunct}{\mcitedefaultseppunct}\relax
\EndOfBibitem
\bibitem[Stukowski(2010)]{Stukowski2010}
Stukowski,~A. Visualization and analysis of atomistic simulation data with
  OVITO--the Open Visualization Tool. \emph{Modelling Simul. Mater. Sci. Eng.}
  \textbf{2010}, \emph{18}, 015012\relax
\mciteBstWouldAddEndPuncttrue
\mciteSetBstMidEndSepPunct{\mcitedefaultmidpunct}
{\mcitedefaultendpunct}{\mcitedefaultseppunct}\relax
\EndOfBibitem
\bibitem[Yue \latin{et~al.}(2021)Yue, Muniz, Andrade, Zhang, Car, and
  Panagiotopoulos]{Yue2021LR}
Yue,~S.; Muniz,~M.~C.; Andrade,~M. F.~C.; Zhang,~L.; Car,~R.;
  Panagiotopoulos,~A.~Z. When do short-range atomistic machinelearning models
  fall short? \emph{J. Chem. Phys.} \textbf{2021}, \emph{154}, 034111\relax
\mciteBstWouldAddEndPuncttrue
\mciteSetBstMidEndSepPunct{\mcitedefaultmidpunct}
{\mcitedefaultendpunct}{\mcitedefaultseppunct}\relax
\EndOfBibitem
\bibitem[Zhang \latin{et~al.}(2022)Zhang, Wang, Muniz, Panagiotopoulos, Car,
  and E]{Zhang2022LR}
Zhang,~L.; Wang,~H.; Muniz,~M.~C.; Panagiotopoulos,~A.~Z.; Car,~R.; E,~W. A
  deep potential model with long-range electrostatic interactions. \emph{J.
  Chem. Phys.} \textbf{2022}, \emph{156}, 124107\relax
\mciteBstWouldAddEndPuncttrue
\mciteSetBstMidEndSepPunct{\mcitedefaultmidpunct}
{\mcitedefaultendpunct}{\mcitedefaultseppunct}\relax
\EndOfBibitem
\bibitem[Rumble(2021)]{CRCRumble}
Rumble,~J.~R. \emph{"Compressibility and Expansion Coefficients of Liquids." in
  CRC Handbook of Chemistry and Physics, 102nd Edition (Internet Version
  2021)}; CRC Press/Taylor \& Francis, Boca Raton, FL., 2021\relax
\mciteBstWouldAddEndPuncttrue
\mciteSetBstMidEndSepPunct{\mcitedefaultmidpunct}
{\mcitedefaultendpunct}{\mcitedefaultseppunct}\relax
\EndOfBibitem
\bibitem[Kresse and Hafne(1993)Kresse, and Hafne]{Kresse1993M}
Kresse,~G.; Hafne,~J. Ab initio molecular dynamics for liquid metals.
  \emph{Phys. Rev. B} \textbf{1993}, \emph{47}, 558--561\relax
\mciteBstWouldAddEndPuncttrue
\mciteSetBstMidEndSepPunct{\mcitedefaultmidpunct}
{\mcitedefaultendpunct}{\mcitedefaultseppunct}\relax
\EndOfBibitem
\bibitem[Kresse and Furthmulle(1996)Kresse, and Furthmulle]{Kresse1996E}
Kresse,~G.; Furthmulle,~J. Efficient iterative schemes for ab initio
  total-energy calculations using a plane-wave basis set. \emph{Phys. Rev. B}
  \textbf{1996}, \emph{54}, 11169--11186\relax
\mciteBstWouldAddEndPuncttrue
\mciteSetBstMidEndSepPunct{\mcitedefaultmidpunct}
{\mcitedefaultendpunct}{\mcitedefaultseppunct}\relax
\EndOfBibitem
\bibitem[Kresse and Furthmulle(1996)Kresse, and Furthmulle]{Kresse1996E2}
Kresse,~G.; Furthmulle,~J. Efficiency of ab initio total energy calculations
  for metals and semiconductors using a plane-wave basis set. \emph{Comput.
  Mater. Sci.} \textbf{1996}, \emph{6}, 15--50,\relax
\mciteBstWouldAddEndPuncttrue
\mciteSetBstMidEndSepPunct{\mcitedefaultmidpunct}
{\mcitedefaultendpunct}{\mcitedefaultseppunct}\relax
\EndOfBibitem
\bibitem[Perdew and Zunge(1981)Perdew, and Zunge]{Perdew1981}
Perdew,~J.; Zunge,~A. Self-interaction correction to density-functional
  approximations for many-electron systems. \emph{Phys. Rev. B} \textbf{1981},
  \emph{23}, 5048--5079\relax
\mciteBstWouldAddEndPuncttrue
\mciteSetBstMidEndSepPunct{\mcitedefaultmidpunct}
{\mcitedefaultendpunct}{\mcitedefaultseppunct}\relax
\EndOfBibitem
\bibitem[Deiseroth \latin{et~al.}(2008)Deiseroth, Kong, Eckert, Vannahme,
  Reiner, Zaiss, and Schlosser]{Deiseroth2008}
Deiseroth,~H.-J.; Kong,~S.-T.; Eckert,~H.; Vannahme,~J.; Reiner,~C.; Zaiss,~T.;
  Schlosser,~M. Li6PS5X: a class of crystalline Li-rich solids with an
  unusually high Li$^+$ mobility. \emph{Angew. Chem., Int. Ed.} \textbf{2008},
  \emph{120}, 767--770\relax
\mciteBstWouldAddEndPuncttrue
\mciteSetBstMidEndSepPunct{\mcitedefaultmidpunct}
{\mcitedefaultendpunct}{\mcitedefaultseppunct}\relax
\EndOfBibitem
\bibitem[Zhou \latin{et~al.}(2022)Zhou, Zhang, and Nazar]{Zhou2022}
Zhou,~L.; Zhang,~Q.; Nazar,~L.~F. Li-rich and halide-deficient argyrodite fast
  ion conductors. \emph{Chem. Mater.} \textbf{2022}, \emph{23},
  9634--9643\relax
\mciteBstWouldAddEndPuncttrue
\mciteSetBstMidEndSepPunct{\mcitedefaultmidpunct}
{\mcitedefaultendpunct}{\mcitedefaultseppunct}\relax
\EndOfBibitem
\bibitem[Adeli \latin{et~al.}(2019)Adeli, Bazak, Park, Kochetkov, Huq, Goward,
  and Nazar]{Adeli2019}
Adeli,~P.; Bazak,~J.~D.; Park,~K.~H.; Kochetkov,~I.; Huq,~A.; Goward,~G.~R.;
  Nazar,~L.~F. Boosting solid-state diffusivity and conductivity in lithium
  superionic argyrodites by halide substitution. \emph{Angew. Chem., Int. Ed.}
  \textbf{2019}, \emph{58}, 8681--8686\relax
\mciteBstWouldAddEndPuncttrue
\mciteSetBstMidEndSepPunct{\mcitedefaultmidpunct}
{\mcitedefaultendpunct}{\mcitedefaultseppunct}\relax
\EndOfBibitem
\end{mcitethebibliography}
\end{document}